\documentclass{aa}

\usepackage{graphicx}
\usepackage[varg]{txfonts}
\usepackage{natbib}

\usepackage{longtable,lscape}

\bibpunct{(}{)}{;}{a}{}{,}
\begin{document}

\title{Galaxies with Wolf-Rayet signatures in the low-redshift
  Universe}
\subtitle{A survey using the Sloan Digital Sky Survey\thanks{Tables
    3--5 are only available in electronic form at the CDS via
    anonymous ftp to cdsarc.u-strasbg.fr (130.79.128.5) or via
    http://cdsweb.u-strasbg.fr/cgi-bin/qcat?J/A+A/}}

\author{J.\ Brinchmann\inst{1,2}\thanks{jarle@strw.leidenuniv.nl} \and
  D.\ Kunth\inst{3} \and 
  F.\ Durret\inst{3}}
\offprints{J. Brinchmann}

\institute{%
  Leiden Observatory, Leiden
 University, PO Box 9513, 2300 RA Leiden, the Netherlands
  \email{jarle@strw.leidenuniv.nl}
\and 
  Centro de Astrof{\'\i}sica da Universidade do Porto, Rua das Estrelas
  S/N, 4150-762 Porto, Portugal
  \and
  Institut d'Astrophysique de Paris,  UMR7095 CNRS, 
  Universit\'{e} Pierre \& Marie Curie, 
  98 bis Boulevard Arago, F-75014 Paris, France}

\date{Received\ldots; accepted\ldots}

  \abstract
   {The availability of large spectroscopic datasets has opened up
     the possibility of constructing large samples of rare objects in a systematic manner. }
   {The goal of this study is to analyse the properties of galaxies
     showing Wolf-Rayet features in their optical spectrum using
     spectra from the Sloan Digital Sky Survey Release 6. With this
     unprecedentedly large sample we aim to constrain the properties
     of the Wolf-Rayet phase and its impact on the surrounding
     interstellar medium.}
   {We carried out very careful continuum subtraction on all galaxies
     with equivalent widths of H$\beta>2$\AA\ in emission and identify
     Wolf-Rayet features using a mixture of automatic and visual
     classification.  We combined this with spectroscopic and
     photometric information from the SDSS and derive metal abundances
     using a number of methods.}
   {We find a total of 570 galaxies with significant Wolf-Rayet (WR)
     features and a further 1115 potential candidates, several times
     more than even the largest heterogeneously assembled
     catalogues. We discuss in detail the properties of galaxies
     showing Wolf-Rayet features with a focus on their empirical
     properties. We are able to accurately quantify the incidence of
     Wolf-Rayet galaxies with redshift and show that the likelihood of
     otherwise similar galaxies showing Wolf-Rayet features increases
     with increasing metallicity, but that WR features are found in
     galaxies of a wide range in morphology. The large sample allows
     us to show explicitly that there are systematic differences in
     the metal abundances of WR and non-WR galaxies. The most striking
     result is that, below EW(\ensuremath{\mathrm{H}\beta})$=100$\AA,
     Wolf-Rayet galaxies show an elevated N/O relative to non-WR
     galaxies. We interpret this as a rapid enrichment of the ISM from
     WR winds.  We also show that the model predictions for WR
     features strongly disagree with the observations at low
     metallicity; while they do agree quite well with the data at
     solar abundances. We discuss possible reasons for this and show
     that models incorporating binary evolution reproduce the low-metallicity results reasonably well.  Finally we combine the WR
     sample with a sample of galaxies with nebular
     \ensuremath{\mbox{He\,\textsc{ii}\,$\lambda 4686$}} to show that,
     at $12 + \log \mathrm{O/H}< 8$, the main sources of
     \ensuremath{\mbox{He\,\textsc{ii}}} ionising photons appears to
     be O stars, arguing for a less dense stellar wind at these
     metallicities, while at higher abundances WN stars might
     increasingly dominate the ionisation budget.}
   {}


   \maketitle

\section{Introduction}
\label{sec:intro}

The presence of features originating from Wolf-Rayet stars in the
spectra of galaxies provide considerable information on the recent
star formation activity in these systems and can be used to study the
properties of the Wolf-Rayet stars and place strong constraints on the
massive end of the initial mass function (IMF). Wolf-Rayet features in galaxy spectra are interesting because the first Wolf-Rayet stars typically start to appear about $2\times 10^6$
years after a star formation episode and disappear within some
$5\times 10^6$ years. Thus they provide a high-resolution temporal
tracer of the recent star formation activity of a galaxy.

Galaxies containing the signatures of Wolf-Rayet stars have been known
for several decades. Beginning with the first detection of Wolf-Rayet
features in the galaxy He 2-10 by \citet{1976MNRAS.177...91A}, a
number of these galaxies have been reported, some in systematic searches 
(e.g. \citealp{1985A&A...142..411K}),
but mostly serendipitously. In fact the first early reports of the
detection of Wolf-Rayet stellar features in galaxy spectra were
incidental results from studies of starbursting low-metallicity
galaxies, typically blue compact dwarfs, aiming at obtaining spectra
of high S/N for deriving the primordial He abundance
\citep{1983ApJ...273...81K}.
\citet{1981A&AS...44..229K}
realised from their first spectra of
NGC 3125 that the presence of Wolf-Rayet stars provides a powerful
constraint on the recent star formation in a galaxy. Soon after these
discoveries, the term ``WR galaxy'' was introduced by 
\citet{1982ApJ...261...64O} and \citet{1991ApJ...377..115C}.
This term should be used with caution, however, since the distance of
the object and the spatial extension of the observation may by referred to
as ``extragalactic HII regions'' that are typical of the outskirts of
spiral galaxies or quite frequently to the nucleus of a powerful
starburst.  The properties and importance of these phenomena in both
cases are completely different (in terms of metallicity, luminosity,
number of WR stars, etc.). As a consequence the WR galaxy definition
hides a loose concept. WR galaxies are found among a very wide variety of
morphological types ranging from low-mass, blue compact low-metallicity dwarf galaxies to massive spirals, luminous mergers, IRAS
galaxies and even LINERS or Seyfert 2 
(\citealp{1995ApJS...98..477H,1997AJ....114...69H,Contini+01}).

The possibility that WR stars are seen in central cluster galaxies has
been reported by \citet{1995MNRAS.276..947A}.  The use of Wolf-Rayet
galaxies to constrain the recent star formation should be seen in the
context of other techniques and indeed in the context of the other
properties of galaxies.  Despite these drawbacks we shall refer to the
WR galaxy nomenclature for simplicity.  Since the compilation of
\citet{1991ApJ...377..115C} that included 37 objects, the most recent
catalogues are from \citet{1999A&AS..136...35S} and
\citet{2000ApJ...531..776G} with more than 130 objects and the study
using SDSS by \citet{2007ApJ...655..851Z} which includes 174 objects
and is currently the largest survey for WR galaxies, although as we
will see only 101 of these satisfy the criteria to be included into
our sample. 

Observationally most, if not all, WR galaxies are identified whenever
their integrated spectra show a broad
\ensuremath{\mbox{He\,\textsc{ii}\,$\lambda 4686$\AA}} emission
feature that is thought to originate in the stellar winds of these WR
stars. Additional features are also found, for instance
N\,\textsc{iii}\,$\lambda 4640${\AA} and/or
C\,\textsc{iii/iv}\,$\lambda 4650$\AA\ as well as
C\,\textsc{iv}\,$\lambda 5808$\AA\ that was shown to be present in
many cases (\citealp{1999A&A...341..399S}) and originate from WN
and WC stars.  Most recent studies have attempted to reproduce the
number of WR stars responsible for the observed stellar emission
features. The model predictions for this are strongly dependent on
metallicity as stellar winds show a strong metallicity dependence
\citep[e.g.][]{2000A&A...360..227N}.  Simple calculations involving
the \ensuremath{\mathrm{H}\beta}\ emission line combined with the
strength of the so-called WR bump at $\lambda$4686\AA\ already give a
hint (\citealp{1981A&A...101L...5K}), but more refined theoretical
evolutionary models predict that at fixed metallicity the WR/O ratio
strongly varies with the age of the starburst
(\citealp{1991A&AS...88..399M,1994A&A...287..803M,1998ApJ...497..618S}).
It is found that this ratio reaches a maximum of 1 for solar
metallicity but decreases to 0.02 when the metallicity decreases to
Z$_\odot$/50. This latter point remains a concern in the sense that it
does not quite reproduce the observational fact that some low
metallicity starburst galaxies such as I Zw 18  contain a WR population
that largely outnumbers model predictions
(\citealp{1997A&A...326L..17L,1998ApJ...507..199D}), forcing one to envisage other
channels for the WR phase (such as the binary channel hypothesis),
revise the models or question the adopted luminosity of the WCE stars
in metal-poor models (\citealp{2004MNRAS.355..728F}).  On the high
metallicity side, \citet{2002A&A...394..443P} in contrast find too low
a I(WRbump)/I(H$\beta$) ratio compared with model predictions,
suggesting again that WR luminosities are not correctly calculated.
The model predictions have recently been contrasted with observed
broad WR emission features by \cite{2004MNRAS.355..728F} and the
application of existing models to a sample of WR galaxies from the
SDSS was shown to imply significant modifications of the IMF
by~\citet{2007ApJ...655..851Z}, which might equally be taken to
indicate significant problems with the current sample of models. We
return to this issue in Section~\ref{sec:models}.

While interesting in their own rights most of the galaxies under
investigation here show very strong star formation activity with many
showing SFR/M$_*$ similar to that of high redshift Lyman-break
galaxies
\citep[e.g.][]{2001ApJ...546..665S,2003ApJ...592..728S,2003ApJ...588...65S}.
Several of these galaxies are found to show potential Wolf-Rayet
spectral features, in particular the \ensuremath{\mbox{He\,\textsc{ii}\,$\lambda$1640\AA}} feature
\citep[e.g.][]{1991ApJ...377L..73L,1997ApJ...481..673L,2000ApJ...545..712K,2003ApJ...588...65S},
hence the present compilation might be suitable as a local comparison
sample.

In addition there is mounting evidence that the progenitors of long
gamma-ray burst (GRB) events are related to the Wolf-Rayet phase
\citep[e.g.][]{2006ARA&A..44..507W} and \citet{2006A&A...454..103H} were able to show that nearby gamma-ray burst host galaxies often show Wolf-Rayet stars, although these do not necessarily coincide with the location of the gamma-ray bursts.  Since studies of Wolf-Rayet galaxies
allow one to place constraints on the models for massive star
evolution they may also provide insight into the properties of GRB
progenitors. However to achieve this it is essential that the sample
of Wolf-Rayet galaxies has a well characterised selection function and
covers a wide range in metallicity and star-burst age. In previous
studies the emphasis has often been on heterogeneous samples, and even
when the sample selection is clearly stated, the samples have been
selected to contain only systems with high equivalent width Balmer
lines.  Metal-rich galaxies are known to be on average more massive
\citep{2004ApJ...613..898T} so this would bias against metal rich
galaxies. 

In this paper we carry out a survey for Wolf-Rayet features in
galaxies covering a very wide range in properties using the Sloan
Digital Sky Survey Data Release 6 \citep[SDSS
DR6,][]{2008ApJS..175..297A}. We introduce the data we use in
Section~\ref{sec:data} and discuss our pre-selection of candidates
with Wolf-Rayet emission features in
Section~\ref{sec:identifying_features}. Our method to measure
Wolf-Rayet features is outlined in Section~\ref{sec:fitting_features}
where we verify that the distribution of line widths is similar to
that seen in Galactic and Magellanic WR stars.
Section~\ref{sec:overall} describes the overall properties of the
sample and place it in the context of the full SDSS. The abundance of
Wolf-Rayet galaxies with redshift is discussed in
Section~\ref{sec:wr_star_abundance} and their emission line properties
in Section~\ref{sec:diagnostic_diagrams}. We discuss evidence of
large-scale pollution of the host galaxy ISM by WR winds in
Figure~\ref{sec:no_abundance_trends}. The modelling of WR features in
galaxies is discussed in Section~\ref{sec:models}. We conclude in
Section~\ref{sec:discussion}.

\section{Data}
\label{sec:data}

We base our search for Wolf-Rayet galaxies on the Sloan Digital Sky
Survey (SDSS, York et al 2000\nocite{2000AJ....120.1579Y}). The SDSS
uses a dedicated 2.5m telescope \citep{2006AJ....131.2332G} and
obtains five-band images using a drift-scan technique
\citep{1998AJ....116.3040G} and spectra using a double-barred fibre
spectrograph with 3'' fibre apertures and a total of 640 fibres per
plate. The tiling algorithm used is discussed in
\citet{2003AJ....125.2276B}, the photometric system by
\citet{1996AJ....111.1748F} and the photometric calibration is
discussed in \citet{2002AJ....123.2121S} and
\citet{2006AN....327..821T}. The photometric stability is monitored
using a dedicated set-up discussed by \citet{2001AJ....122.2129H} and
objects are detected using the Photo pipeline discussed in Lupton
(2008\nocite{Lupton-Submitted}). The astrometric precision is better than 0.1
arcseconds for the objects considered here
\citep{2003AJ....125.1559P}.


We take as starting point the SDSS Data Release 6
\citep{2008ApJS..175..297A} which contains a total of 1,271,680
spectra. Our analysis method is optimised for normal galaxy spectra so
we limit our sample to spectra that are classified by the SDSS
pipeline to be a galaxy spectrum (\texttt{SPECTROTYPE='GALAXY'}).
This excludes all QSOs and essentially all Type 1 Seyferts. Note also
that we \emph{do} include duplicate observations, a total of 91,825, and will use these
below.  These initial cuts reduce our sample to 796,912 spectra.

%
%
We also require that the equivalent width of \ensuremath{\mathrm{H}\beta}\ in emission is
$>2$\AA. This leads to a total of $N=307,210$ spectra. We do not impose a signal-to-noise (S/N) cut, but 96\% of these spectra
have \ensuremath{\mathrm{H}\beta}\ detected at $\mathrm{S/N}>3$. For the analysis of the ionisation
properties of the galaxies we need to apply a S/N cut and define a
sub-sample where the spectra have $\mathrm{S/N}>3$ in \ensuremath{\mathrm{H}\beta}, [O\,\textsc{iii}]5007, \ensuremath{\mathrm{H}\alpha}\ and
[N\,\textsc{ii}]6584, after the uncertainty estimates are adjusted as discussed
in section~\ref{sec:empirical_adjustment}. This sub-sample contains
224,939 spectra.

The minimum requirement of EW(\ensuremath{\mathrm{H}\beta})$>2$\AA, ensures that almost all
spectra under consideration have significant emission lines present
and/or a weak continuum.  One might worry that this could lead us to
miss a significant number of spectra with Wolf-Rayet features. By
extrapolating the trend of number of spectra with WR features versus
EW(\ensuremath{\mathrm{H}\beta}) shown in Figure~\ref{fig:wr_distrib2} below we estimate that
less than 10 (0.002\%) spectra with Wolf-Rayet features are excluded
by this cut.

While the SDSS database is a unique resource, it is well worth
pointing out that it is not optimally suited for the study of low
redshift galaxies. This is in part because the low wavelength cut-off
of the spectrograph is $\sim 3800$\AA, which means that the important
{[O\,\textsc{ii}]} 3727 line falls outside the spectral range for $z<0.02$. As
pointed out by \citet{2004ApJS..153..429K} this can to some extent be
remedied by using the {[O\,\textsc{ii}]} 7320,7330 quadruplet instead, but this is a
much weaker line. Furthermore, the spectroscopic target selection is
done in the $r$-band and while this ensures a fairly good sampling of
the galaxy population based on their stellar content, it will include
fewer strongly star-forming systems, such as the galaxies showing
Wolf-Rayet features discussed here, than, for example, a $B$-band selection.

Finally, the SDSS spectroscopic target selection is done based on an
analysis of the SDSS images using the \texttt{Photo} pipeline, which
detects objects using a sophisticated segmentation routine. This works
very well in general, but for the strongly star forming systems we are
interested in here, the galaxies might be split into several separate objects by the segmentation process.

A careful re-analysis of the images is beyond the scope of this paper
but is discussed by \citet{2005ApJ...631..208B}. What concerns us is
that the spectra frequently target the brightest H\,\textsc{ii}\
regions in nearby galaxies, often significantly displaced from the galaxy centre and we will return to this issue below.


The spectra of all target galaxies have been re-analysed using the
two-step procedure outlined in Appendix~\ref{sec:pipelining}. This is
based on the pipeline discussed by \citet[][hereafter
T04]{2004ApJ...613..898T} and the resulting fits are then refined to
provide flux and equivalent width measurements for 40 emission lines,
as well as continuum indices and other derived parameters.

\subsection{Empirical adjustment of uncertainty estimates}
\label{sec:empirical_adjustment}

The SDSS pipeline provides an uncertainty estimate for each pixel in a
spectrum including internal error sources such as Poissonian noise
from the sky and detectors.  These uncertainty estimates do, however,
not include external error sources such as uncertainties in the
overall flux calibration and in the continuum subtraction. To
accurately check and adjust the uncertainty estimates one needs
duplicate observations of galaxies.

The SDSS targets some galaxies twice to allow for systematic checks of
this kind, as first exploited by \citet{2003ApJ...584..210G}. The SDSS
DR6 contains 86,156 duplicate observations of galaxies, and we use
these to adjust the standard uncertainty estimates.

To compare two spectra of the same galaxy we calculate:
\begin{equation}
  \label{eq:d_flux}
  \Delta_N  = \frac{f^{\mathrm{first}} - f^{\mathrm{second}}}{%
    \sqrt{\sigma_{\mathrm{first}}^2 + \sigma_{\mathrm{second}}^2}},
\end{equation}
where $f$ stands for the line flux and $\sigma$ represents the associated formal
uncertainty. Assuming Gaussian noise, $\Delta_N$ should be a normal
random variable with a standard deviation of 1 if the error estimates are
accurate. We use the difference between the observed spread and that
expected to determine correction factors to the uncertainty estimates.
The typical corrections to the uncertainty estimates are 20--50\% for
forbidden lines and a factor of 2--2.5 for Balmer lines and for {[O\,\textsc{ii}]}
3727. 


\subsection{Internal dust attenuation}
\label{sec:internal_attenuation}

The spectra are corrected for Galactic attenuation using the
\citet{1998ApJ...500..525S} dust maps before running the fitting
pipeline. To adjust for internal attenuation we use the standard
approach of comparing the ratio
\ensuremath{\mathrm{H}\alpha}/\ensuremath{\mathrm{H}\beta}\ to the
expected Case B value.  We use the Case B value appropriate for the
value of $T_e$ estimated as discussed in the following section.  We
assume a simple, $\tau \propto \lambda^{-1.3}$, dust law as advocated
by \citet{2000ApJ...539..718C}. The details of the dust correction do
not influence the results below, nearly identical results would be
obtained if a fixed
\ensuremath{\mathrm{H}\alpha}/\ensuremath{\mathrm{H}\beta}\ ratio was
assumed.

While this recipe for dust correction is sufficient for 
nebular emission lines, it is not entirely clear whether it is
appropriate to assume the same dust attenuation for the Wolf-Rayet
features as for the nebular lines.  In the following we will assume that the
attenuation at H$\beta$ and the blue bump is the same. This is likely
to give an upper limit to the attenuation of the Wolf-Rayet bump.

\subsection{Abundance measurements}
\label{sec:oxygen_abundance}

Since our sample spans a wider range in galaxy properties than is
normally encountered in Wolf-Rayet galaxy studies it is more
complicated to estimate element abundances in a uniform way for all
galaxies. At low metallicity it is common to deduce the electron
temperature, $T_e$, from the
{[O\,\textsc{iii}]4363}/{[O\,\textsc{iii}]4959,5007} ratio and derive
the oxygen abundance from the {[O\,\textsc{ii}]}\ and
{[O\,\textsc{iii}]}\ lines --- the so-called ``direct'' method or
$T_e$ method. At higher metallicity one might use one of several
``strong-line'' methods which are calibrated on photoionisation models
\citep[e.g.][]{1989agna.book.....O,2001MNRAS.323..887C,2002ApJS..142...35K}.
However it is now well-known that there is a systematic offset between
strong-line method and the $T_e$ estimates of oxygen abundance
\citep[e.g.][]{2000ApJ...537..589K,2003ApJ...591..801K,2000A&A...362..325P,2007ApJ...656..186B}.
Thus there is no generally accepted method to estimate oxygen
abundances for all abundance regimes and significant differences can
be found between methods as discussed in detail by
\citet{2008arXiv0801.1849K}.

We therefore tabulate a number of different abundance estimators for
our objects to allow us to check how our results depend on the chosen
estimator, we have adopted one representative of the three main
classes of estimators, the direct method, the strong-line method and
the empirically calibrated estimators.  For all low-metallicity
objects with detected {[O\,\textsc{iii}]\,$\lambda 4363$\AA}, we use the $T_e$
method based on the formulae presented by \citet{2006A&A...448..955I}
to estimate element abundances. For all galaxies we also calculate
oxygen abundances using the Bayesian method discussed in T04 and B04;
we will refer to these as CL01 abundances (a comparison with other
methods is shown in T04). We also estimate oxygen abundances using the
N2 method proposed by \citet[hereafter PP]{2004MNRAS.348L..59P} and
will refer to these as PP abundances in the following.

We have also verified that the results below do not change
significantly if the PP O3N2 estimator, the
{[N\,\textsc{ii}]}/{[O\,\textsc{ii}]}\ estimator advocated by
\citet{2002ApJS..142...35K} or the P-method devised by
\citet{2000A&A...362..325P} are used instead, although slight changes
do occur.
 
Unless otherwise stated we will in the following adopt $T_e$
abundances for low-metallicity systems and CL01 abundances for the
rest as a reference estimate. Specifically we adopt $T_e$ estimators
when they give O/H$<8.3$ and the S/N in {[O\,\textsc{iii}]4363} is
$>5$. Otherwise we assign the CL01 oxygen abundances adjusted by
subtracting off a linearly varying offset which is $-0.05$ at $12 + \log
\mathrm{O/H}=8.2$ and 0.3 dex at $12 + \log \mathrm{O/H}=8.5$.  We
will refer to this as the \emph{mixed} abundance estimator.

We arrived at this simple correction by comparing CL01 abundances for
the H\,\textsc{ii}\ regions in \citet{2003ApJ...591..801K} with $T_e$
abundances, requiring a continuous mass-metallicity relation and also
taking into account the offsets seen in the comparison with $T_e$
estimators by \citet{2007A&A...462..535Y}.

\section{Identifying Wolf-Rayet features}
\label{sec:identifying_features}

The main Wolf-Rayet features seen in the optical spectra of galaxies
are two broad emission features: the blue bump around
4600--4680\textrm{\AA}\ and the red bump around
5650--5800\textrm{\AA}. As mentioned in the introduction, the blue
bump is primarily composed of lines from N\,\textsc{iii},
N\,\textsc{v}, C\,\textsc{iii/iv}\ and a broad
\ensuremath{\mbox{He\,\textsc{ii}}}\ emission line at 4686\AA.

These broad features do vary significantly in appearance, reflecting
the strong variation in the relative emission line strength and
velocity broadening seen in individual Wolf-Rayet stars. As a
consequence the accurate measurement of the flux in WR bumps requires
some care. Furthermore a number of nebular emission lines are often
superposed on the Wolf-Rayet features making the disentangling of the
individual fluxes rather challenging.

To identify candidate Wolf-Rayet galaxies we calculate the excess flux
above the best-fit continuum in regions around the main Wolf-Rayet
features.  The features are typically weak and depend sensitively on
the continuum estimation so a visual inspection is a necessary final
step in this procedure to deal with false positives.  However, it is
necessary to have a semi-automated method to identify candidate
Wolf-Rayet features in large, $N>10^5$, spectroscopic surveys.

Our approach is to first estimate the continuum doing a non-negative
least-squares combination of single burst template spectra from the
\citet[][, BC03]{2003MNRAS.344.1000B} library. We use a special
version provided by Bruzual \& Charlot (priv. comm.), updated with
empirical stellar spectra from the MILES spectral library
\citep{2006MNRAS.371..703S}. The method is essentially the same as
that used in T04 and B04 but no smooth component is added to the
spectral fit as this would often obliterate any signs of a Wolf-Rayet
component. We note that this procedure is sensitive to the library of
stellar spectra used. In particular the BC03 models with the STELIB
library (\citep{2003A&A...402..433L}) do show residual calibration
features around \ensuremath{\mathrm{H}\beta}\ (also noted by
\citet{2007MNRAS.381..263A}) which serve to introduce features close
to the expected location of the WR bumps and hence severely increase
the number of false positives).

From the continuum subtracted spectrum we define the excess in the
blue WR feature around \ensuremath{\mbox{He\,\textsc{ii}4686}} (the blue bump) as 
\begin{equation}
  \label{eq:heii_excess}
  \ensuremath{e_{\mbox{blue}}} = F_{\textrm{He\,\textsc{ii}}} - F_{\textrm{He\,\textsc{ii} continuum}},
\end{equation}
where $F$ is the summed flux in windows around the blue bump, corrected for nebular
emission. We do a similar calculation for the red bump but have not
used this for ranking.

\begin{table}[tbp]
  \centering
  \caption{The WR filters used for spectrophotometry in this work}
  \begin{tabular}{lccc}\hline
\multicolumn{1}{c}{Name} & \multicolumn{1}{c}{Central wavelength
  (\textrm{\AA}) } & \multicolumn{1}{c}{Width (\textrm{\AA})}\\ \hline\hline
He\,\textsc{ii}           & 4705 & 100 \\
He\,\textsc{ii} continuum & 4517 & 50 \\
                & 4785 & 50 \\
C\,\textsc{iv}            & 5810 & 100 \\
C\,\textsc{iv} continuum  & 5675 & 50 \\
                & 5945 & 50 \\ \hline
  \end{tabular}
  \label{tab:filters}
\end{table}

\begin{figure}
  \centering
  \caption{The filter response functions adopted in this work.}
  \includegraphics[angle=90,width=88mm]{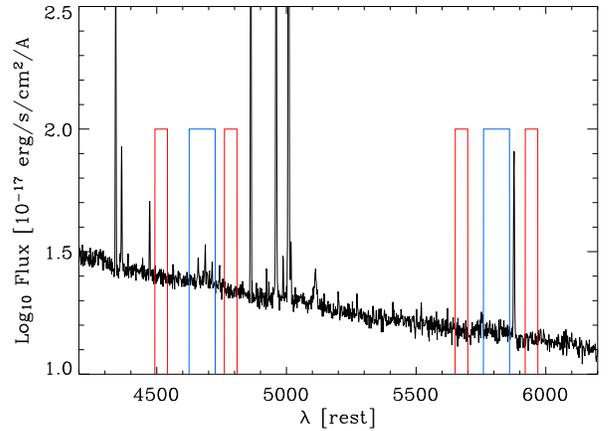}
  \label{fig:filters}
\end{figure}

The filters are square and are summarised in Table~\ref{tab:filters}
and illustrated in Figure~\ref{fig:filters}. The red lines show the
two continuum filters and the blue the ``bump'' filters. Note that the
filter measuring the continuum consists of two parts to minimise the
effect of a non-flat continuum. To determine the optimal filter
functions we cross-correlated the \citet[][hereafter
S99]{1999A&AS..136...35S} catalogue with the SDSS DR6 and used those
galaxies where the SDSS spectrum showed clear Wolf-Rayet features to
determine what filter functions give us the best discrimination
between WR and non-WR galaxies. For each filter we calculate $F$ by
summing up the flux within the filter range of the spectrum after
subtracting the continuum and nebular emission lines. We estimate the
uncertainty on this flux by adding the uncertainty estimates of the
spectrum in quadrature.

This results in a list of galaxies with possible Wolf-Rayet features
and is similar to the approach used to identify Wolf-Rayet stars in
imaging surveys
\citep[e.g.][]{1985ApJ...291..685A,2004A&A...419L..17C}. We emphasise
that this approach is only used to find \emph{candidate} WR galaxies.

Due to the large parent sample, the number of galaxies with elevated
$\ensuremath{e_{\mbox{blue}}}$ due to noise, residual sky lines, instrumental failures and
poor continuum fitting is comparable to or even higher than the number
of genuine WR galaxies. As a consequence we have therefore carefully
examined 11,241 spectra sorted according to $\ensuremath{e_{\mbox{blue}}}$. These were
assigned to four basic classes based on the appearance of the
\emph{blue} bump:
\begin{description}
  \item[Class 3] Very clear Wolf-Rayet features. Typically a broad
   component to \ensuremath{\mbox{He\,\textsc{ii}\,$\lambda 4686$}} as well as N\,\textsc{iii}\,$\lambda 4640$ are
   seen. 
  \item[Class 2] Convincing Wolf-Rayet features seen, but either noisier
   or not obvious before continuum subtraction. 
 \item[Class 1] Possible Wolf-Rayet features but generally too noisy
   or too dependent on the continuum subtraction to be useable.
   Galaxies that show a single, apparently broad, \ensuremath{\mbox{He\,\textsc{ii}}}{\,$\lambda
     4686$} line but no further Wolf-Rayet features are normally
   assigned to this class unless a clear identification of a broad
   component can be done.
  \item[Class 0] No Wolf-Rayet features seen. 
\end{description}
In the following we will consider Classes 2 and 3 as being galaxies
with Wolf-Rayet features and except for Section~\ref{sec:overall}
below we will not distinguish between these two classes.

There are two crucial issues that must be taken into account when
classifying the spectra visually. Firstly the sensitivity of the
Wolf-Rayet features to the continuum fitting must be assessed for weak
features. We do this by redoing the continuum fit using different
wavelength ranges --- typically contrasting a fit using
3800\AA--7000\AA\ with one using only the spectrum between 4000\AA\ 
and 5000\AA. If a feature changes significantly between these two
continuum fits it is normally assigned Class 0. The other problematic
issue has to do with nebular emission. This is discussed in detail in
the following section, but for the visual classification the main
problem is when a nebular \ensuremath{\mbox{He\,\textsc{ii}\,$\lambda 4686$}} line is superposed
on a broader line. The SDSS spectra often do not have sufficiently
high S/N to disentangle these two components, and unless other
supporting features such as N\,\textsc{iii}\,$\lambda 4640$ can be seen, the
galaxy will be assigned to either Class 1 or Class 0, depending on the
strength of the broad component as judged by the rigourous fits
described below.

\section{Fitting Wolf-Rayet features}
\label{sec:fitting_features}

The existing literature on Wolf-Rayet galaxies usually estimates the
bump luminosity by fitting a single Gaussian to the spectrum after
removal of emission lines, either subtracted or masked
\citep[e.g.][]{2000ApJ...531..776G}.  This has been an
acceptable approach because the resolution of the spectra have often
been fairly low. With the SDSS spectra it is clear that this
is not always a good solution as a single Gaussian often provides a
poor fit to the overall detected features. This was also realised
by~\citet{2007ApJ...655..851Z} in their study of WR galaxies in the
SDSS and they therefore calculated the flux integrating all the flux
between 4600{\AA} and 4750{\AA} after correction for nebular
emission. This does remove all distinction of WN and WC features
however, and it is not an optimal approach in terms of S/N since
regions with no WR features are included in the calculation. Finally, the
subtraction of nebular lines cannot be viewed as a separate problem
from fitting the broad features as there are significant degeneracies
between these.

We have therefore adopted a more rigourous approach: we fit the
relevant nebular emission lines jointly with Gaussians for each
feature which can be expected to be present based on the observations
of WR stars in the local Universe. In the blue we fit:
\begin{enumerate}
\item Nebular lines: [Fe\,\textsc{iii}]\,$\lambda 4559$, [Fe\,\textsc{iii}]\,$\lambda 4669$, \ensuremath{\mbox{He\,\textsc{ii}\,$\lambda 4686$}}, [Fe\,\textsc{iii}]\,$\lambda 4702$, [Ar\,\textsc{iv}]\,$\lambda 4711$, He\,\textsc{i}\,$\lambda 4714$. The line width is fixed to that
  determined from the joint fit to the strong emission lines in the
  spectrum. The [Ar\,\textsc{iv}]\,$\lambda 4740$ line is far enough
  from the features of interest that we do not need to include it
  in the fit.
\item Wolf-Rayet features: The blue bump is primarily composed of N\,\textsc{v},
  N\,\textsc{iii}, C\,\textsc{iii/iv}\ blends as well as the \ensuremath{\mbox{He\,\textsc{ii}\,$\lambda 4686$}} line. The N\,\textsc{v}\
  doublet is fit as a single feature centred at 4610\AA. The N\,\textsc{iii}\
  feature is centred around 4640\AA, whereas the C\,\textsc{iii/iv}\ feature is
  assumed to be centred at 4650\AA\ and a broad component to \ensuremath{\mbox{He\,\textsc{ii}\,$\lambda 4686$}}
  is also included in the fit.
\end{enumerate}

\begin{figure}
  \centering
  \includegraphics[angle=90,width=88mm]{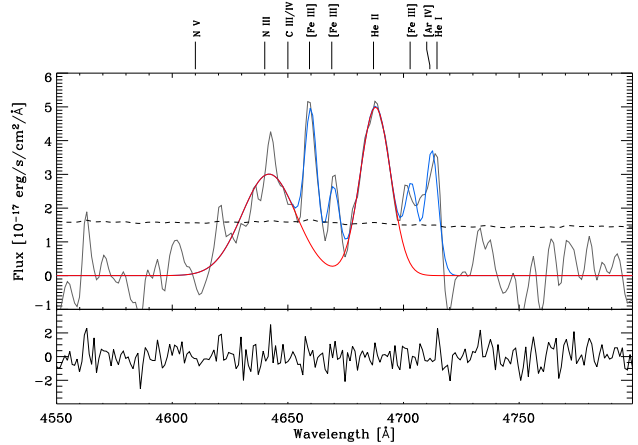}
  \caption{An example of a fit to the blue bump for a WR galaxy. The
    solid black line shows the continuum subtracted spectrum
    smoothed with a 3\AA\ Gaussian. The dashed black line shows
    the noise level assuming uncorrelated errors in the pixels. The
    red line shows the fit to the blue bump features. In this
    particular fit there is a N\,\textsc{iii}\ component as well as a broad
    component to \ensuremath{\mbox{He\,\textsc{ii}}}. The blue line shows the narrow emission lines
    fitted in this spectrum. The bottom panel shows the residual after
    subtracting the best fit model from the spectrum, normalised by the
    uncertainty estimate in each pixel. This residual spectrum is
    consistent with pure Gaussian noise.}
  \label{fig:bbump}
\end{figure}

An example of a blue bump fit is shown in Figure~\ref{fig:bbump}. The
red line traces the WR features fit, whereas the blue line shows the
nebular emission lines that must be included in the fit. When working with this sample it became evident that
there were several [Fe\,\textsc{iii}]\ lines that normally are not taken into
account, particularly at 4669\AA\ and 4702\AA\ which will bias any
fits to the bump features high if not removed. It is also obvious from this figure
that a single Gaussian is a poor approximation to the bump shape.

In the red there are fewer nebular emission lines superposed on the broad WR features
and we fit:
\begin{enumerate}
\item Nebular emission lines: {[N\,\textsc{ii}]\,$\lambda 5755$} and He\,\textsc{i}\,$\lambda 5876$.
\item Wolf-Rayet features: one feature centred at 5696\AA\ is fit as
  well as a doublet at 5803, 5815\AA. 
\end{enumerate}

When carrying out the fit, the line fluxes are constrained to be
non-negative. This might cause a slight bias in the derived
fluxes at very low flux values, but was found to be required to ensure physically meaningful fits. 
We furthermore limit the overall wavelength shift of the blue and
red features to be $\left|\Delta \lambda\right|<3$\AA. This has no
influence on the results.

However we also found that we needed to provide an upper limit to the
width of the Wolf-Rayet features. This upper limit does affect the
\emph{relative} contributions of the different features to the overall
bump although the integrated flux is not significantly changed.  We
have generally adopted 20\AA\ (FWHM $\sim 47$\AA) to ensure an
adequate fit. This corresponds to $\sigma \sim 1280$\,km/s ($\sim
3000$\,km/s FWHM) and we have found this to be a reasonable default
upper limit to the width of the individual Wolf-Rayet features for all
our galaxies. For reference only 2 stars in the compilation of
\citet[][]{2006MNRAS.368.1822H} exceed this velocity width.  All fits
were visually inspected and for 58 we redid the fit manually to better
reproduce the shape of the bump. Figure~\ref{fig:spectrum_samples}
shows a collection of spectra ordered by increasing
EW(\ensuremath{\mathrm{H}\beta}). It is noticeable that the bumps show
a range of morphologies reflecting the broad mix of WN and WC stars
seen in these spectra.

\begin{figure}
  \centering
  \includegraphics[angle=90,width=88mm]{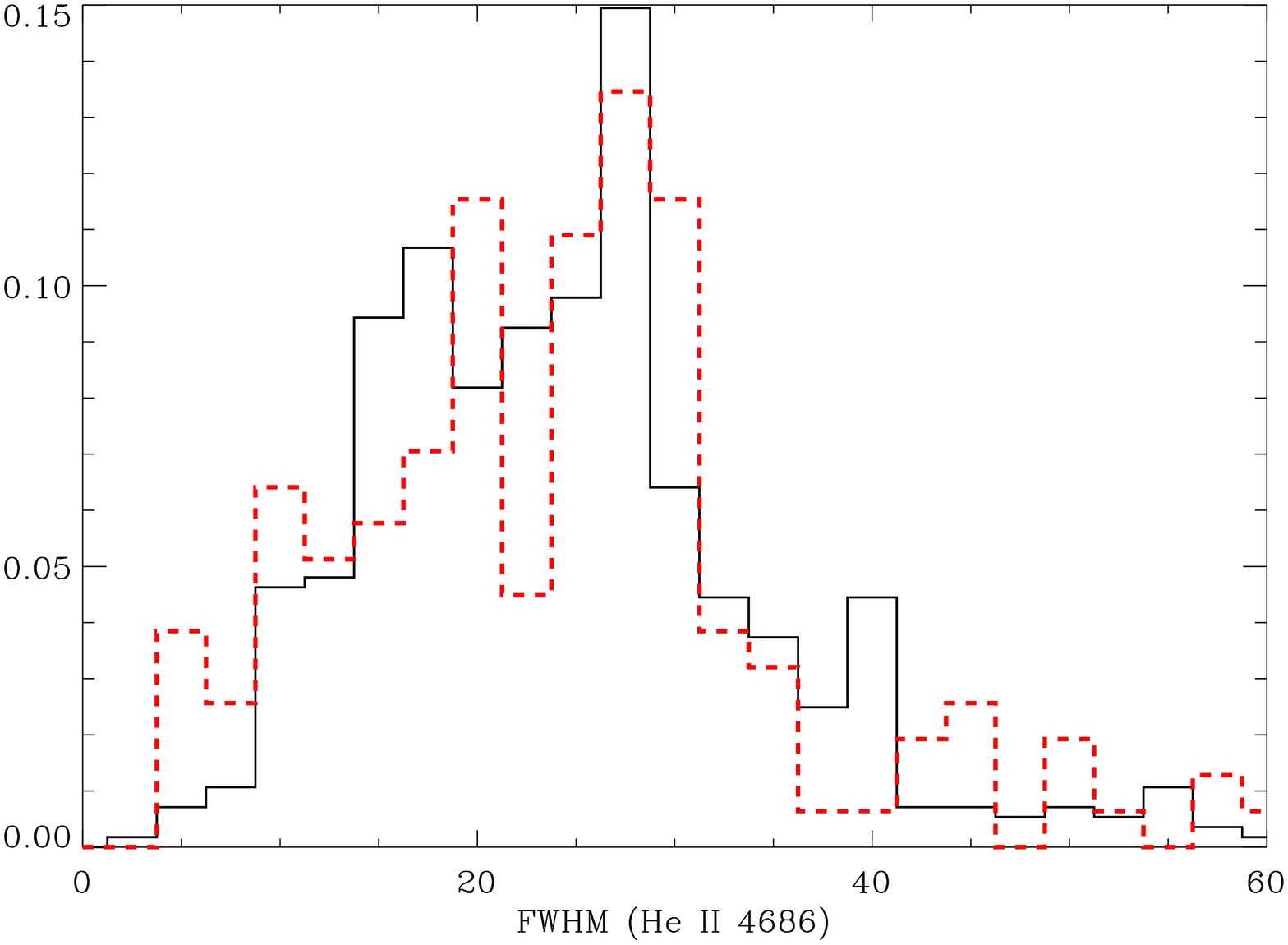}
  \caption{The distribution of widths of the \ensuremath{\mbox{He\,\textsc{ii}\,$\lambda 4686$}} line measured
    in Galactic, LMC and SMC WR stars (dashed line) compared to the
    widths determined by the fit to our Wolf-Rayet candidate galaxies
    (solid line).} 
  \label{fig:width_comparison}
\end{figure}

\begin{figure*}
 \centering
 \includegraphics[width=160mm]{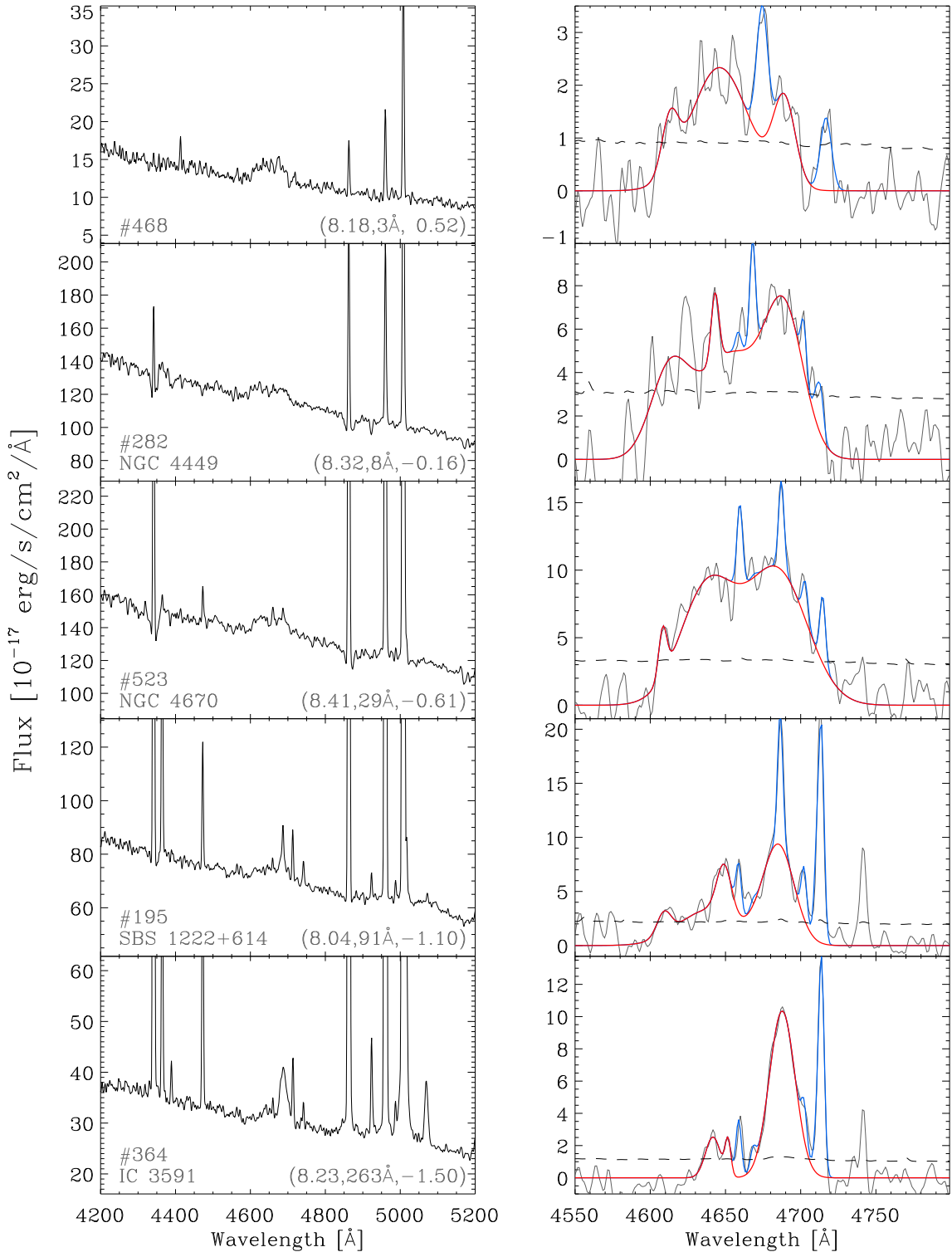}
 \caption{A montage of five WR galaxy spectra ordered by
   EW(\ensuremath{\mathrm{H}\beta}) increasing by a factor of 2
   downwards. The left column shows the full spectrum zoomed in around
   the blue bump and the right column the continuum subtracted
   spectrum similarly to that shown in Figure~\ref{fig:bbump}. The
   number of the spectrum in our catalogue as well as any common names
   are indicated in the lower left corner of the full spectrum
   plot. The three numbers in the parenthesis in the lower right
   corner show 12+Log O/H, EW(H$\beta$) and Log L(Blue bump)/L(H$\beta$).
 } 
 \label{fig:spectrum_samples}
\end{figure*}

We can test these fits in a model independent manner by
realising that the intrinsic line widths of the Wolf-Rayet features are
considerably larger than those caused by galactic rotation, so we can
reasonably assume that the distribution of the FWHM of the \ensuremath{\mbox{He\,\textsc{ii}\,$\lambda 4686$}} line
measured by our fits should follow closely the distribution found for
Galactic, LMC and SMC Wolf-Rayet stars\footnote{These three samples of WR stars sample reasonably well the overall metallicity distribution of our WR galaxies so this comparison should be meaningful.}. To do this we combined the
measurements for Galactic and LMC WN stars in
\citet{1996MNRAS.281..163S} and for the LMC and SMC in
\citet{2006A&A...449..711C}. Where two measurements of the same star
existed we took a straight average.

The results of this comparison are shown in
Figure~\ref{fig:width_comparison}.  Clearly there is very good
agreement between the two distributions and this can be taken as a
reasonable consistency check. It also indicates that we can make use
of the width of the \ensuremath{\mbox{He\,\textsc{ii}}}\ feature to study the variation of the width
of WR features with galaxy properties.

Note that we do seem to have fewer narrow \ensuremath{\mbox{He\,\textsc{ii}}}\ features than seen in
the stars. This is to be expected, both because the narrow features
are more difficult to disentangle from the nebular lines, and because
when averaged over a population the lines tend to be broadened due to
peculiar motions and because on average the WR stars with broad lines appear to be more luminous (CH06).

To ascertain the reliability of the deblending of the individual
features in the blue we carried out a number of simulations where we
added noise to a set of realistic simulated spectra with Wolf-Rayet
features superposed. This showed that the fits have significant
degeneracies between the individual features, in particular between
the N\,\textsc{iii}\ and C\,\textsc{iii/iv}\ features which are closely spaced, but also
with the superposed nebular lines.  
 For the
purposes of this paper we will mostly make use of the total blue and
red bump fluxes rather than the individual features. This does limit our comparisons to models somewhat but it will turn out to be sufficient for the purposes of this paper.

\begin{figure}
  \centering
  \includegraphics[width=88mm]{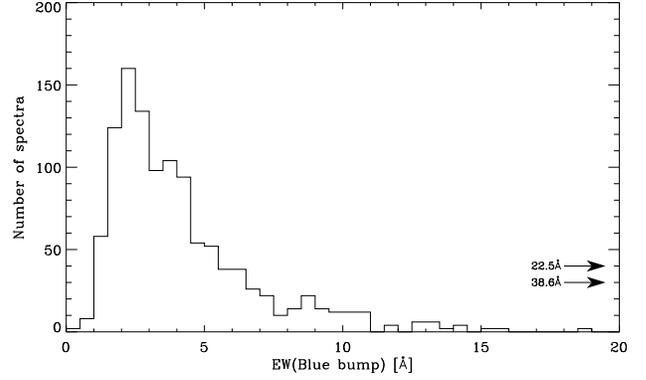}
  \caption{The distribution of the equivalent width of the blue bump measured four our Class 2 and Class 3 spectra. Note the strong drop at $\mathrm{EW}<1$\AA. Two spectra fall outside the plotted region as indicated by arrows.} 
  \label{fig:ew_distrib}
\end{figure}

The procedure adopted for classification of WR galaxies has not made
any explicit cuts on the equivalent width or S/N of the bump features.
Figure~\ref{fig:ew_distrib} shows the resulting distribution of the
equivalent width of the blue bump. This shows that the spectra have an
effective minimum EW of 1\AA, and this is also constant with redshift.
It is therefore reasonable to take this as the \textit{a posteriori}
completeness limit for our sample.

\section{Overall sample properties}
\label{sec:overall}

\subsection{Spectral classification and overall inventory}
\label{sec:classifications}

We follow B04 in separating the galaxies in our sample based on their
location in the \citet[][BPT]{1981PASP...93....5B} ionisation diagram.
We will refer to galaxies whose emission line spectrum is dominated by
star formation as the SF class, and to those dominated by an active
galactic nucleus (AGN) as the AGN class. Galaxies that fall between
these two classes are assigned to the Composite class, see B04 for
details.

As discussed in B04 \citep[see also e.g.
T04,][]{2003MNRAS.346.1055K,2006MNRAS.372..961K} the reliable
classification of emission line galaxies requires a S/N in each of the lines \ensuremath{\mathrm{H}\beta},
{[O\,\textsc{iii}]\,$\lambda 5007$}, \ensuremath{\mathrm{H}\alpha}\ \&
{[N\,\textsc{ii}]\,$\lambda 6584$} of at least $S/N=3$. For a few
galaxies the {[O\,\textsc{iii}]\,$\lambda 5007$} line is truncated by
the SDSS pipeline reductions; for these {[O\,\textsc{iii}]\,$\lambda
  4959$} is used instead. Finally a few galaxies have no
{[N\,\textsc{ii}]\,$\lambda 6584$} detected at $\mathrm{S/N}>3$ and
for classification purposes and abundance determinations we assign
these the 3$\sigma$ upper limit to the line flux. After these steps a
total of 9,500 (3.1\%) galaxies cannot be classified using emission
line diagnostic and are placed in the Unclassified category. For
galaxies with $\mathrm{EW}(\ensuremath{\mathrm{H}\beta})>5$\,\AA, this
is entirely due to instrumental problems and we have verified that
this does not lead to any systematic biases between the parent sample
and the WR galaxy sample. For the galaxies with weaker
\ensuremath{\mathrm{H}\beta}\ there will be a trend that the fraction
of unclassifiable galaxies increases from $\sim 4\%$ at
$\mathrm{EW}(\ensuremath{\mathrm{H}\beta})=7$\,\AA\ to $\sim 50\%$ at
$\mathrm{EW}(\ensuremath{\mathrm{H}\beta})=2$\,\AA. Hence any
comparisons of the abundance of WR galaxies with the parent population
at low equivalent widths must take this into account if appropriate.

\begin{table}
  \centering
  \caption{The summary of the Wolf-Rayet sample.}
  \begin{tabular}{lrr}
\hline
Sub-sample    & Number   & Number  \\
             & All $z$  & $0.005 < z < 0.22$ \\ \hline\hline
     Class 2+3 &  570 &  452\\
        Class 2 &  373 &  317\\
        Class 3 &  197 &  135\\
        Class 1 & 1115 & 1058\\
   Star-forming &  453 &  346\\
            AGN &   42 &   38\\
      Composite &   68 &   67\\
   Unclassified &    7 &    1\\
\hline\hline
      \multicolumn{3}{c}{With \texttt{SCIENCEPRIMARY=1}}  \\ \hline\hline
      Class 2+3 &  531 &  419\\
        Class 2 &  344 &  293\\
        Class 3 &  187 &  126\\
        Class 1 &  972 &  919\\
   Star-forming &  422 &  321\\
            AGN &   38 &   34\\
      Composite &   64 &   63\\
   Unclassified &    7 &    1\\
\hline
  \end{tabular}
  \label{tab:summary_of_sample}
\end{table}

The final sample is summarised in Table~\ref{tab:summary_of_sample}
and given in full Table~\ref{tab:sample_id} which is available online.
The luminosities and equivalent widths of the blue and red bumps as
well as the width of the Gaussian fit to the He\,\textsc{ii}\,$\lambda
4686$ line are given in Table~\ref{tab:sample_bumps} and the
abundances derived from the spectra in Table~\ref{tab:sample_abund}.
The full sample consists of 570 spectra from the SDSS. This is more
than twice the number of all previously known Wolf-Rayet galaxies, and
a factor of $\sim 3$ more than the largest rigourously selected
sample of WR galaxies from a single study, surpassing the 174 galaxies
in the Zhang et al (2007) study. We should comment that in
fact only 101 of the objects from that survey are classified as
class 2 or 3 by us and a further 36 as class 1.  For the final 37 we
are unable to reproduce the identification of broad Wolf-Rayet
features by Zhang et al.

\onllongtabL{3}{%
\begin{landscape}
\begin{longtable}{rrcrrrrrrl}
  \caption{\label{tab:sample_id} The positions and identification of the%
 sample of galaxies with Wolf-Rayet features.}\\
  \hline\hline
  Running ID                &
  \multicolumn{1}{c}{Z} &
  SciPri\footnote{The SDSS \texttt{SCIENCEPRIMARY} parameter} &
  \multicolumn{2}{c}{$\alpha$ (J2000) $\delta$}            &
  WR          &
  BPT          &
  \multicolumn{1}{c}{Spectrum ID}         &
  Index\footnote{Zero-based index in the tables provided at the MPA-JHU site (\texttt{http://www.mpa-garching.mpg.de/SDSS}.)}               &
  \multicolumn{1}{c}{Other names} 
\\
&   & & &
                      &
                      &
                      &
\multicolumn{1}{c}{Plate--MJD--FiberID} &
&
 \\
\hline
\endfirsthead
\caption{continued.}\\
\hline\hline
 Running ID                &
  \multicolumn{1}{c}{Z} &
  SciPri\footnote{The SDSS \texttt{SCIENCEPRIMARY} parameter} &
  \multicolumn{2}{c}{$\alpha$ (J2000) $\delta$}            &
  WR          &
  BPT          &
  \multicolumn{1}{c}{Spectrum ID}         &
  Index\footnote{Zero-based index in the tables provided at the MPA-JHU site (\texttt{http://www.mpa-garching.mpg.de/SDSS}.)}               &
  \multicolumn{1}{c}{Other names}
 \\
& & & &
                      &
                      &
                      &
  \multicolumn{1}{c}{Plate--MJD--FiberID} &
&
 \\
\hline
\endhead
\hline
\endfoot
WR000 &  0.05160 & 0 & +09:43:33.90 & +01:06:59.48 &  2 &        SF &    0266-51630-407 &    315 &                                 SHOC 269 \\
WR001 &  0.00633 & 1 & +09:49:54.14 & +00:36:58.65 &  3 &        SF &    0267-51608-384 &    817 &                                 NGC 3023 \\
WR002 &  0.04478 & 1 & +10:09:03.22 & -00:13:37.53 &  3 &        SF &    0270-51909-265 &   2308 &                                   \ldots \\
WR003 &  0.02149 & 1 & +10:21:53.20 & +00:17:43.22 &  2 &        SF &    0271-51883-583 &   3083 &                             CGCG 009-008 \\
WR004 &  0.02148 & 1 & +10:26:21.13 & +00:41:05.84 &  2 &      Comp &    0272-51941-542 &   3557 &                             CGCG 009-034 \\
WR005 &  0.02836 & 1 & +10:31:45.72 & +00:23:16.61 &  3 &        SF &    0273-51957-464 &   3963 &                                   \ldots \\
WR006 &  0.01855 & 1 & +10:39:24.38 & -00:23:21.43 &  3 &        SF &    0274-51913-187 &   4248 &                                  IC 0633 \\
WR007 &  0.03763 & 1 & +10:50:31.24 & -00:47:51.62 &  3 &        SF &    0275-51910-015 &   4643 &                                  HN 1278 \\
\end{longtable}
\end{landscape}
}

\onllongtab{4}{%
\begin{landscape}
\begin{longtable}{lrrrrrrr}
  \caption{\label{tab:sample_bumps} The luminosities and equivalent
    widths of the Wolf-Rayet features in the spectra.}\\
  \hline\hline
  Running ID                &
  \multicolumn{1}{c}{$\log L_{BB}^\mathrm{obs}$} &
  \multicolumn{1}{c}{$\log L_{RB}^\mathrm{obs}$} &
  \multicolumn{1}{c}{$\log L_{BB}^\mathrm{dered}$} &
  \multicolumn{1}{c}{$\log L_{RB}^\mathrm{dered}$} &
  \multicolumn{1}{c}{EW(BB)} &
  \multicolumn{1}{c}{EW(RB)} &
  $\sigma$(He\,\textsc{ii}\,$\lambda 4686$)\footnote{Values in parenthesis indicate where the width are forced by the upper limit to the width adopted.} \\
 & (log $10^{40}$\,erg/s) & (log $10^{40}$\,erg/s) &
 (log $10^{40}$\,erg/s) & (log $10^{40}$\,erg/s)&
\multicolumn{1}{c}{(\AA)} & \multicolumn{1}{c}{(\AA)} &
 \multicolumn{1}{c}{(\AA)} \\
\hline

\endfirsthead
\caption{continued.}\\
\hline\hline
  Running ID                &
  \multicolumn{1}{c}{$\log L_{BB}^\mathrm{obs}$} &
  \multicolumn{1}{c}{$\log L_{RB}^\mathrm{obs}$} &
  \multicolumn{1}{c}{$\log L_{BB}^\mathrm{dered}$} &
  \multicolumn{1}{c}{$\log L_{RB}^\mathrm{dered}$} &
  \multicolumn{1}{c}{EW(BB)} &
  \multicolumn{1}{c}{EW(RB)} &
  $\sigma$(He\,\textsc{ii}\,$\lambda 4686$)\footnote{Values in parenthesis indicate where the width are forced by the upper limit to the width adopted.} \\
 & (log $10^{40}$\,erg/s) & (log $10^{40}$\,erg/s) &
 (log $10^{40}$\,erg/s) & (log $10^{40}$\,erg/s)&
\multicolumn{1}{c}{(\AA)} & \multicolumn{1}{c}{(\AA)} &
 \multicolumn{1}{c}{(\AA)} \\
\hline
\endhead
\hline
\endfoot
WR000 & $ 3.65 \pm  0.99$ & \ldots & $ 3.88 \pm  0.99$ & \ldots & $ 2.64 \pm  0.20$ & \ldots & $10.15 \pm  5.93$  \\
WR001 & $ 0.30 \pm  0.91$ & $-0.17 \pm  0.81$ & $ 0.34 \pm  0.91$ & $-0.14 \pm  0.81$ & $ 2.29 \pm  0.24$ & $ 1.34 \pm  0.18$ & $10.42 \pm  1.95$  \\
WR002 & $ 3.88 \pm  0.49$ & $ 3.35 \pm  0.32$ & $ 4.39 \pm  0.49$ & $ 3.74 \pm  0.32$ & $ 5.79 \pm  0.46$ & $ 2.06 \pm  0.25$ & $ 3.66 \pm  0.85$  \\
WR003 & $ 2.72 \pm  0.48$ & \ldots & $ 3.14 \pm  0.48$ & \ldots & $ 2.97 \pm  0.25$ & \ldots & $ 8.33 \pm  3.44$  \\
WR004 & $ 2.40 \pm  0.79$ & \ldots & $ 3.01 \pm  0.79$ & \ldots & $ 1.69 \pm  0.24$ & \ldots & $ 7.84 \pm  2.37$  \\
WR005 & $ 3.12 \pm  0.42$ & \ldots & $ 3.34 \pm  0.42$ & \ldots & $ 2.68 \pm  0.30$ & \ldots & \multicolumn{1}{r}{$(15)$}  \\
WR006 & $ 2.17 \pm  0.52$ & \ldots & $ 2.30 \pm  0.52$ & \ldots & $ 2.96 \pm  0.19$ & \ldots & $ 7.24 \pm  1.40$  \\
WR007 & $ 3.50 \pm  0.54$ & \ldots & $ 3.80 \pm  0.54$ & \ldots & $ 3.36 \pm  0.29$ & \ldots & $ 9.90 \pm  3.81$  \\
\end{longtable}
\end{landscape}
}

\onllongtabL{5}{%
\begin{landscape}
\begin{longtable}{rrrrrrrl}
  \caption{\label{tab:sample_abund} The gas-phase abundance estimates
    for the star-forming Wolf-Rayet galaxies.}\\
  \hline\hline
  Running ID                &
  \multicolumn{4}{c}{12 + Log O/H} &
  12 + Log N/H &
  12 + Log Ar/H &
  12 + Log Ne/H \\
&
   Mixed & $T_e$ & PP & CL01 & & & \\[7pt]
\hline
\endfirsthead
\caption{continued.}\\
\hline\hline
  Running ID                &
  \multicolumn{4}{c}{12 + Log O/H} &
  12 + Log N/H &
  12 + Log Ar/H &
  12 + Log Ne/H \\
&
   Mixed & $T_e$ & PP & CL01 & & & \\[7pt]
\hline
\endhead
\hline
\endfoot
WR000 & $8.35 \pm 0.04$ & \ldots & $8.38 \pm 0.19$ & $8.65 \pm 0.04$ & \ldots & \ldots & \ldots \\
WR001 & $8.12 \pm 0.02$ & $8.12 \pm 0.02$ & $8.13 \pm 0.19$ & \ldots & $6.85 \pm 0.11$ & $5.77 \pm 0.06$ & $7.49 \pm 0.08$ \\
WR002 & $9.03 \pm 0.04$ & \ldots & $8.71 \pm 0.19$ & $9.33 \pm 0.04$ & \ldots & \ldots & \ldots \\
WR003 & $8.93 \pm 0.01$ & \ldots & $8.67 \pm 0.19$ & $9.23 \pm 0.01$ & \ldots & \ldots & \ldots \\
WR005 & $8.69 \pm 0.01$ & \ldots & $8.62 \pm 0.19$ & $8.99 \pm 0.01$ & \ldots & \ldots & \ldots \\
WR006 & \ldots & \ldots & $8.42 \pm 0.19$ & \ldots & \ldots & \ldots & \ldots \\
WR007 & $8.81 \pm 0.01$ & \ldots & $8.63 \pm 0.19$ & $9.11 \pm 0.01$ & \ldots & \ldots & \ldots \\
WR008 & $8.30 \pm 0.02$ & $8.30 \pm 0.02$ & $8.24 \pm 0.19$ & $8.47 \pm 0.08$ & $6.94 \pm 0.14$ & $5.90 \pm 0.08$ & $7.58 \pm 0.15$ \\
\end{longtable}
\end{landscape}
}

Out of the 570 spectra, a total of 531 have \texttt{SCIENCEPRIMARY=1}
and can thus be used for statistical studies. A total of 419 of these
have $z>0.005$ which we will occasionally adopt as our lower redshift
limit because deviations from the Hubble flow are limited and the
photometry from the SDSS is more reliable
\citep[see][]{2005ApJ...631..208B} at these redshifts.

Combining our sample with the compilation of S99 leads to a total of
641 unique Wolf-Rayet galaxies, including galaxies classified as Class
1 would increase this to 1,613 unique sources, a total of 1,778
spectra. To put this in perspective it is worth pointing out that only
227 Wolf-Rayet stars in the Milky Way are listed in the compilation by
\citet{2001NewAR..45..135V}, and with 135 stars in the LMC and 9 in
the SMC listed by van der Hucht, the total sample of Wolf-Rayet
\emph{galaxies} here is comparable in size to the samples of
individual Wolf-Rayet \emph{stars} studied in detail. Thus we expect
that the present study will provide a useful complement to the studies
of individual stars in nearby galaxies.

\subsection{Physical properties of the host galaxies}
\label{sec:physical_prop}

\begin{figure}
  \centering
    \includegraphics[width=88mm]{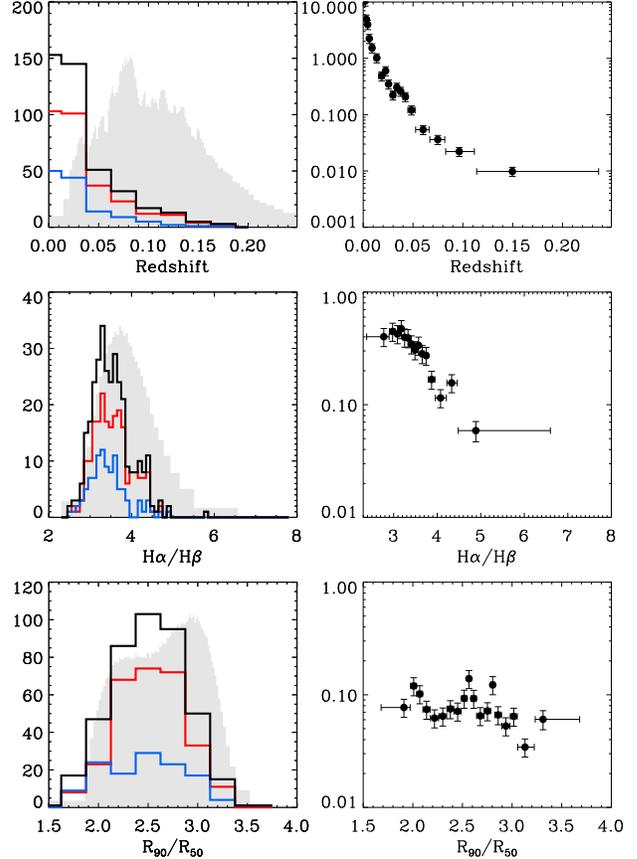}
  \caption{The overall properties of the Wolf-Rayet candidate galaxies
    compared with the properties of the SDSS parent sample as a whole
    (shaded grey histograms) ---  duplicates have been
    excluded. The top row shows the redshift distribution of the SDSS
    as a filled light-grey histogram on the left, with the
    distribution of Class 3 objects in blue, Class 2 objects in red and
    the two classes combined as the black overplotted histogram. The
    right hand panel shows the percentage of galaxies at a given
    redshift that show Wolf-Rayet features. The second row shows the
    same but for the dust sensitive \ensuremath{\mathrm{H}\alpha}/\ensuremath{\mathrm{H}\beta}\ ratio and the bottom row
    the concentration parameter.}
  \label{fig:wr_distrib1}
\end{figure}

Figure~\ref{fig:wr_distrib1} shows the redshift distribution of the
SDSS parent sample as the shaded grey histogram in the top left panel.
The bins each contain 5000 galaxies and the histogram is normalised to
have the same peak value as the black histogram. The redshift
distributions of Class 2 and Class 3 objects together with the
redshift distribution of the two taken together are superimposed. What
is obvious is that the galaxies showing Wolf-Rayet features have a
redshift distribution that is strongly shifted towards low redshift
compared with the SDSS as a whole. This is natural because more
distant galaxies must be intrinsically more luminous to fall within
the selection limits of the SDSS and this makes the detection of
Wolf-Rayet features more difficult.

\begin{figure}
  \centering
    \includegraphics[width=88mm]{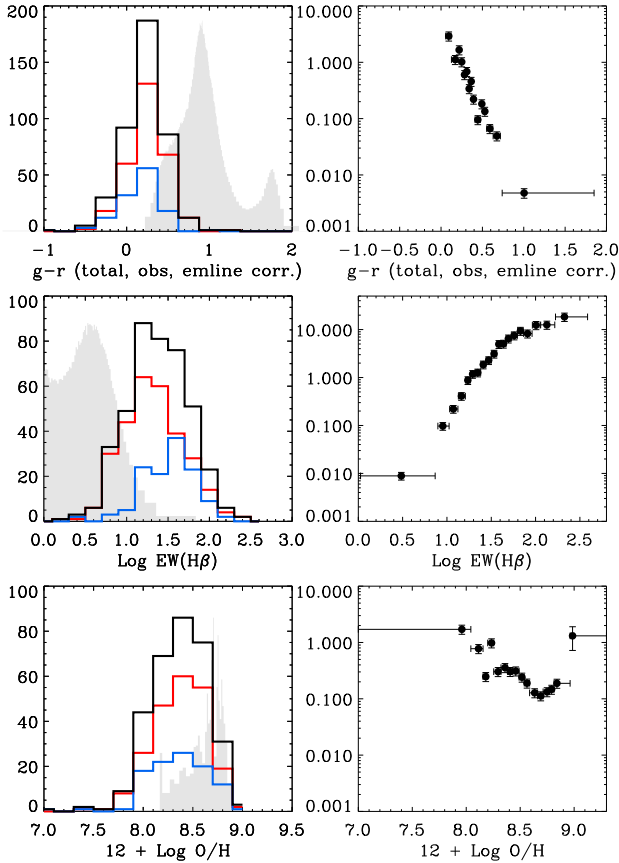}
  \caption{Similar to Figure~\ref{fig:wr_distrib1}. The top row shows
    the total $g-r$ colour, corrected for the emission line
    contribution to the flux as discussed in the text. The middle row
  shows the EW(\ensuremath{\mathrm{H}\beta}) and the bottom row the oxygen abundance --- note that only galaxies classified as SF are included in the final panel.} 
  \label{fig:wr_distrib2}
\end{figure}

This is more explicitly illustrated in the top right panel which shows
the percentage of all galaxies in the SDSS that show Wolf-Rayet
features as a function of redshift. The bins here are chosen to
contain 30 galaxies with Wolf-Rayet features each. Clearly galaxies
with Wolf-Rayet features make up a significant number of the very
low-$z$ spectra in the SDSS.

The middle row shows the same information but the distributions are
now those of the dust-sensitive \ensuremath{\mathrm{H}\alpha}/\ensuremath{\mathrm{H}\beta}\ ratio and only galaxies that
fall within the SF class are included because it is only for these
galaxies that the \ensuremath{\mathrm{H}\alpha}/\ensuremath{\mathrm{H}\beta}\ ratio can be easily interpreted in terms of
dust attenuation. It is clear that Wolf-Rayet galaxies typically have
lower dust extinction than typical SDSS galaxies.

Finally the bottom row shows the distributions against the
concentration parameter, $R_{90}/R_{50}$, where $R_{50}$ and $R_{90}$
are the radii that enclose 50\% and 90\% of the light respectively out to
the Petrosian radius of the galaxies. This is known to correlate
with visual morphology \citep{2001AJ....122.1238S}. It is noteworthy
that galaxies with Wolf-Rayet features appear to have a
very broad range in morphology (right-hand panel) from the most
diffuse to the most concentrated galaxies.

It is reassuring that in the plot for \ensuremath{\mathrm{H}\alpha}/\ensuremath{\mathrm{H}\beta}\ and concentration, the
distributions of Class 2 and Class 3 sources are very similar, as one
would expect since these quantities ought not to strongly affect the
detectability of Wolf-Rayet features.

Figure~\ref{fig:wr_distrib2} shows the location of WR
galaxies in $g-r$ colour in the top row. As is clear from the
right-hand panel, the fraction of galaxies that show Wolf-Rayet
features drops precipitously at $g-r>0$. This $g-r$ colour is
calculated from the SDSS model colours with no $k$-correction and it
has been corrected from emission line contamination assuming that the
contribution from emission line flux found within the fibre is
appropriate for the galaxy as a whole.

The middle row of Fig.~\ref{fig:wr_distrib2} shows the distribution of
EW(\ensuremath{\mathrm{H}\beta}). It is readily seen that the fraction of galaxies with
Wolf-Rayet features increases rapidly with EW(\ensuremath{\mathrm{H}\beta}), reaching an
apparent plateau with nearly 20\% of galaxies with $EW(\ensuremath{\mathrm{H}\beta})>100$\AA\
showing Wolf-Rayet bumps.

The bottom panel shows the distribution with respect to metallicity
for the full sample. This includes only the SF class objects. It shows that a smaller fraction of galaxies with high metallicity shows Wolf-Rayet features than objects with lower metallicity, although we should caution that this particular plot does depend on how the metallicities of the objects are assigned. At first glance this might appear to say that Wolf-Rayet stars are less common at high metallicity, but that
ignores the multi-variate nature of the present sample, and we
address the interesting question of the abundance of Wolf-Rayet stars
at different metallicities using a multi-dimensional view in section~\ref{sec:wr_star_abundance} below. 

In addition to the preceding discussion which places Wolf-Rayet
galaxies in the context of the overall galaxy population, it is also
of interest to ask whether there are systematic differences between
galaxies showing Wolf-Rayet features and similar galaxies without WR
features. The bottom line of this exercise is that WR galaxies and
similar galaxies without WR features do not differ in dust attenuation
but that there is a slight tendency for WR galaxies to be more
concentrated than galaxies with similar mass and star formation
activity without WR features.

To reach these conclusions we adopted a similar approach to
\citet{2006MNRAS.367.1394K}. For each WR galaxy we find all SDSS
galaxies whose stellar mass is within 0.1 dex of the WR galaxy, the
D4000$_N$ is within 0.05 and EW(\ensuremath{\mathrm{H}\beta}) is
within a factor of 3 of that of the WR galaxy. If any of these cuts
select less than 1000 galaxies we loosen the constraints to the first
1000 objects. We then calculate the difference between the properties
of the WR galaxies and these similar SDSS galaxies.

\subsection{Host galaxy morphology and location of WR regions}
\label{sec:wr_hosts_and_morphologies}

As mentioned above a number of the spectra studied here are those of
non-nuclear regions in the galaxies. This is the case for
approximately 30\% of the spectra, the rest correspond to the central
regions of the galaxies. In the case of mergers and some types of
irregular galaxies it is difficult to determine a centre, this is true
for $\sim 5\%$ of the spectra. However most spectra do originate well
within the main body of the galaxies, only 93 spectra (16\%) really
originate from the outskirts of the host galaxy. These are invariably
bright H\,\textsc{ii}-regions, in fact a total of 202 (35\%) of the spectra can
be identified with a clearly delimited star forming region, ranging
from H\,\textsc{ii}-regions to massive super-star clusters spanning almost 50\%
of the galaxy.

The host galaxies span a range of morphologies and dynamical
situations. There are a number of clear mergers, at least 7\% of the
spectra, and $\sim 11\%$ originate from galaxies with a strong
bar. Thus one might worry that these classes have star bursts of
different effective ages and hence that including them all in one
group might bias or wash-out any trends. To check this we have tested
a large number of relationships and seen how they vary with the
location of galaxies and with the morphological properties of the
galaxies.  We do find that oxygen abundance distribution of spectra
originating in the outskirts of their parent galaxies are slightly
shifted to lower metallicity as compared to that for the central
region, as one might expect given the well-known metallicity gradients
in galaxies. Apart from this there are no detectable dependencies on
any relationships/trends with the regions of the galaxies the spectra
originate from or with the galaxy morphologies. We will therefore
ignore this in the following.

\section{The abundance of WR stars with metallicity}
\label{sec:wr_star_abundance}

Models for WR star formation
predict that the number of Wolf-Rayet stars increases with increasing
metallicity \citep{2005A&A...429..581M} due in large part to a strong
metallicity dependence on the stellar winds
\citep{2005A&A...442..587V}. From an observational point of view it is
also established that there is a decline in the number of WR stars as
compared to O-stars, N(WR)/N(O), with declining metallicity
\citep[see][for an overview]{2007ARA&A..45..177C}. However, the range
of metallicities is limited and the spread around the relation is not
established and the stellar parameters are typically hard to constrain
without involving modelling \citep[e.g.][]{2006A&A...457.1015H}.
In this section we will focus on empirical measures of the abundance
of WR stars, postponing a comparison with models to
section~\ref{sec:models}.

\begin{figure}[tbp]
  \centering
  \includegraphics[angle=90,width=88mm]{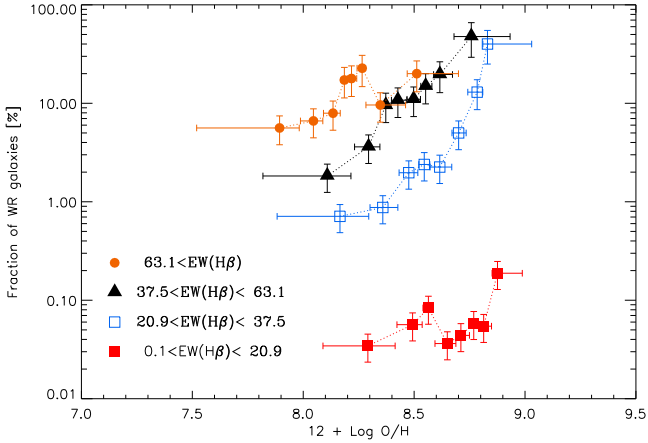}
  \caption{The fraction of SF galaxies that show Wolf-Rayet features as a
    function of oxygen abundance in four different EW(\ensuremath{\mathrm{H}\beta}) bins. The
    bins in EW(\ensuremath{\mathrm{H}\beta}) each contain 80 Class 2 \& 3  galaxies, except the
    highest bin which contains 81. The points
    are calculated in bins containing 10 Class 2\&3 galaxies each and
    the error bars reflect Poissonian uncertainties. Only SF galaxies
    are included and the oxygen abundance estimator is the mixed one.} 
  \label{fig:wr_abundance_vs_ewhb_oh}
\end{figure}

Figure~\ref{fig:wr_abundance_vs_ewhb_oh} shows the fraction of
SF galaxies showing WR features as a function of oxygen abundance in four
bins in EW(\ensuremath{\mathrm{H}\beta}). It shows a very clear increase in the fraction of
galaxies showing WR features with metallicity with a clear indication of a
\emph{non-linear} relationship.  The comparison is done in bins of
EW(\ensuremath{\mathrm{H}\beta}) to take out the correlation between oxygen abundance and
EW(\ensuremath{\mathrm{H}\beta}) in the SDSS sample. It is, however, clear that the trend is
similar for different bins in EW(\ensuremath{\mathrm{H}\beta}) and it turns out that they are independent of the abundance estimator chosen.

\begin{figure}[tbp]
  \centering
  \includegraphics[angle=90,width=88mm]{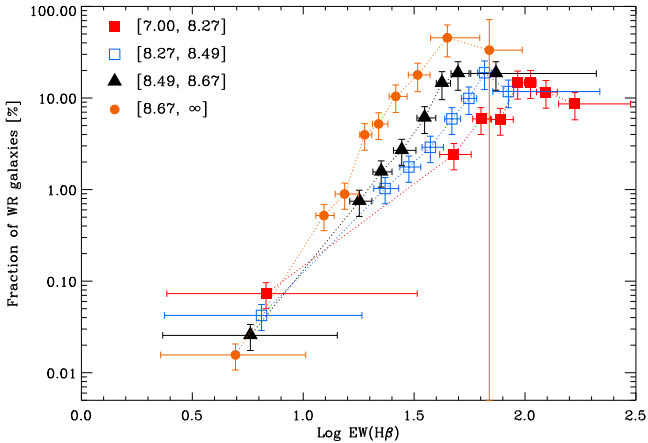}
  \caption{Similar to Figure~\ref{fig:wr_abundance_vs_ewhb_oh}, this
    figure shows the the fraction of SF galaxies that contain
    Wolf-Rayet features as a function of EW(\ensuremath{\mathrm{H}\beta}) in four bins of
    oxygen abundance. }  
  \label{fig:wr_abundance_vs_ewhb}
\end{figure}

The figure shows a clear relationship between the oxygen abundance of
a galaxy and the likelihood that it harbours significant numbers of
Wolf-Rayet stars. It is also clear that there is a correlation with EW(\ensuremath{\mathrm{H}\beta}). This is made more
explicit in Figure~\ref{fig:wr_abundance_vs_ewhb} which shows the
relative abundance of WR galaxies as a function of EW(\ensuremath{\mathrm{H}\beta}) in four
bins of oxygen abundance. It is clear that galaxies with the same
EW(\ensuremath{\mathrm{H}\beta}) are more likely to have WR features at higher
metallicity.

\begin{figure}
  \centering
  \includegraphics[angle=90,width=88mm]{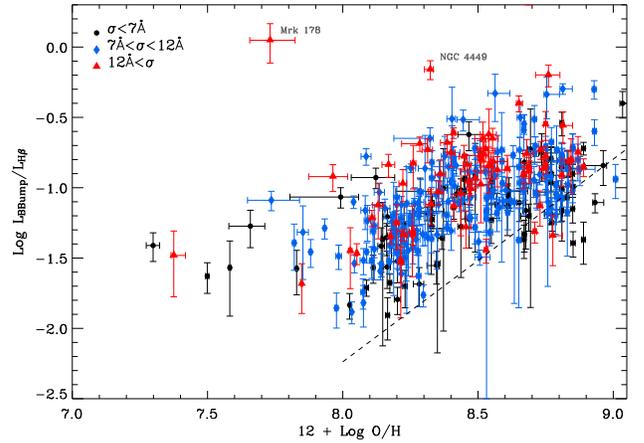}
  \caption{The ratio of the blue bump luminosity to the luminosity of
    the \ensuremath{\mathrm{H}\beta}\ as a function of metallicity.
    There is a clear trend towards lower maximum bump luminosities
    relative to \ensuremath{\mathrm{H}\beta}\ at lower metallicity but
    it appears to flatten out at $12 + \log \mathrm{O/H}<8$.  The
    dashed line shows the detection limit based on the data in
    Figure~\ref{fig:ew_distrib}.  The points are colour coded
    according to the width of the
    \ensuremath{\mbox{He\,\textsc{ii}\,$\lambda 4686$}} line. The red
    points show the location of the galaxies with the widest bumps,
    the black the narrowest. There is no statistically significant
    difference between these classes.}
  \label{fig:wr_vs_hb_mixed}
\end{figure}

It is also intriguing that all the curves appear to reach a maximum
value and then turn over. This is expected to happen when the
EW(\ensuremath{\mathrm{H}\beta}) samples burst ages that are short
relative to the time of start of the WR phase (after $\sim 2$Myr).
While EW(\ensuremath{\mathrm{H}\beta}) is often used as an age
indicator \citep{1986A&A...156..111C,2007ApJ...655..851Z} it is
well-known that this might be a poor approximation both due to different dust attenuation of the lines and the continuum
and to an older underlying population
\citep[e.g.][]{1999A&A...349..765M}. For our sample this is means that
EW(\ensuremath{\mathrm{H}\beta}) in general is \emph{not} a reliable
age indicator, but for a given metallicity range it might be useful as
a relative age indicator and/or burst strength indicator. The
fraction appears to reach a maximum at $\log
\mathrm{EW}(\ensuremath{\mathrm{H}\beta})>1.7$. For a short
($<1$\,Myr) burst this corresponds to an age of 5--6 Myr. Since
10--20\% of all spectra at this EW show WR features, we conclude that
the WR features are clearly visible in 0.5--1 Myr of this time,
although we should caution that the age estimates are strongly
dependent on the assumed properties of the burst.

An alternative formulation of this result can be found by adopting the
method already used by \citet{1981A&A...101L...5K} to determine the
ratio of WR to O stars \citep[see also][]{2007ApJ...655..851Z}. This
requires a comparison between metal content and the flux of the blue
bump divided by the \ensuremath{\mathrm{H}\beta}\ flux. This is a
rough, but model independent, estimate of the number of WR stars per O
stars. This follows because to first order, the flux in the blue bump
is proportional to the number of WR stars. In addition the luminosity
of the \ensuremath{\mathrm{H}\beta}\ line is roughly proportional to
the number of O stars within the area surveyed and the ratio is
approximately independent of dust attenuation (see discussion in
section~\ref{sec:internal_attenuation}). Thus this quantity should
give a reasonable estimate of the relative number of WR and O stars in
a model-independent way. It has recently become common to convert this
to N(WR)/N(O+WR) in a model dependent way
\cite[e.g.][]{2000ApJ...531..776G,2004MNRAS.355..728F} but in view of
the uncertainties in the current models discussed below we prefer not
to do this at this point, and return to it in section~\ref{sec:origin} below.

\begin{figure*}[tbp]
  \centering
  \includegraphics[angle=90,width=170mm]{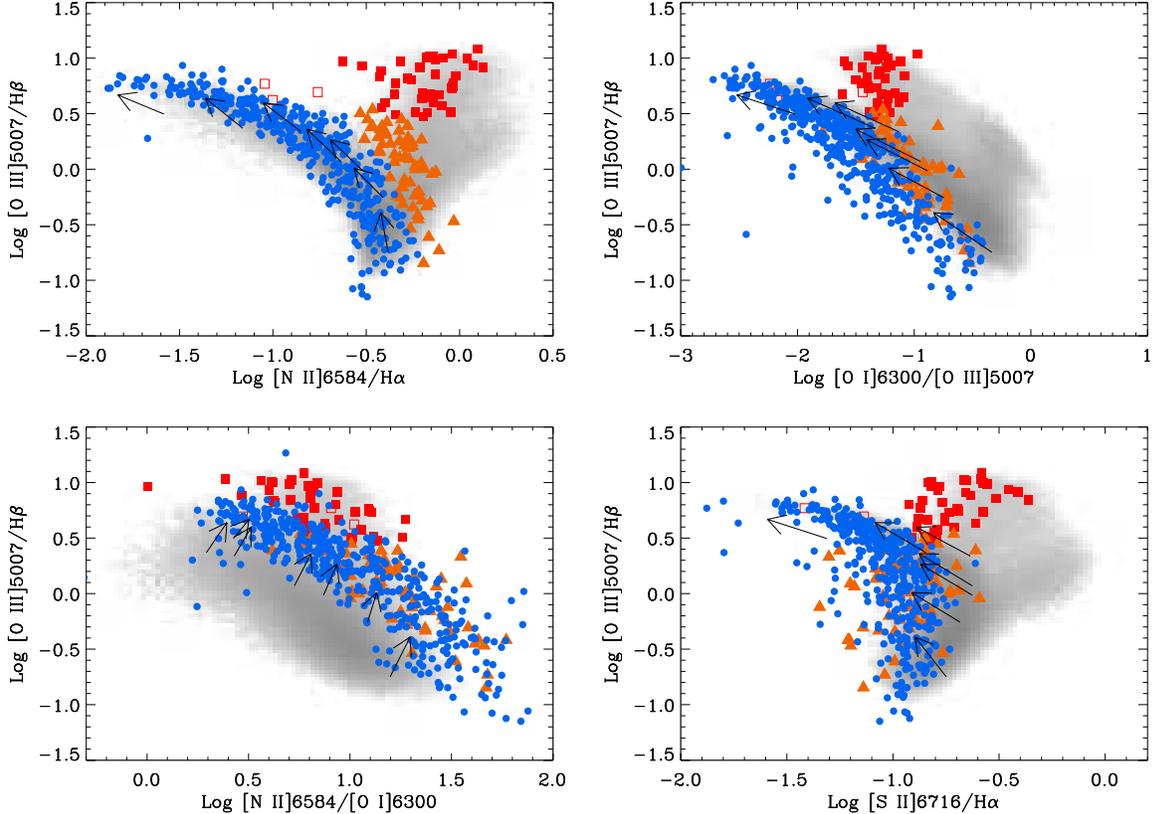}
  \caption{Top left: The BPT diagram for the SDSS with Wolf-Rayet
    galaxies overplotted. The top right panel shows the
    [O\,\textsc{iii}]\,$\lambda 5007$/\ensuremath{\mathrm{H}\beta}\
    versus [O\,\textsc{i}]\,$\lambda 6300$/[O\,\textsc{iii}]\,$\lambda
    5007$ diagram and the bottom left two further diagnostic diagrams.
    In all panels the grey-scale shows the logarithm of the number of
    galaxies in bins on the $\log {[N\,\textsc{ii}]\,\lambda
      6584}/\ensuremath{\mathrm{H}\alpha}$ vs.\ $\log
    {[O\,\textsc{iii}]\,\lambda 5007}/\ensuremath{\mathrm{H}\beta}$
    plane. The galaxies plotted are chosen to have S/N$>3$ in all
    these four lines as well as [O\,\textsc{i}]\,$\lambda 6300$ and
    [S\,\textsc{ii}]\,$\lambda 6717$. The points show the location of
    our candidate WR galaxies. The blue filled circles show the
    galaxies classified as SF, the orange triangles that of the
    composite galaxies and the filled red squares the location of
    AGNs. The open red squares show the location of Composite and AGNs
    that have $\log [N\,\textsc{ii}]\,\lambda
    6584/\ensuremath{\mathrm{H}\alpha}<-0.7$. These objects resemble
    SF galaxies more than AGN.  The small black arrows show the effect
    of increasing the ionisation parameter, $U$, by 0.5 dex, see text
    for details.}
  \label{fig:bpt_w_WR}
\end{figure*}

Figure~\ref{fig:wr_vs_hb_mixed} shows the results of this exercise.
The figure plots galaxy oxygen abundance along the $x$-axis.  There is
an apparent correlation between the metallicity and the ratio of the
blue bump to \ensuremath{\mathrm{H}\beta}\ luminosities but the lack
of points in the bottom right is due to incompleteness. This is shown
by the dashed line which indicates the location of a galaxy with an
EW(Blue bump) of 1\AA\ given the observed distribution of
EW(\ensuremath{\mathrm{H}\beta}) at that metallicity in the full SDSS
DR6. The spread of points at a given metallicity is several times the
observational uncertainty and is entirely consistent with being due to
the fact that the star-bursts are not all coeval.

However the decrease in the maximum value attained by the ratio of
blue bump luminosity to \ensuremath{\mathrm{H}\beta}\ luminosity
\emph{is} real and reflects a decrease in the total luminosity of the
Wolf-Rayet phase with decreasing luminosity. This trend does however
not extend to the very lowest metallicities and at metallicities below
$12+\log \mathrm{O/H} <8$ there appears to be a flattening off.

As mentioned by CH06 the WN stars in the SMC are known to have weaker
and narrower lines than in more metal rich environments
\citep{1989ApJ...341..113C} so we also indicate by different symbols
and colours the location of galaxies with different
\ensuremath{\mbox{He\,\textsc{ii}}}\ 4686 widths. There is no
statistically significant difference between these classes but it is
clear that the systems with the strongest bumps relative to
\ensuremath{\mathrm{H}\beta}\ all show broad
\ensuremath{\mbox{He\,\textsc{ii}}}\ lines. In particular the extreme
galaxy Mrk 178 which has the strongest WR features relative to
\ensuremath{\mathrm{H}\beta}\ of any galaxy in the sample, shows broad
WR lines and an intriguingly low metallicity. We will discuss
  this object in some detail in Section~\ref{sec:models}.

\section{The ionisation conditions of the interstellar medium of WR
  galaxies} 
\label{sec:diagnostic_diagrams}

It is of considerable interest to understand the properties of the
interstellar medium (ISM) of galaxies with very strong star formation
activity since this might give important insights into the ISM of
actively star forming galaxies at all redshifts \citep[e.g.][]{2008MNRAS.385..769B,2008arXiv0801.1670L}.

Here we ask whether galaxies that harbour significant populations of
Wolf-Rayet stars differ systematically in their emission line
properties from other star forming galaxies. We use the BPT diagram
shown in the top-right panel of Figure~\ref{fig:bpt_w_WR} as a
starting point. This shows the distribution of all SDSS galaxies with
$\mathrm{S/N}>3$ in \ensuremath{\mathrm{H}\beta}, [O\,\textsc{iii}]\,$\lambda
5007$, [O\,\textsc{i}]\,$\lambda 6300$, \ensuremath{\mathrm{H}\alpha},
[N\,\textsc{ii}]\,$\lambda 6584$ and [S\,\textsc{ii}]\,$\lambda 6717$
as a grey scale 2D distribution --- the grey-scale shows the logarithm
of the number of galaxies in each bin. The galaxies showing WR
signatures are overplotted with different symbols indicating different
emission line classifications with red squares indicating AGNs, orange
triangles Composite objects and the filled blue circles Wolf-Rayet
galaxies in the SF class. The open red squares indicate five galaxies
lying well above the main sample of galaxies, in fact close to the
location of the $z\sim 2$ galaxies studied by
\citet{2006ApJ...644..813E}. While these galaxies formally lie in the
region of Composite or AGN galaxies, i.e.\ above the locus of
photoionisation models for H\,\textsc{ii}-regions
\citep{2001ApJ...556..121K}, some of these will turn out to show
characteristics of star forming systems. Since it is well-known that
photoionisation models tend to underpredict the
[O\,\textsc{iii}]\,$\lambda 5007$/\ensuremath{\mathrm{H}\beta}\ ratio
for low metallicity H\,\textsc{ii}-regions
\citep[e.g.][]{2006ApJS..167..177D}, we have chosen to identify these
objects with a different symbol in Figure~\ref{fig:bpt_w_WR}

The top right panel of Figure~\ref{fig:bpt_w_WR} shows the
[O\,\textsc{iii}]\,$\lambda 5007$/\ensuremath{\mathrm{H}\beta}\ versus
[O\,\textsc{i}]\,$\lambda 6300$/[O\,\textsc{iii}]\,$\lambda 5007$
ratio. While the AGN are distributed in a similar manner to the main
bulk of AGN, the SF galaxies with WR signatures seem to be offset from
the bulk of the SDSS galaxies. Since the [O\,\textsc{i}]\,$\lambda
6300$ line predominantly originates in the neutral ISM, this might
indicate that the WR galaxies have a significantly more ionised ISM
compared to other SF galaxies.

This offset is even clearer in the bottom right panel, which has
[N\,\textsc{ii}]\,$\lambda 6584$/[O\,\textsc{i}]\,$\lambda 6300$ on the x-axis. Clearly
the SF galaxies showing WR signatures occupy almost the same region as
the AGN and composite galaxies. This is might be interpreted as
a suppressed [O\,\textsc{i}]\,$\lambda 6300$ shifting the galaxies rightwards
from the SF sequence.

A similar offset can be seen in the [O\,\textsc{iii}]\,$\lambda 5007$/\ensuremath{\mathrm{H}\beta}\ 
versus [S\,\textsc{ii}]\,$\lambda 6716$/\ensuremath{\mathrm{H}\alpha}\ diagram shown in
the bottom right panel. [S\,\textsc{ii}]\ is also significantly produced in
the neutral ISM and it is again clear that there is a systematic
offset between the WR galaxies and the bulk of galaxies in the SDSS.

On these diagrams we have indicated the effect of changing the
ionisation parameter, $U$, by 0.5 dex by the black arrows, for six
different metal abundances based on the CL01 models. It is clear that
the offsets we see are consistent with a higher effective ionisation
parameter in the Wolf-Rayet galaxies than in the SDSS as a whole. This
is in good agreement with the findings of
\citet[][BCP08]{2008MNRAS.385..769B} and \citet{2008arXiv0801.1670L}
for strongly star-forming galaxies in the SDSS in general --- see
BPC08 for more discussion. It is worth pointing out that such simple
scaling arguments are insufficient to reliably distinguish between an
increased ionisation parameter or a non-negligible escape of ionising
photons from the region sampled by the spectrum
\citep[e.g.][]{1996A&A...312..365B}. A more thorough analysis would
require further data and is outside the scope of the present paper.

However the trends taken together do imply that there is no
significant contribution to the emission lines from shocks based on
the models of \citet{1995ApJ...455..468D} in good agreement with the
results of BPC08 who find that strongly starbursting galaxies in the
SDSS do not show any significant contribution to their emission line
fluxes from shocks.

\section{The origin of He\,{\small II} nebular emission}
\label{sec:heii_nebular}

\begin{figure}
  \centering
    \includegraphics[angle=90,width=88mm]{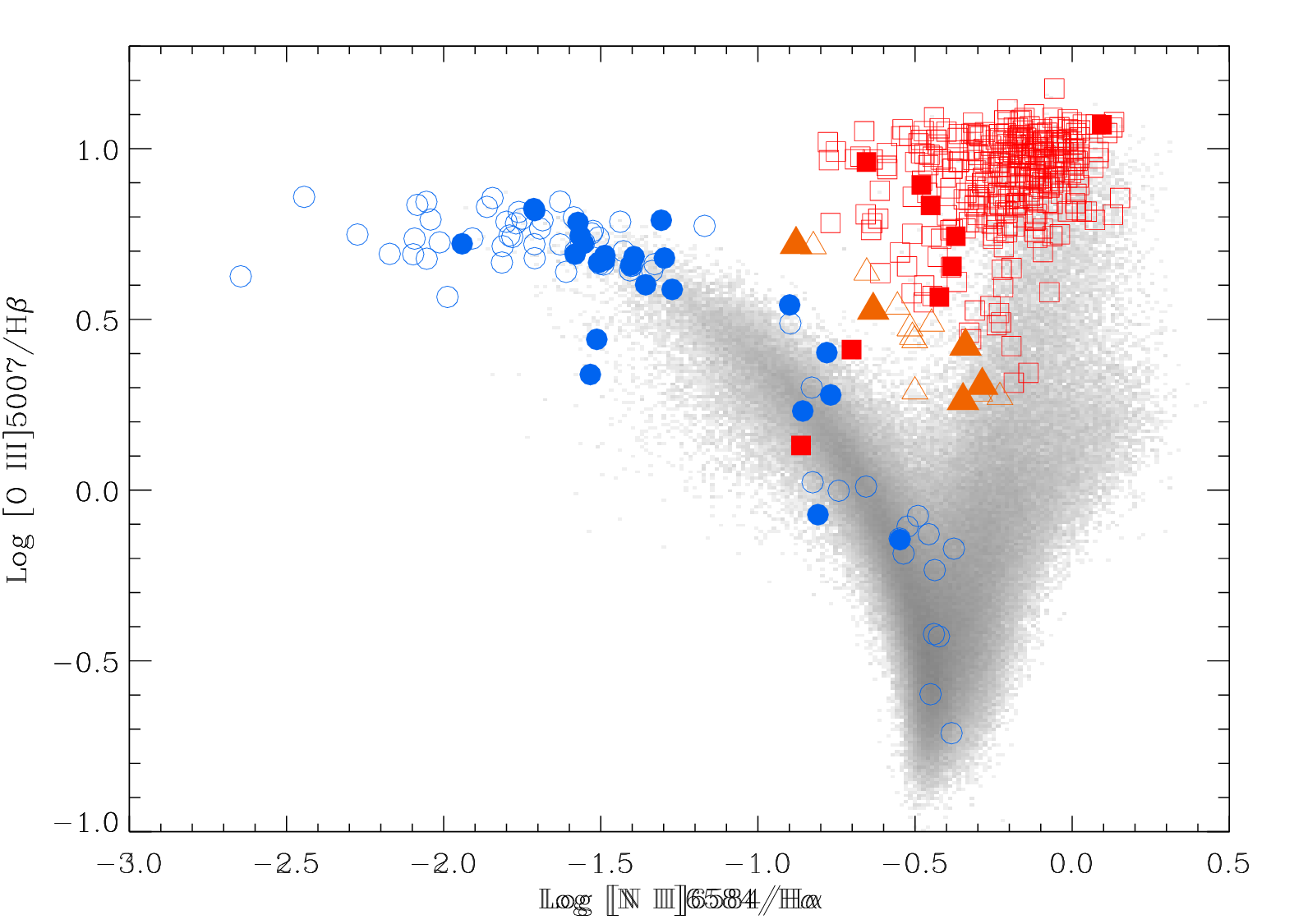}
  \caption{The distribution of galaxies showing nebular
    \ensuremath{\mbox{He\,\textsc{ii}\,$\lambda 4686$}} with S/N$>3.5$ in the BPT diagram. The
    filled symbols mark objects that also show WR signatures in their
    spectra. The symbols and colours are otherwise as in
    Figure~\ref{fig:bpt_w_WR}.}
  \label{fig:heii_bpt}
\end{figure}

\begin{figure}
  \centering
  \includegraphics[angle=90,width=88mm]{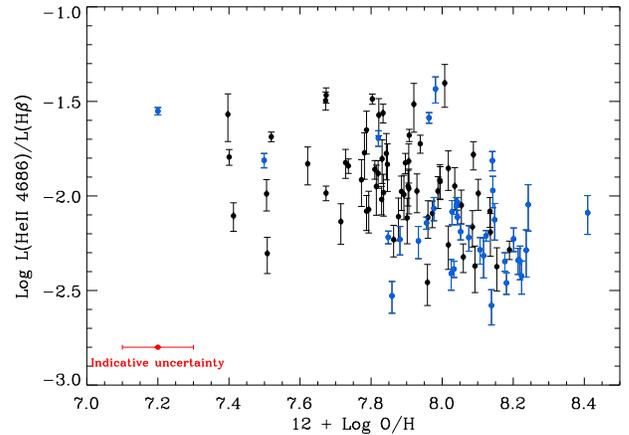}
  \caption{The \ensuremath{\mbox{He\,\textsc{ii}}}/\ensuremath{\mathrm{H}\beta}\ ratio as a function of metallicity for the SDSS DR6. Only galaxies with \ensuremath{\mbox{He\,\textsc{ii}\,$\lambda 4686$}} detected at S/N$>3.5$ and classified as SF galaxies are included. The blue symbols show those galaxies with detected WR features. The uncertainties in the oxygen abundance have been suppressed for clarity but an uncertainty of 0.1 dex is indicated in the lower left.}
  \label{fig:he_hb_vs_oh}
\end{figure}

\begin{figure*}
  \centering
  \includegraphics[width=170mm]{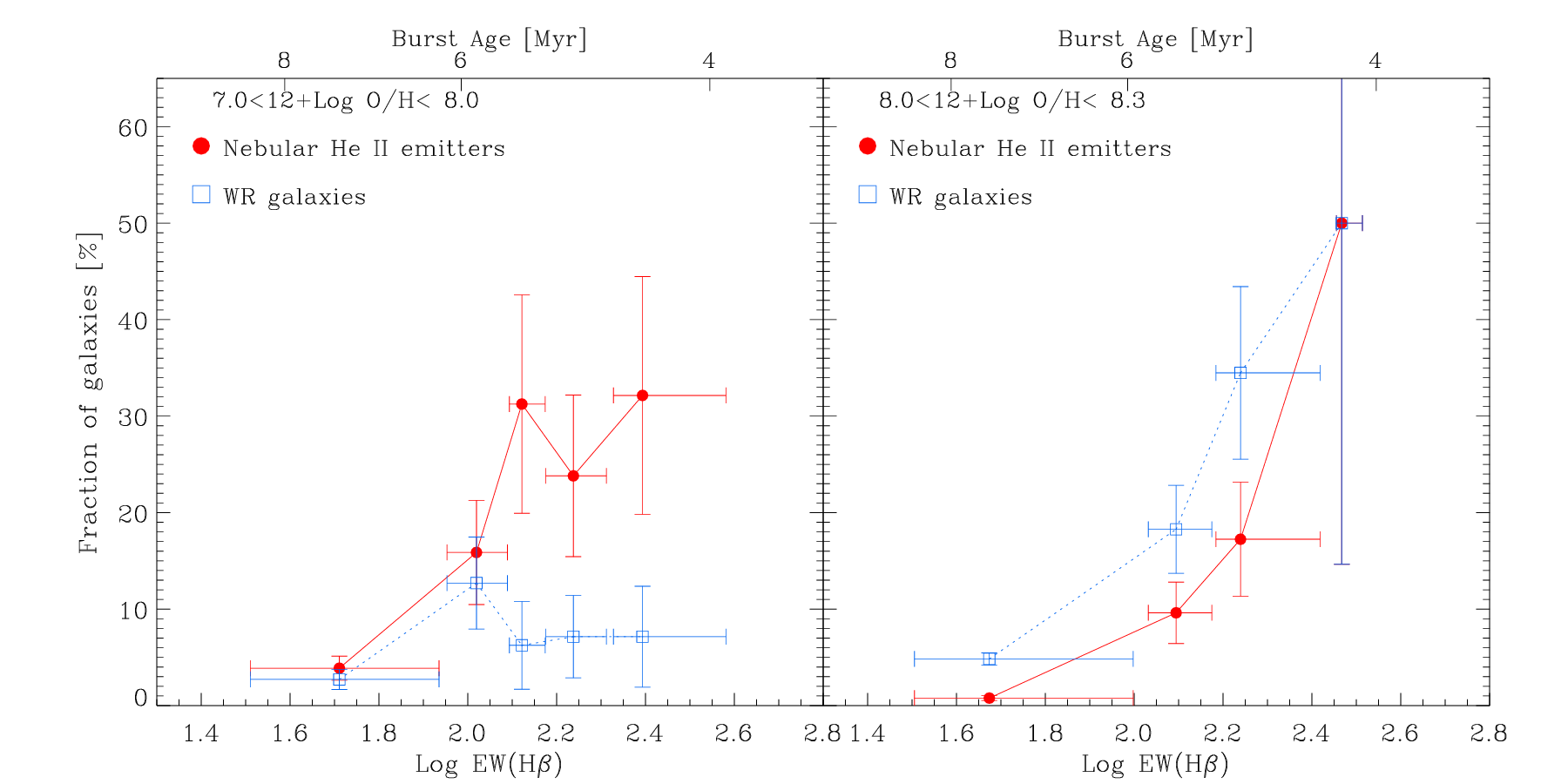}
  \caption{In the left panel the solid line shows the fraction of star-forming galaxies with $7 < 12 + \log \mathrm{O/H} < 8$ that show nebular \ensuremath{\mbox{He\,\textsc{ii}\,$\lambda 4686$}} in their spectra as a function of EW(\ensuremath{\mathrm{H}\beta}). The error bars are Poissonian and each bin contains 10 \ensuremath{\mbox{He\,\textsc{ii}\,$\lambda 4686$}} emitters.  The upper x-axis shows the burst age in Myr corresponding to the EW(\ensuremath{\mathrm{H}\beta}), see text for details. The dashed line connecting open squares shows how the fraction of WR galaxies in the same abundance range varies with EW(\ensuremath{\mathrm{H}\beta}). The right panel shows the same for the abundance range $8 < 12 + \log \mathrm{O/H} < 8.3$.   }
  \label{fig:fraction_of_he}
\end{figure*}

The previous section indicated an increased ionisation parameter in
the WR galaxies but equivocal evidence of a harder ionisation field
in these sources. It has been suggested in the past  that the hard
radiation field of Wolf-Rayet stars causes the nebular
\ensuremath{\mbox{He\,\textsc{ii}\,$\lambda 4686$}} occasionally seen
in H\,\textsc{ii}\ galaxies \citep[e.g.][]{1996ApJ...467L..17S}. Guseva et al (2000) carried out a careful
examination of this issue but were not able to show conclusively that
the cause of the nebular \ensuremath{\mbox{He\,\textsc{ii}}}\ was
Wolf-Rayet galaxies. Here we re-examine the issue, using the small set
of SDSS galaxies that show nebular
\ensuremath{\mbox{He\,\textsc{ii}\,$\lambda 4686$}} emission. If we
focus on those that show \ensuremath{\mbox{He\,\textsc{ii}}}\ at
S/N$>7 (3.5)$ we find a total of 318 (1461) galaxies with nebular
\ensuremath{\mbox{He\,\textsc{ii}}}. Out of these 288 (1392) are at
$z>0.01$ and most of these, 269 (1222), appear to by AGN dominated
with only 15 (81) being dominated by star-formation. These are plotted
on top of the BPT diagram in Figure~\ref{fig:heii_bpt} where it is
clear that the star-forming sources with nebular
\ensuremath{\mbox{He\,\textsc{ii}\,$\lambda 4686$}} have low
{[N\,\textsc{ii}]}/\ensuremath{\mathrm{H}\alpha}. The solid symbols
indicate the location of objects that also show WR features in their
spectra.

Figure~\ref{fig:he_hb_vs_oh} shows the
\ensuremath{\mbox{He\,\textsc{ii}\,$\lambda
    4686$}}/\ensuremath{\mathrm{H}\beta}\ diagram for star-forming
galaxies in the SDSS DR6. Only galaxies that have
\ensuremath{\mbox{He\,\textsc{ii}\,$\lambda 4686$}} detected at
$\mathrm{S/N}>3.5$ have been included. Note that the SF galaxies with
$\ensuremath{\mbox{He\,\textsc{ii}}}/\ensuremath{\mathrm{H}\beta}>0.1$
have unreliable \ensuremath{\mbox{He\,\textsc{ii}}}\ and/or
\ensuremath{\mathrm{H}\beta}\ measurements.

The figure shows a correlation between O/H and
$\ensuremath{\mbox{He\,\textsc{ii}}}/\ensuremath{\mathrm{H}\beta}$
such that
$\ensuremath{\mbox{He\,\textsc{ii}}}/\ensuremath{\mathrm{H}\beta}
\propto -0.8 \log \mathrm{O/H}$ over the range $7.6<\log
\mathrm{O/H}<8.0$. This is apparently not due to an increased number
of WR stars at low metallicity: the blue points in
Figure~\ref{fig:he_hb_vs_oh} show the location of galaxies with
detected WR features, and these do not appear to dominate at low
metallicity.

We also notice that there appears to be a levelling off at
$\log \mathrm{O/H}<7.6$. At lower metallicity there is a rather broad
range in
\ensuremath{\mbox{He\,\textsc{ii}}}/\ensuremath{\mathrm{H}\beta}, but
no continuation of the trend seen at higher metallicity. We note that
this is not likely to be a selection effect because such strong
\ensuremath{\mbox{He\,\textsc{ii}}}\ lines are easy to detect. The
physical reasons for this flattening off are unclear and require more
modelling of winds and stellar atmospheres in low metallicity stars.

A similar result was reached by Guseva et al who concluded that in
those galaxies not classified as WR galaxies, the WR feature was too
weak to detect. Based on these results it is not clear whether this is
the right interpretation or whether low density stellar winds at low
metallicity lead to a higher flux of
\ensuremath{\mbox{He\,\textsc{ii}}}\ ionising photons
\citep{2002MNRAS.337.1309S,2007MNRAS.381..418H}. In this case not only
WR stars but also regular O stars are expected to contribute to the
\ensuremath{\mbox{He\,\textsc{ii}}}-ionising flux.

The increase in He\,\textsc{ii}-ionising flux, relative to H-Ionising flux could also have a contribution from other sources. Shocks produce nebular He\,\textsc{ii}, but in star forming regions they are mostly caused by colliding winds but as the stellar winds are expected to be weaker at lower metallicities \citep[e.g.][]{2005A&A...442..587V} this would not explain the trend seen. Likewise if the increased He\,\textsc{ii}-ionising flux is caused by X-ray binary evolution, one would require an increased binary fraction with reduced metallicity and in absence of any strong evidence of this we do not view this as a likely explanation. This leaves us with the two scenarios that either the main contribution to the He\,\textsc{ii} ionisation is caused by Wolf-Rayet stars, especially WC stars, or that the main contributors are more massive stars such as O or WN stars.

The key difference between these two scenarios is the time-scale. If the
dominant source of ionising flux is WC stars, the onset of nebular
\ensuremath{\mbox{He\,\textsc{ii}\,$\lambda 4686$}} would be later
than in the case of WN or O stars being the primary sources.

Figure~\ref{fig:fraction_of_he} shows the fraction of galaxies with
nebular \ensuremath{\mbox{He\,\textsc{ii}\,$\lambda 4686$}} as a
function of EW(\ensuremath{\mathrm{H}\beta}) in two different
abundance regimes. The abundance ranges were chosen to have 50
\ensuremath{\mbox{He\,\textsc{ii}}} emitters in each and we use
EW(\ensuremath{\mathrm{H}\beta}) as a proxy for age of the burst.
However we also indicate the age corresponding to a given
EW(\ensuremath{\mathrm{H}\beta}) assuming that the burst has a
duration of 1 Myr on the top x-axis. The dashed line shows the
fraction of WR galaxies as a function of
EW(\ensuremath{\mathrm{H}\beta}).


From this figure it appears that at low metallicity the WR
galaxies, and by implication the WR stars, are not the main cause of
nebular \ensuremath{\mbox{He\,\textsc{ii}\,$\lambda 4686$}} emission
because the time-evolution is noticeably different. This 
indicates that regular O stars at low metallicity indeed show a
significant emission of photons with $\lambda < 228$\,\AA\
\citep[c.f.][]{2002MNRAS.337.1309S,2007MNRAS.381..418H}. On the other
hand at higher metallicity the two show qualitatively similar trends
with the WR lines shifted to lower EW(\ensuremath{\mathrm{H}\beta}).
This is consistent with the sources producing the majority of the
$<228$\,\AA\ photons having shorter life-times than the WR phase as a
whole and that they originate at an early stage in the evolution of a
star burst. It is not possible on the basis of these data to make a statement on the relative importance of  WN or O stars to the ionisation of \ensuremath{\mbox{He\,\textsc{ii}}}.

One might worry that the fraction of galaxies harbouring nebular He\,\textsc{ii} emission does not reach 100\%, while the argument above would indicate that all sufficiently young star-forming regions should show nebular He\,\textsc{ii} emission. The reason for this apparent discrepancy is the S/N requirement for detecting nebular He\,\textsc{ii}. All the galaxies for which we have detection of nebular He\,\textsc{ii} with S/N$>3$, have a median S/N in their spectra higher than 10 and show a fairly flat distribution in the S/N of their spectra. In contrast, the S/N distribution for spectra with EW(H$\beta$)$>30$\,\AA\ is sharply peaked towards low S/N. This does mean that the absolute vertical scale in Figure~\ref{fig:fraction_of_he} is suspect. The appropriate way to deal with this would be to enforce a particular distribution in S/N, but the sample is not  big enough to do this rigourously. However when simply drawing the comparison sample from a distribution in S/N similar to that of the galaxies showing nebular He\,\textsc{ii}, the fraction of galaxies with nebular He\,\textsc{ii} does tend to 100\% at high EW(H$\beta$). A similar conclusion is reached when looking at the 2D distribution of S/N in the spectra versus EW(H$\beta$).

Thus we conclude that our data are consistent with the hypothesis that \emph{all} systems with $\log \mathrm{EW(H\ensuremath{\beta})} \sim 2.5$ show nebular He\,\textsc{ii} as long as the S/N of the spectrum allows its detection. We also conclude that at low metallicity the ionisation of He\,\textsc{ii} is most likely dominated by O stars, although there could be a contribution of WN stars that we are unable to detect because their features are very weak in the optical. At metallicities higher than $\approx 20$\% solar the data are consistent with the major source of ionising radiation being WN stars but a significant contribution from O stars is possible.

\section{N/O abundance trends and local enrichment}
\label{sec:no_abundance_trends}

Since Wolf-Rayet stars have strong winds they should influence their
immediate surroundings. However observational evidence of this on large scales has
been equivocal. The comprehensive study by \citet{1996ApJ...471..211K}
found no clear difference in the ISM abundance of WR galaxies and
other star-burst galaxies. However a number of studies have found
spatial variations in the N/O ratio in NGC 5253
\citep{1997ApJ...477..679K,2007ApJ...656..168L}. These studies point
to the possibility that winds from Wolf-Rayet stars can mix with the ISM on a relatively short time-scale, in contrast to supernova explosions that require $> 10^8$ years for the eject to cool enough to allow efficient mixing with the ISM.

\begin{figure}
  \centering
  \includegraphics[width=88mm]{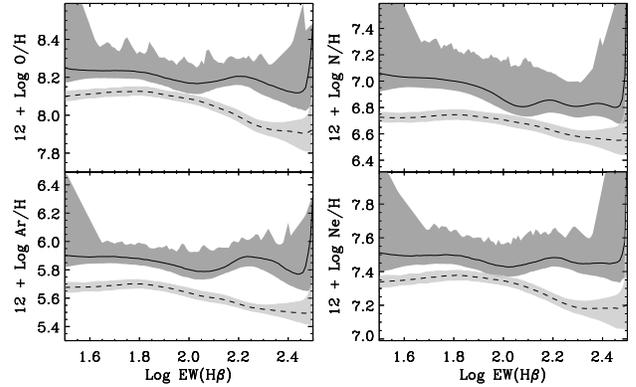}
  \caption{Top left: The oxygen abundance as a function of EW(\ensuremath{\mathrm{H}\beta}) for galaxies with WR features in their spectra (solid line and dark shading) and for non-WR galaxies (dashed line and light shading). The lines show the median and the shading the 68\% confidence limit on the median  (see  text for details of the calculation). The top-right panel shows the  same for 12+log N/H, the bottom left the same for Ar/H and the bottom right the same for Ne/H.} 
  \label{fig:abundance_vs_ewhb}
\end{figure}

In Figure~\ref{fig:abundance_vs_ewhb} we show the abundance trends for WR galaxies versus that of non-WR galaxies for O/H, N/H, Ar/H and Ne/O. It is clear that WR galaxies are in general more metal rich at a given EW(\ensuremath{\mathrm{H}\beta}) and this is likely a reflection of WR stars being more abundant at higher metallicities as shown in Figure~\ref{fig:wr_abundance_vs_ewhb_oh} than WR galaxies. We will therefore focus on abundances relative to oxygen in the following, and as Ar and Ne will turn out to follow oxygen approximately we will also focus
mainly on the nitrogen abundance.

\begin{figure}
  \centering
  \includegraphics[width=88mm]{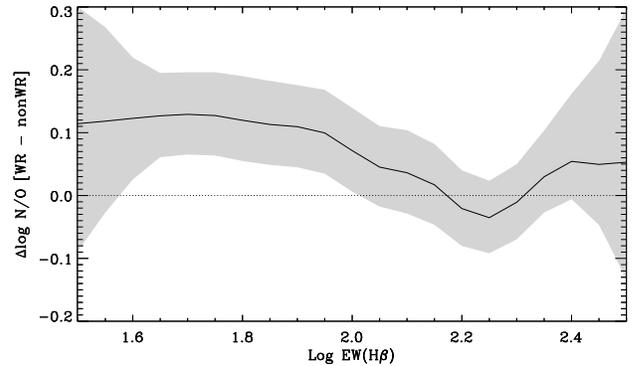}
  \caption{The difference in N/O between Wolf-Rayet galaxies and
    similar galaxies showing no sign of Wolf-Rayet stars as a function
    of EW(\ensuremath{\mathrm{H}\beta}). The solid line shows the
    median trend and the the shaded area indicates the 68\% confidence
    limit on the median (see  text for details of the calculation).
    The dotted line indicates the zero level.}  
  \label{fig:nh_vs_ewhb}
\end{figure}

The solid line in Figure~\ref{fig:nh_vs_ewhb} shows the median
difference in N/O for Wolf-Rayet galaxies and similar galaxies showing
no sign of WR stars as a function EW(\ensuremath{\mathrm{H}\beta}) with the shaded region
showing the 1$\sigma$ confidence interval on the median at each value
of EW(\ensuremath{\mathrm{H}\beta}). We see a clear excess N/O for WR galaxies at
$\mbox{EW(\ensuremath{\mathrm{H}\beta})} < 100$\AA. Quantitatively we find $\left\langle \Delta
  \log (\mathrm{N/O}) \right\rangle = 0.133 \pm 0.035$ for a 3$\sigma$-clipped
mean with errors from bootstrapping. 

To calculate this figure we selected for each WR galaxy a set
of similar galaxies that show no WR features. These similar galaxies
where chosen to have oxygen abundance within 0.1 dex of the WR
galaxy and EW(\ensuremath{\mathrm{H}\beta}) to within a factor of 3 that of the WR
galaxies. We then calculate the difference in N/O between the WR
galaxies and their similar galaxies. In view of the difference in
O/H for WR and non-WR galaxies seen in
Figure~\ref{fig:abundance_vs_ewhb} it is clear that there will be a
difference in N/O between the two classes just due to secondary
enrichment of nitrogen if we had not limited our comparison to
similar galaxies.

We include all galaxies that are classified as star-forming, have $\mathrm{S/N}>5$ in
{[O\,\textsc{iii}]\,$\lambda 4363$} and are not duplicate observations of the same
region and  limiting
ourselves to spectra for which we can calculate abundances using the
$T_e$ method. To estimate confidence limits we bootstrap this procedure 9999 times and
for each iteration we draw a new realisation of the relevant
observables using their estimated uncertainties, although for EW(\ensuremath{\mathrm{H}\beta}) we
use a flat uncertainty of 2\AA, as the error on high EW(\ensuremath{\mathrm{H}\beta}) lines is
dominated by continuum subtraction uncertainties. We emphasise that
since the same algorithm is used to calculate abundances regardless of
whether a spectrum shows WR features or not, the result is robust to our abundance calculations.

The immediate result from Figure~\ref{fig:nh_vs_ewhb} is that at
$\mathrm{EW}(\ensuremath{\mathrm{H}\beta})>100$, the non-WR and WR galaxies appear to show nearly 
identical values for N/O. This is consistent with these being very
young bursts where the WR stars have not yet had a chance to enrich
the surrounding ISM to a noticeable degree. However at lower EW(\ensuremath{\mathrm{H}\beta})
there is clear and significant difference between the WR galaxies and
the non-WR galaxies in the sense that the former has a higher N/O at
the same EW(\ensuremath{\mathrm{H}\beta}).  This shows the results for the median trends and it
is perhaps even more striking that for $\log \mathrm{N/O}>-1.3$ and
$1.6 < \log \mathrm{EW}(\ensuremath{\mathrm{H}\beta}) < 2.4$ approximately 25\% of the spectra
show WR features, whereas for $\log \mathrm{N/O}<-1.6$ in the same
range in EW(\ensuremath{\mathrm{H}\beta}), only 4\% of the galaxies show WR features. This is a
striking difference and clearly associates an increased N/O with the
WR phenomenon.

There are at least two possible explanations for this. The most immediate is that we are seeing
an effect of WR winds on the surrounding ISM. This process was already
suggested by \citet{1986PASP...98.1005P} and appears to be the
most likely reason for the trends seen.

The second possibility is that this might be a selection effect. WN
stars are expected to show CNO in equilibrium proportions in their
atmosphere, and as pointed out by \citet{2000A&A...356..191C} this
leads to the typical WN star to have an earlier type, or in other
words to have higher ionisation N lines in their spectra. The
practical consequence for us is that it might be more difficult to
classify these stars because the N\,\textsc{iii}\,$\lambda 4640$ line will
become weak. In this case we would be biased against detecting WR
features in systems with lower N abundance. Again this ought to lead
to an offset between the WR and non-WR spectra that is constant with
EW(\ensuremath{\mathrm{H}\beta}) in disagreement with what is seen in
Figure~\ref{fig:nh_vs_ewhb}. 


\subsection{Wind pollution of the ISM}
\label{sec:wind_pollution}

In the preceding we emphasised the role of nitrogen. This is in part
because this is expected to be the most sensitive to stellar wind
ejecta, but we have also carried out a similar study for Ne/O and Ar/O
and we find no comparable increase in these ratios for the WR
galaxies, although the uncertainties are larger. This is expected if
the nitrogen enhancement is due to a wind but would not be expected if
the enhancement was due to enrichment from SNe. As pointed already by
\citet{1986PASP...98.1005P,1992MNRAS.255..325P} and emphasised by
\citet{1997ApJ...477..679K} one should also expect an elevated He/H if
the observed region has been polluted by the ejecta of WR
stars. We are not able to confirm this --- the two samples have
identical He/H to within the uncertainties. Is this consistent with
the idea that the increased N abundance is due to the WR wind?

To answer this we make the simple assumption that the element
abundances found in  ring nebulae around Galactic WR stars measured
by \citet{1992A&A...259..629E} are reasonably representative of the
heavy element yields of WR winds. These observations indicate that the
expected average increase in log He/H in WR ejecta is 0.18\,dex whereas
the increase in log N/O is $\sim 0.85$\,dex so one must expect a
higher sensitivity to changes in N/O than in He/H and our lack of
difference in He/H between the two samples is fully consistent with
the expectations.

It is also important to estimate whether the observed increase in N/O
is expected with what one expects based on the output of WR winds.
Starting by estimating the mass of ionised hydrogen from the
luminosity of \ensuremath{\mathrm{H}\beta},
$L_{\ensuremath{\mathrm{H}\beta}}$, through
\citep[e.g.][]{2003adu..book.....D}:
\begin{equation}
  \label{eq:m_ionised}
  M_{\mathrm{ionised}} = \frac{m_\mathrm{H}*L_{\ensuremath{\mathrm{H}\beta}}}{1.235\times
    10^{-25} T_4^{-0.86} n_e},
\end{equation}
where $m_\mathrm{H}$ is the mass of the hydrogen atom, $T_4$ the
electron temperature in units of 10,000\,K and $n_e$ the electron
density. 

\begin{figure}
  \centering
  \includegraphics[angle=90,width=88mm]{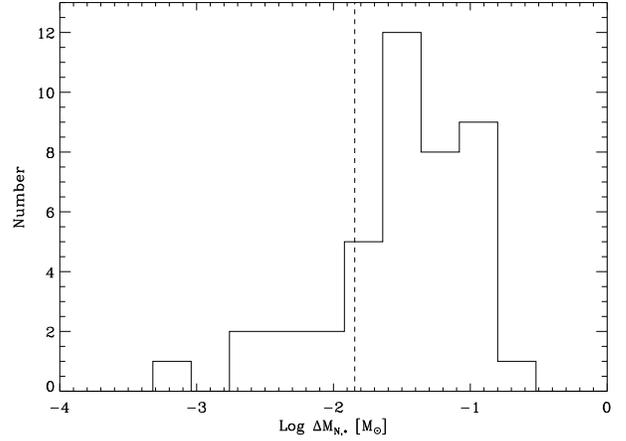}
  \caption{The distribution of mass of nitrogen ejected in each system
    per star, $\Delta M_{N,*}$, see text for details. The
    dashed vertical line shows the mass of nitrogen expected from a
    mass loss of     $10^{-5}M_\odot/$\,yr over 2 Myr with the
    abundance pattern consistent with Galactic WR ring nebulae.} 
  \label{fig:m_ejected}
\end{figure}

To quantify the nitrogen enrichment we used the sample of similar
galaxies discussed above to calculate
\begin{equation}
  \label{eq:delta_N}
  \Delta \left(\frac{N}{H} \right) =
  \left(\frac{N}{H}\right)_{\mathrm{WR}}  -
  \left(\frac{N}{H}\right)_{\mathrm{ref}},
\end{equation}
where $(N/H)_{\mathrm{ref}}$ is the average nitrogen abundance in the
galaxies similar to the WR galaxy. This way we can calculate the mass
of nitrogen required to explain the offsets
\begin{equation}
  \label{eq:mn_excess}
  \Delta M_N = \frac{m_N}{m_H} M_{\mathrm{ionised}} \Delta
  \left(\frac{N}{H}\right).
\end{equation}

This gives the mass of nitrogen for each system. To estimate the mass
of nitrogen ejected per star we need an estimate of the number of WR
stars that have contributed to the enrichment process. This is
non-trivial to estimate so we make the simple assumption that the
present Wolf-Rayet population is a good estimate of the stars involved
in the enrichment process. In effect we assume that the WR stars
observed are the result of the enrichment process.  Given the
approximate nature of these estimates we approximate the number of WR
stars by dividing the blue bump luminosity, $L_{\mathrm{BB}}$, by a
fiducial luminosity, $L_{\mathrm{ref}}$, taken to be $5\times
10^{35}$\,erg/s, giving $N_{\mathrm{WR}}^L = L_{\mathrm{BB}} /
L_{\mathrm{ref}}$.

We would like to compare these quantities to the nitrogen ejected in
Wolf-Rayet ring nebulae. In the Milky Way \citet{1992A&A...259..629E}
found $\mathrm{N/O} \sim -0.47$ \citep[see also][]{1999AJ....117.1433C}. This gives the total mass
nitrogen ejected by stellar winds with a chemical composition like
that of Galactic WR ring nebulae
\begin{equation}
  \label{eq:m_ring}
  M_N^{\mathrm{ring}} = \left(\frac{m_N}{m_H}\right) \left(\frac{O}{H}
    \right) M_{\mathrm{ejected}}
    \left(\frac{N}{O}\right)_{\mathrm{ring}},
\end{equation}
where $M_{\mathrm{ejected}}$ is the total mass ejected from a star. As
a rough approximation we take this to be $10^{-5} M_\odot$/yr over a
period of 2 Myr, or equivalently, a total mass loss of 20\,$M_\odot$
per star. We have here ignored the depletion of oxygen found in ring
nebulae by \citet{1992A&A...259..629E} which is less obvious in the
resolved studies of NGC 5253.

We then approximate the amount of N ejected per star, $\Delta
M_{N,*}$, by $\Delta M_N$ divided by $N_{\mathrm{WR}}^L$. We plot this
in Figure~\ref{fig:m_ejected} and the median $M_N^{\mathrm{ring}}$ is
indicated by the dashed line. Given the rough estimates the agreement
is encouraging. Indeed, our calculations assumed that only the
currently visible WR population has contributed to the enrichment process,
but there might indeed have been additional sources in the past which
would bring the two estimates into better agreement. While
  there could also be a contribution from OB stars, given their
  observed elevated  surfaced abundance
  \citep[e.g.][]{2008A&A...479..541H,2007A&A...466..277H}, it is
  not clear that this would cause a systematic offset at low EW(H$\beta$).

\section{Modelling Wolf-Rayet features}
\label{sec:models}

In the preceding we have carried out an empirical study of our sample
and the trends seen are in qualitative agreement with the trends seen
when studying individual Wolf-Rayet stars in nearby galaxies. However
it is well-known \citep[e.g.][]{1997A&A...326L..17L,2000ApJ...531..776G,2004MNRAS.355..728F,2007ApJ...655..851Z} that there is a significant
discrepancy between the predictions of models for Wolf-Rayet stars and
the observations at low metallicity. This has led a number of authors
to explore models with modified IMFs, the
most recent study being that of Zhang et al (2007). Given the scant observational
evidence in favour of a variable IMF we feel it is more appropriate at
this point to assess the weaknesses and strengths of models for WR
features before appealing to a change in the IMF.

To gain a more thorough understanding of the properties of the
galaxies in our sample we need to model the luminosity of the
Wolf-Rayet features using stellar evolutionary models.  We base
ourselves on the methodology of \citet[][SV98]{1998ApJ...497..618S}, who
presented a model for the spectral evolution of WR stars including
their emission lines. The basic framework of this model has since been
included in the Starburst 99 stellar population synthesis code
\citep{1999ApJS..123....3L} with refined stellar spectra from
\citet{2002MNRAS.337.1309S}. We have adopted this code with the
modifications discussed below, to produce instantaneous burst models
for the evolution of Wolf-Rayet features.

When comparing measured WR features to models it is common to
  use very simple star formation histories, either single-burst or
  constant star formation \citep[e.g. SV98,
  ][]{2000ApJ...531..776G,2004MNRAS.355..728F,2005A&A...429..581M,2007ApJ...655..851Z}.
  While appealing because of its simplicity there are two main
  problems with the approach. The first is that it is commonly assumed
  that EW(\ensuremath{\mathrm{H}\beta}) can be used as an age indicator for the bursts in
  question. While it is often acknowledged that this is an uncertain
  approximation it is then adopted for comparison to models
  \citep[e.g.][]{2007ApJ...655..851Z}. However for large samples of WR
  galaxies spanning a wide range in metallicity and size this might
  lead to strong biases.
  
  The second problem is that while an assumption of an instantaneous
  burst is a reasonable approximation for the analysis of
  H\,\textsc{ii}-regions, it is less ideal for the analysis of our data since we
  probe larger regions of the galaxy where the star formation history
  is likely to be more complex. Thus we have combined these
  predictions from Starburst 99 with the
  \citet[][BC03]{2003MNRAS.344.1000B} models to generate predictions
  for the different Wolf-Rayet emission lines for a wide range of star
  formation histories.

SV98 based their models on the best empirical compilations of
Wolf-Rayet line fluxes available at the time, and while this has stood
the test of time quite well, a number of refinements have been
published recently. We have updated the line fluxes used by SV98 with
the results of the study by \citet{2006A&A...449..711C} who found
that lower metallicity systems show a lower \ensuremath{\mbox{He\,\textsc{ii}}}\ luminosity and who
provide line luminosities for a more uniformly analysed sample of WN
stars. As discussed in Brinchmann, Pettini \& Charlot (2008; BPC08), we have
adopted their WN class 5 as representative for WNE stars. For WC stars
we have adopted the same luminosities in SV98 as reference.

\ensuremath{\mbox{He\,\textsc{ii}}}\ lines are also produced by O\,\textsc{i}f stars. At high
metallicity these do not contribute greatly to the \ensuremath{\mbox{He\,\textsc{ii}}}\ line
luminosity, contributing just a few \% around solar metallicity.
However at low metallicity the Wolf-Rayet phase is less prominent and
coupled with the reduced line luminosity this does mean that contribution
to the \ensuremath{\mbox{He\,\textsc{ii}4686}} line for O\,\textsc{i}f stars is more
important. The luminosity of \ensuremath{\mbox{He\,\textsc{ii}}}\ from these stars was provided by
P.\ Crowther (2007, priv.\ comm.) and is given in Table~1 in BPC08.

However even with these improvements in modelling, the fact remains
that the predictions of the models rely on a number of steps that
are not necessarily robust. The assignment of observed Wolf-Rayet
properties to stellar tracks is somewhat uncertain and
as pointed out by BPC08 there is a significant spread in the observed
\ensuremath{\mbox{He\,\textsc{ii}}}\ 4686\ luminosities even at near constant metallicity and hence
predictions are sensitive to the properties of the most extreme stars
in any star formation episode.

Furthermore the exact stellar tracks followed by massive stars suffer
from uncertainties, both due to rotation
\citep[e.g.][]{2003A&A...404..975M,2005A&A...429..581M} and to wind loss and binary evolution \cite[e.g][]{2003A&A...400...63V,2007ApJ...662L.107V,2008MNRAS.tmp..111E}. We focus on single-star models for now and use the
tracks of the Geneva High Mass loss tracks of
\citet{1994A&AS..103...97M} since full sets of tracks including
rotation are not yet available. However as the minimum masses for the
onset of the WR phase are relatively similar in these tracks and those
including rotation by \citet{2005A&A...429..581M}, we do not expect
substantial differences, a fact supported by the study of
\citet{2007ApJ...663..995V}.

\subsection{Comparison to models}
\label{sec:models_vs_data}

The luminosity of the WR features and their strength relative to other
spectral properties in our spectra depend on both age, star formation
history and metallicity. To simplify comparisons we have grouped our
spectra into four metallicity bins according to the oxygen abundance
estimate, similar to the bins adopted by \citet{2000ApJ...531..776G}.
We assign galaxies with $12 + \log \mathrm{O/H}<8.1$ to the models
with $Z=0.2 Z_\odot$. For the most metal poor galaxies this model
metallicity is too high, but we will see that even with this
conservative model, the discrepancies between models and data are
substantial. The $Z=0.4 Z_\odot$ models are compared to galaxies with
$8.1<12 + \log \mathrm{O/H}<8.4$ and the range $8.4< 12 + \log
\mathrm{O/H}<8.8$ is compared to solar metallicity tracks, assuming a
solar abundance of $\mathrm{O/H}=8.65$ \citep{2004A&A...417..751A}. For galaxies
with higher oxygen abundances we compare to the predictions of
$Z=1.5Z_\odot$ models which are obtained from the $Z=Z_\odot$ and
$Z=2.5Z_\odot$ models by linear interpolation in $\log Z$.

\begin{figure*}
  \centering
  \includegraphics[angle=90,width=170mm]{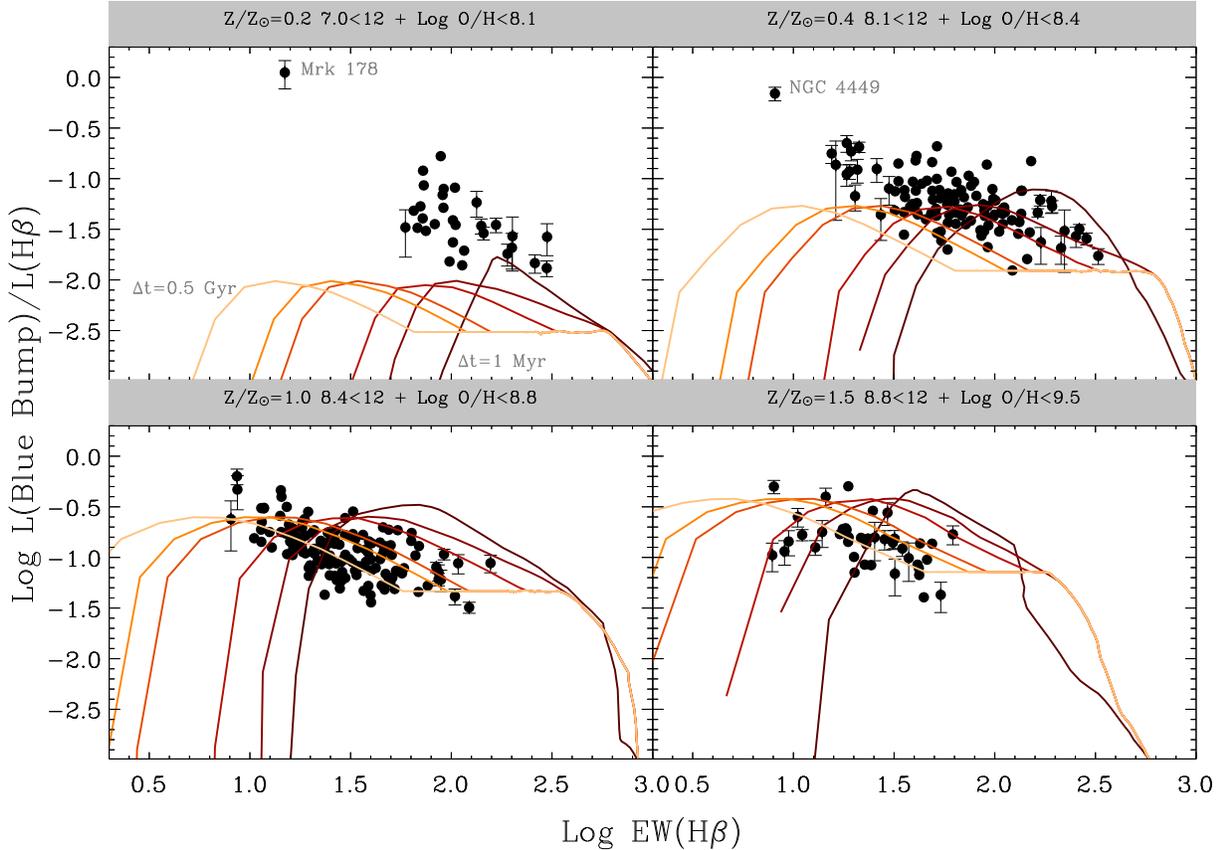} 
  \caption{The blue bump luminosity relative to the \ensuremath{\mathrm{H}\beta}\ luminosity as
    a function of EW(\ensuremath{\mathrm{H}\beta}) in four different metallicity bins. The
    range in oxygen abundance used for each bin is indicated in the
    grey box above each panel. The stellar metallicity adopted for the
    theoretical tracks is also indicated in this box and are
    $Z/Z_\odot=0.2$, $Z/Z_\odot=0.4$, $Z=Z_\odot$ and $Z/Z_\odot=1.5$
    respectively. The coloured lines show the tracks of the models
    described in the text. Moving from left to right they are for
    burst duration 0.5 Gyr down to 1 Myr in steps of a factor of 2.}  
  \label{fig:lbbump_lhb_vs_ewhb_model_comp}
\end{figure*}

\begin{figure*}
  \centering
  \includegraphics[angle=90,width=170mm]{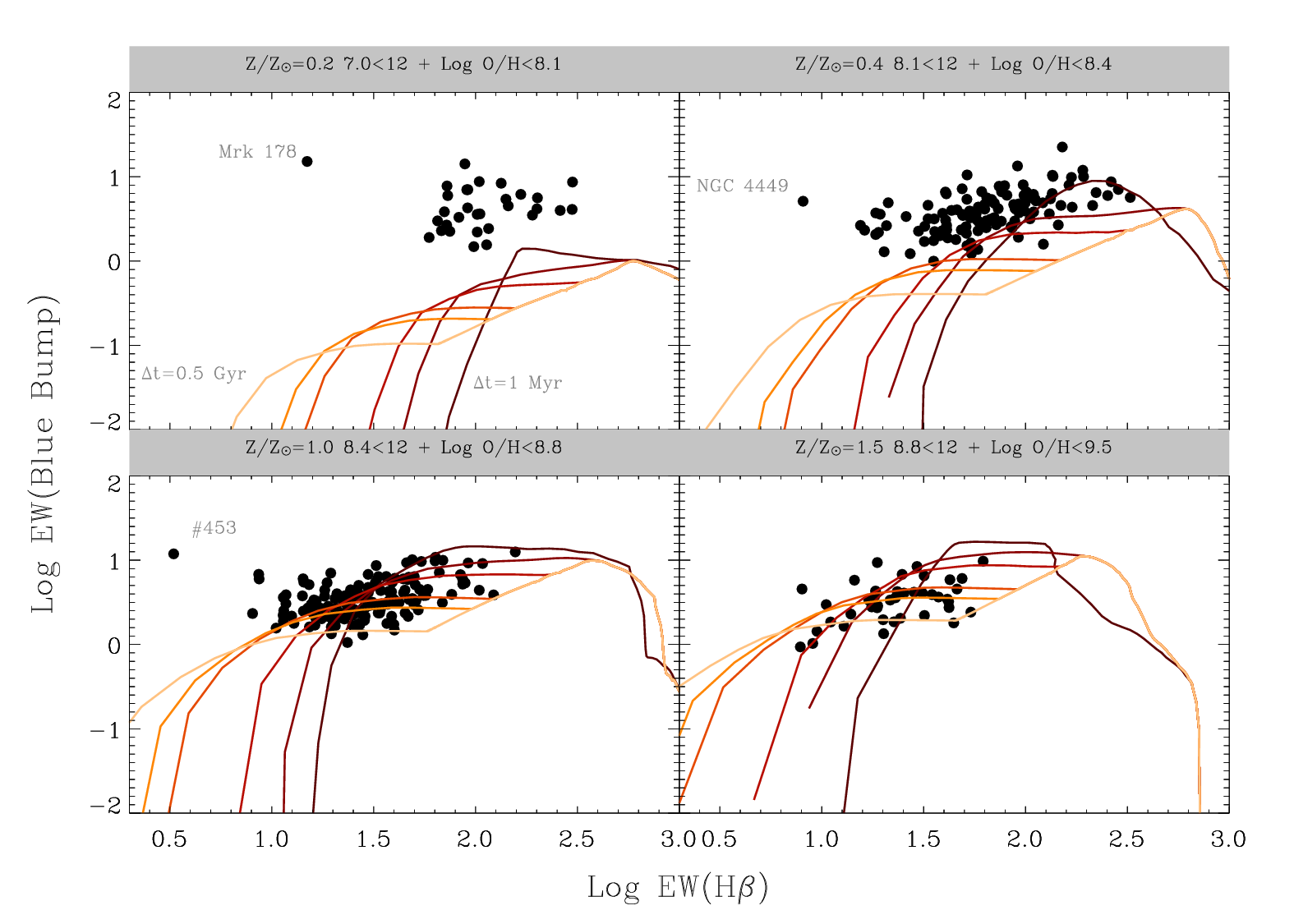}
  \caption{The equivalent width of the blue bump as a function of the
    equivalent width of the \ensuremath{\mathrm{H}\beta}\ line in different metallicity
    ranges. The model tracks are as in
    Figure~\ref{fig:lbbump_lhb_vs_ewhb_model_comp}.}  
  \label{fig:ew_bb_vs_ewhb_model_comp}
\end{figure*}

Figure~\ref{fig:lbbump_lhb_vs_ewhb_model_comp} shows a comparison of
the ratio of the blue bump luminosity to the
\ensuremath{\mathrm{H}\beta}\ luminosity as a function of the
equivalent width of \ensuremath{\mathrm{H}\beta}\ (see also
Figure~\ref{fig:wr_vs_hb_mixed}). The different panels correspond to
the different metallicity bins as indicated above each panel. The
uncertainty estimates are shown when there are less than 5 galaxies in
bins of 0.1 dex in $\log $EW(\ensuremath{\mathrm{H}\beta}).

To illustrate the dependence on star formation history we show tracks
for different burst durations, $\Delta t$. The left-most track is for
$\Delta t=0.5$\,Gyr and the right-most track for $\Delta t=1$\,Myr
with the ones in between differing by a factor of 2 in $\Delta t$. A
similar effect would be seen when adding an underlying stellar
population.  $L(\mathrm{blue bump})/L(\ensuremath{\mathrm{H}\beta})$
increases with time until the end of the burst when both lines fade
away.

It is obvious that the models do not cover the range of the data in
the lowest metallicity bin, and indeed for the most extreme galaxy in
the sample, Mrk 178, the discrepancy is close to 2 orders of
magnitude. It is however interesting to note that the gradient with
respect to EW(\ensuremath{\mathrm{H}\beta}) is fairly similar in the
model and in the data; this could be interpreted to say that the
\emph{time-scale} of the WR evolution is approximately correct but
that the absolute number of WR stars is strongly underpredicted.

The 40\% solar models are in better agreement with the data,
especially when it is recalled that the curves can be translated to
the left by adding an underlying older population or if there is different attenuation of the continuum and lines \citep[e.g.][]{1997AJ....113..162C}. It is however clear
that even with this caveat the current models do underpredict the
strength of the WR features to a certain degree.

For solar and super-solar metallicities the agreement is good between
models and data as one might expect since this is where the
observational data on Wolf-Rayet stars is best and where the
evolutionary models for stars have been better tested. Indeed it is
clear that where the models are well-constrained by observations in
the LMC and Milky Way the agreement between models and data is
satisfactory but for lower metallicity systems it is increasingly
poor.

A similar conclusion can be reached by looking at the equivalent width
of the blue bump relative to EW(\ensuremath{\mathrm{H}\beta}) as shown in
Figure~\ref{fig:ew_bb_vs_ewhb_model_comp}. This figure uses the same
metallicity bins and model tracks. However in contrast to the
preceding figure, an addition of an underlying population shifts the
model tracks down and to the left. Thus it is hard to reconcile even
the 40\% solar metallicity models with the data. Otherwise the agreement between
data and models is as seen before.

\subsection{The origin of the discrepancies}
\label{sec:origin}

The mismatches between models and data have been known for some time
now
\citep[e.g.][]{1997A&A...326L..17L,2000ApJ...531..776G,2006A&A...449..711C}
although they have usually not been made as explicit because previous
studies have formulated the mismatch using derived quantities such as
the number of WR stars, $N(\textrm{WR})$. However in view of what we
have seen above we feel it is inappropriate to attempt to derive the
number of WR stars using this kind of approach for the full sample, so
we have focused on a comparison in the observed frame.

As mentioned above, previous studies have often attempted to explain
the discrepancy between observations and models by appealing to a
change in the IMF. While we cannot rule this out with the current data
we note that there is a lack of independent observational support for
a significant change in the IMF with metallicity so we have opted here
to fix the IMF.

This leaves us with a significant, up to an order of magnitude,
discrepancy between models and data at low metallicity. This is of
similar magnitude to that identified by \citet{1997A&A...326L..17L} in
their study of I Zw 18 and recently confirmed by
\citet{2006A&A...449..711C}. Since cosmological GRBs appear to originate in low metallicity hosts \citep[e.g.][]{2007ApJ...666..267P} and potentially
from Wolf-Rayet progenitors, resolving this issue is of major
importance not only from the point of view of Wolf-Rayet star
formation but also to understand the possible progenitor population
for GRBs.

We emphasise here that the problem is in reproducing the amount of
Wolf-Rayet stars relative to OB stars
(Figure~\ref{fig:lbbump_lhb_vs_ewhb_model_comp}) and relative to the
full underlying stellar population
(Figure~\ref{fig:ew_bb_vs_ewhb_model_comp}). Since we make the
assumption that the IMF is constant there are three possible sources
for the discrepancy \citep[see also][ for a similar
discussion]{2007ARA&A..45..177C}. 

Firstly, it could be that the maximum
line luminosities of the WR lines starts to increase again at the
lowest metallicities. Since this requires a reversal of the trend
towards weaker lines at lower metallicities 
\citep{2006A&A...449..711C} this appears very unlikely and indeed
probably unphysical.

The second possibility is that the life-time of the Wolf-Rayet phase is
longer at low metallicity than in the models used here. This appears
to be the case for models including rotation as seen in the work of
\citet{2007ApJ...663..995V}. However it is not nearly sufficient to
bridge the gap between models and data at low metallicity.

The final possibility is related to the previous, namely to produce
more WR stars than currently favoured by the models. As discussed by
\citet{2007ARA&A..45..177C} this can either come from changes in the
stellar tracks when rotation is introduced or when binarity is
included in the evolutionary codes. It is well known that including
rotation in the stellar evolution lowers the minimum mass required to
form a Wolf-Rayet star \citep{2005A&A...429..581M}, but the
differences to the Geneva high mass-loss tracks used here are small at
metallicities of 40\% solar and above. 

The effect of binaries on the evolution of massive stars has been studied quite extensively although the number of relatively poorly known parameters have meant that it is rarely included in population synthesis codes \citep[see][ for two
exceptions]{2007MNRAS.380.1098H,2003A&A...400...63V}. However this is another promising source for an increased WR population although initial results of surveys for
binarity among the WR population in the Magellanic clouds did not find
an increased binarity at low metallicity
\citep{2003MNRAS.338.1025F,2003MNRAS.338..360F}.

Quantitatively the requirement from
Figure~\ref{fig:lbbump_lhb_vs_ewhb_model_comp} is that to achieve a
match, except for NGC 4449, for the $Z=0.4Z_\odot$ case we need to
increase the predicted blue bump luminosity by a factor of 2. This is
in fact in good agreement with the increase in WR/O ratio for constant
star formation rate, between the tracks we use here and the Geneva
tracks with rotation used by \citet[][their Figure
7]{2007ApJ...663..995V}. Thus we do not consider the discrepancy
between models and data for the 40\% solar case to be of much
concern.

\begin{figure}
  \centering
  \includegraphics[angle=90,width=88mm]{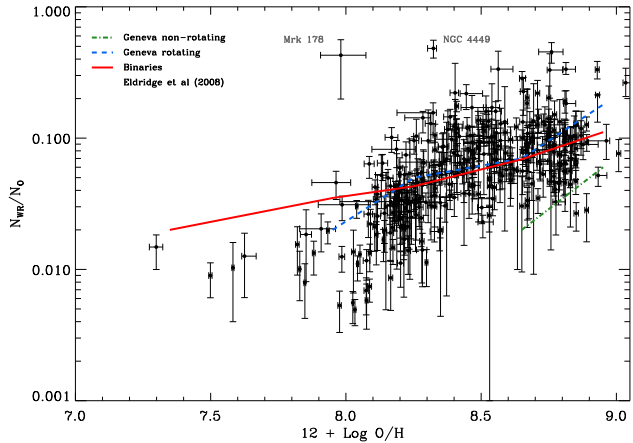}
  \caption{The ratio of the number of WR stars to that of O stars as a
  function of oxygen abundance, where we have assumed that the solar
  oxygen abundance is 8.65. The solid line shows the predictions of
  the binary model by~\citet{2008MNRAS.tmp..111E}, the dashed line the
  Geneva tracks for rotating stars and the dash-dotted line for
  non-rotating Geneva tracks, both from~\citet{2005A&A...429..581M}.}  
  \label{fig:nwr_no_vs_oh}
\end{figure}

However the situation is rather different for the lowest
metallicity. We would need to scale the predictions for the luminosity
of the blue bump at $Z=0.2Z_\odot$ by a factor of 7--10 to match the
data, with the exception Mrk 178. This is considerably more than the
difference due to rotation shown by \citet{2007ApJ...663..995V}, although newer versions of their models may reduce this deficiency \citep{2008arXiv0802.2805M}. This
might indicate that a binary channel for producing  Wolf-Rayet stars
is important at very low metallicity as suggested by
\citet{1997A&A...326L..17L}. This was investigated by \citet{2003A&A...400...63V} who showed that the inclusion of binaries led to a prolonged WR phase. More recently \citet{2008MNRAS.tmp..111E} have shown explicitly how this leads to a increase in N(WR)/N(O) relative to single star models. At low metallicity this becomes very important because single star models are less efficient at producing Wolf-Rayet stars.

In absence of complete evolutionary tracks for these recent models it
is not possible to compare to the data shown in
Figure~\ref{fig:lbbump_lhb_vs_ewhb_model_comp}
and~\ref{fig:ew_bb_vs_ewhb_model_comp}. However we can make a first
comparison to the models by calculating
$N_{\mathrm{WR}}/N_{\mathrm{O}}$ using the data in
Figure~\ref{fig:wr_vs_hb_mixed}. To do this we need to estimate the
average blue bump luminosity per WR star and the average H$\beta$
luminosity per O star. We do this by fitting a grid of models to the
observed colours and EW(H$\beta$) as described in
section~\ref{sec:fitting}. The conversion ratio is well-constrained,
although we do caution that there might be systematic uncertainties,
in particular we might overestimate the luminosity of WR stars at low
luminosity.

We plot the inferred $N_{\mathrm{WR}}/N_{\mathrm{O}}$ in
Figure~\ref{fig:nwr_no_vs_oh}. This is very similar to
Figure~\ref{fig:wr_vs_hb_mixed} as the average luminosities vary only
weakly with time. The well-known decline with metallicity
\citep[e.g.][]{2007ARA&A..45..177C} is seen again, but there is clear
evidence of a flattening of the relation at low metallicity. This
appears to be a robust result as 
$N_{\mathrm{WR}}/N_{\mathrm{O}}$ might be underestimated at low
metallicity if the trend for declining WR luminosity at lower
metallicity \citep{2006A&A...449..711C} continues at metallicities
below that of the SMC.

The solid line in the figure shows the values predicted by the models
of stellar evolution including massive binaries by
\citet{2008MNRAS.tmp..111E}, assuming a solar oxygen abundance of
8.65. The dashed and dashed-dotted lines show predictions for rotating
and non-rotating single stars from \citet{2005A&A...429..581M}. All predictions are for models with constant star formation and changing the star formation history will move the curves. In view of this and the general model uncertainties, it is reasonable to view agreement within a factor of two as satisfactory.

It is clear that extending the dashed lines to lower metallicity would lead
to significant underprediction of WR features and the models without
rotation are very strongly ruled out in agreement with various
previous studies. The model with binaries on the other hand appears to
reproduce the data quite well even at low metallicity.  However given
that binaries and rotation both give similar predictions at higher
metallicity, it is not clear what the relative importance of rotation
and binaries will be.

Finally we point out that while the N(WC)/N(WN) ratio provides
  a potentially very useful constraint on models for Wolf-Rayet
  stars \citep[e.g.][]{1999A&A...341..399S,2005A&A...429..581M}, we
  have opted to postpone a discussion of this ratio for 
  later work as measurements of red bump fluxes are significantly less
  secure than that of the blue bump.

\begin{figure*}[tbp]
  \centering
  \includegraphics[width=170mm]{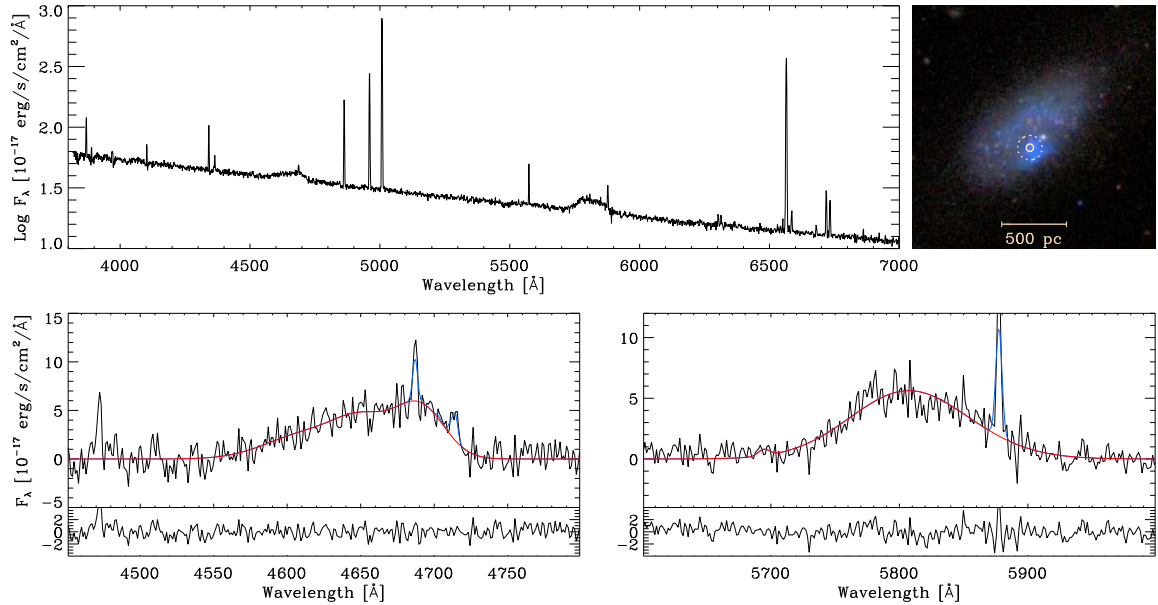}
  \caption{Top row: On the left, the rest-frame SDSS spectrum of Mrk
  178 in units of $\log F_\lambda$, the blue and red bump are both
  very prominent, note also the relatively weak emission lines.  On
  the right a colour image of Mrk 178 from the SDSS with the location
  of spectroscopic observations indicated by the circles. The outer,
  dashed, circle is 10" in diameter while the inner circle shows the
  size of the SDSS fibre, 3" in diameter. The  bottom row shows the
  continuum subtracted spectra around the blue bump, on the left, and
  the red bump on the right, with the residuals after subtracting the
  fit to the nebular and WR lines indicated in the panel below each as
  in Figure~\ref{fig:bbump}.} 
  \label{fig:mrk178}
\end{figure*}

In the preceding we did not discuss the highly significant outliers,
NGC 4449, our object \#199 and Mrk 178, which we will discuss in some
more detail here. Mrk 178 is a starbursting dwarf galaxy in the Canes
Venatici group and we adopt a distance of 3.89 Mpc for this object
based on the tip of the Red Giant Branch distance from
\citet{2003A&A...398..467K}. The oxygen abundance of $\mbox{12 + log
  O/H} = 7.73 \pm 0.08$  and the equivalent width of $15.4 \pm
0.4$\AA\ are in excellent agreement with Guseva et al (2000) who found
$7.82\pm 0.06$ and $15.92 \pm 0.5$\AA\ respectively from an
independent observation of the galaxy. 

Mrk 178 has one major star forming region to its south where the SDSS
spectrum, shown in the top row of Figure~\ref{fig:mrk178}, is also
obtained. The prominence of the WR features is obvious and as the
bottom row of the figure shows, these are very broad, featureless
bumps. The strength of the red bump shows clearly that the main source
is WC stars as pointed out by Guseva et al as well. The immediate
impression from Figure~\ref{fig:mrk178} as compared to
Figure~\ref{fig:spectrum_samples} is the prominence of the WR features
relative to the nebular lines. Indeed, $\log L(\mathrm{Blue
  Bump})/L(\ensuremath{\mathrm{H}\beta}) \approx 0.05$ would convert into a log N(WR)/N(O)$\sim
0.5$, much higher than expected from any model. 

However, the luminosity of the blue \& red bumps is only $\sim
10^{37}$\,erg/s, so we expect only a few Wolf-Rayet stars to be sampled. Thus
the cause of the offset is almost certainly just that the SDSS fibre,
which projects to a size of $56$\,pc, by chance samples a region that
is overabundant in Wolf-Rayet stars. This appears also to be the case
for NGC 4449, and in this case we also have several spectra additional
of star forming regions in NGC 4449, none of which show an offset in
the relationships plotted above.

\section{Summary}
\label{sec:discussion}

We have presented here a study of galaxies with Wolf-Rayet features in
their spectra for a carefully selected sample spanning an
unprecedented wide range of physical properties. It more than doubles
the number of known Wolf-Rayet galaxies and has a well understood
selection function. This has allowed use to carry our a number of
empirical studies of the abundance of WR stars with metallicity and
evolutionary state. 

We have shown that by fitting Wolf-Rayet features carefully we can
recover the distribution of line widths in Wolf-Rayet stars and we
have argued that this shows that we can accurately recover the flux of
the Wolf-Rayet features. This has resulted in a sample of WR
galaxies with a completeness limit of $\mathrm{EW(Blue bump)} \approx
1$\,\AA\ and $\mbox{EW(\ensuremath{\mathrm{H}\beta})} > 2$\,\AA. 

We find that the abundance of Wolf-Rayet stars is a strong function
both of the oxygen abundance of the galaxy as well as of the star
formation intensity as measured by the equivalent width of
\ensuremath{\mathrm{H}\beta}. Intriguingly we find that above $\mathrm{EW}(\ensuremath{\mathrm{H}\beta})\sim 200$\,\AA\
the fraction of galaxies showing signs of WR stars appear to start to
decline. While the EW(\ensuremath{\mathrm{H}\beta}) is a questionable age indicator for the
spectra in our sample, this does appear to be consistent with current
theoretical predictions for the onset of Wolf-Rayet formation at
1--2\,Myr. 

We also find that galaxies that show Wolf-Rayet features in their
spectra have a nitrogen abundance that is $\approx 0.1$\,dex higher
than systems that do not show Wolf-Rayet features. We have argued that
this appears to be the result of pollution of the ISM from Wolf-Rayet
winds. The observed increases in N/O are consistent with the result of
typical Wolf-Rayet winds releasing N into the ISM over a period of a
few Myr.  The present study is limited by the number of galaxies with
high quality nitrogen abundance measurements and it would be very
interesting to extend this study to galaxies with higher oxygen
abundance and less intense star formation and to empirically determine
the region of influence of the WR winds 

Finally we have also examined whether Wolf-Rayet stars are responsible
for the ionisation of \ensuremath{\mbox{He\,\textsc{ii}}} causing the nebular \ensuremath{\mbox{He\,\textsc{ii}}}{\,$\lambda
  4686$}. We were able to show that the time-scales for the WR phase
and the nebular \ensuremath{\mbox{He\,\textsc{ii}}} are different at low metallicity and that the
dominant contribution of the ionisation of \ensuremath{\mbox{He\,\textsc{ii}}} here is likely to
be massive O stars. By inference, the winds of low metallicity massive
O stars must be weaker than high metallicity equivalents to allow for
the escape of $\lambda < 228$\,\AA\ photons. At higher metallicity we
find that it is very likely that WN stars are contributing
significantly to the production of \ensuremath{\mbox{He\,\textsc{ii}}} ionising radiation. While
these are model-independent inferences it is clear that it would be
very valuable to compare these results with detailed models of high
mass stars at low metallicity to understand the evolution of wind
strengths with metallicity.

\begin{acknowledgements}
  
  We would like to thank St{\'e}phane Charlot, G.\ Comte, Thierry
  Contini, Paul Crowther, Tim Heckman, Max Pettini, D.\ Rosa-Gonzalez,
  E.\ and R.\ Terlevich, D.\ Schaerer, G.\ Meynet and J. Eldridge for fruitful discussions and help on various
  topics in this paper. In particular we gratefully acknowledge feedback on an earlier version of the  paper from G.\ Meynet, D.\ Schaerer and J.\ Eldridge. We warmly acknowledge assistance from and
  discussions with Gilles Missonnier and thank G.\ Bruzual and S.\ 
  Charlot for providing an updated version of their models prior to
  publication. JB acknowledges the receipt of FCT grant
  SFRH/BPD/14398/2003 and support from FCT grant
  PTDC/CTE-AST/66147/2006. The project has a web page at
  \texttt{http://www.strw.leidenuniv.nl/~jarle/WRinSDSS/}. 

Funding for the Sloan Digital Sky Survey (SDSS) and SDSS-II has
been provided by the Alfred P. Sloan Foundation, the Participating
Institutions, the National Science Foundation, the U.S. Department
of Energy, the National Aeronautics and Space Administration, the
Japanese Monbukagakusho, and the Max Planck Society, and the
Higher Education Funding Council for England. The SDSS Web site is
\texttt{http://www.sdss.org/}. 

The SDSS is managed by the Astrophysical Research Consortium (ARC)
for the Participating Institutions. The Participating Institutions
are the American Museum of Natural History, Astrophysical
Institute Potsdam, University of Basel, University of Cambridge,
Case Western Reserve University, The University of Chicago, Drexel
University, Fermilab, the Institute for Advanced Study, the Japan
Participation Group, The Johns Hopkins University, the Joint
Institute for Nuclear Astrophysics, the Kavli Institute for
Particle Astrophysics and Cosmology, the Korean Scientist Group,
the Chinese Academy of Sciences (LAMOST), Los Alamos National
Laboratory, the Max-Planck-Institute for Astronomy (MPIA), the
Max-Planck-Institute for Astrophysics (MPA), New Mexico State
University, Ohio State University, University of Pittsburgh,
University of Portsmouth, Princeton University, the United States
Naval Observatory, and the University of Washington. 

This research has made use of the NASA/IPAC Extragalactic Database
(NED) which is operated by the Jet Propulsion Laboratory, California
Institute of Technology, under contract with the National Aeronautics
and Space Administration. The research has also made extensive use of
the VizieR catalogue access tool, CDS, Strasbourg, France as well as
NASA's Astrophysics Data System Bibliographic Services. The analysis
has also made use of the Perl Data Language (PDL,
\texttt{http://pdl.perl.org}), the Interactive Data Language (IDL) and
the R project for Statistical Computing
(\texttt{http://www.r-project.org}).

\end{acknowledgements}


\begin{thebibliography}{119}
\expandafter\ifx\csname natexlab\endcsname\relax\def\natexlab#1{#1}\fi

\bibitem[{{Adelman-McCarthy} {et~al.}(2008){Adelman-McCarthy}, {Ag{\"u}eros},
  {Allam}, {Allende Prieto}, {Anderson}, {Anderson}, {Annis}, {Bahcall},
  {Bailer-Jones}, {Baldry}, {Barentine}, {Bassett}, {Becker}, {Beers}, {Bell},
  {Berlind}, {Bernardi}, {Blanton}, {Bochanski}, {Boroski}, {Brinchmann},
  {Brinkmann}, {Brunner}, {Budav{\'a}ri}, {Carliles}, {Carr}, {Castander},
  {Cinabro}, {Cool}, {Covey}, {Csabai}, {Cunha}, {Davenport}, {Dilday}, {Doi},
  {Eisenstein}, {Evans}, {Fan}, {Finkbeiner}, {Friedman}, {Frieman},
  {Fukugita}, {G{\"a}nsicke}, {Gates}, {Gillespie}, {Glazebrook}, {Gray},
  {Grebel}, {Gunn}, {Gurbani}, {Hall}, {Harding}, {Harvanek}, {Hawley},
  {Hayes}, {Heckman}, {Hendry}, {Hindsley}, {Hirata}, {Hogan}, {Hogg}, {Hyde},
  {Ichikawa}, {Ivezi{\'c}}, {Jester}, {Johnson}, {Jorgensen}, {Juri{\'c}},
  {Kent}, {Kessler}, {Kleinman}, {Knapp}, {Kron}, {Krzesinski}, {Kuropatkin},
  {Lamb}, {Lampeitl}, {Lebedeva}, {Lee}, {Leger}, {L{\'e}pine}, {Lima}, {Lin},
  {Long}, {Loomis}, {Loveday}, {Lupton}, {Malanushenko}, {Malanushenko},
  {Mandelbaum}, {Margon}, {Marriner}, {Mart{\'{\i}}nez-Delgado}, {Matsubara},
  {McGehee}, {McKay}, {Meiksin}, {Morrison}, {Munn}, {Nakajima}, {Neilsen},
  {Newberg}, {Nichol}, {Nicinski}, {Nieto-Santisteban}, {Nitta}, {Okamura},
  {Owen}, {Oyaizu}, {Padmanabhan}, {Pan}, {Park}, {Peoples}, {Pier}, {Pope},
  {Purger}, {Raddick}, {Re Fiorentin}, {Richards}, {Richmond}, {Riess}, {Rix},
  {Rockosi}, {Sako}, {Schlegel}, {Schneider}, {Schreiber}, {Schwope}, {Seljak},
  {Sesar}, {Sheldon}, {Shimasaku}, {Sivarani}, {Smith}, {Snedden}, {Steinmetz},
  {Strauss}, {SubbaRao}, {Suto}, {Szalay}, {Szapudi}, {Szkody}, {Tegmark},
  {Thakar}, {Tremonti}, {Tucker}, {Uomoto}, {Vanden Berk}, {Vandenberg},
  {Vidrih}, {Vogeley}, {Voges}, {Vogt}, {Wadadekar}, {Weinberg}, {West},
  {White}, {Wilhite}, {Yanny}, {Yocum}, {York}, {Zehavi}, \&
  {Zucker}}]{2008ApJS..175..297A}
{Adelman-McCarthy}, J.~K., {Ag{\"u}eros}, M.~A., {Allam}, S.~S., {et~al.} 2008,
  \apjs, 175, 297

\bibitem[{{Allen} {et~al.}(1976){Allen}, {Wright}, \&
  {Goss}}]{1976MNRAS.177...91A}
{Allen}, D.~A., {Wright}, A.~E., \& {Goss}, W.~M. 1976, \mnras, 177, 91

\bibitem[{{Allen}(1995)}]{1995MNRAS.276..947A}
{Allen}, S.~W. 1995, \mnras, 276, 947

\bibitem[{{Armandroff} \& {Massey}(1985)}]{1985ApJ...291..685A}
{Armandroff}, T.~E. \& {Massey}, P. 1985, \apj, 291, 685

\bibitem[{{Asari} {et~al.}(2007){Asari}, {Cid Fernandes}, {Stasi{\'n}ska},
  {Torres-Papaqui}, {Mateus}, {Sodr{\'e}}, {Schoenell}, \&
  {Gomes}}]{2007MNRAS.381..263A}
{Asari}, N.~V., {Cid Fernandes}, R., {Stasi{\'n}ska}, G., {et~al.} 2007,
  \mnras, 381, 263

\bibitem[{{Asplund} {et~al.}(2004){Asplund}, {Grevesse}, {Sauval}, {Allende
  Prieto}, \& {Kiselman}}]{2004A&A...417..751A}
{Asplund}, M., {Grevesse}, N., {Sauval}, A.~J., {Allende Prieto}, C., \&
  {Kiselman}, D. 2004, \aap, 417, 751

\bibitem[{{Baldwin} {et~al.}(1981){Baldwin}, {Phillips}, \&
  {Terlevich}}]{1981PASP...93....5B}
{Baldwin}, J.~A., {Phillips}, M.~M., \& {Terlevich}, R. 1981, \pasp, 93, 5

\bibitem[{{Binette} {et~al.}(1996){Binette}, {Wilson}, \&
  {Storchi-Bergmann}}]{1996A&A...312..365B}
{Binette}, L., {Wilson}, A.~S., \& {Storchi-Bergmann}, T. 1996, \aap, 312, 365

\bibitem[{{Blanton} {et~al.}(2003){Blanton}, {Lin}, {Lupton}, {Maley}, {Young},
  {Zehavi}, \& {Loveday}}]{2003AJ....125.2276B}
{Blanton}, M.~R., {Lin}, H., {Lupton}, R.~H., {et~al.} 2003, \aj, 125, 2276

\bibitem[{{Blanton} {et~al.}(2005){Blanton}, {Lupton}, {Schlegel}, {Strauss},
  {Brinkmann}, {Fukugita}, \& {Loveday}}]{2005ApJ...631..208B}
{Blanton}, M.~R., {Lupton}, R.~H., {Schlegel}, D.~J., {et~al.} 2005, \apj, 631,
  208

\bibitem[{{Bresolin}(2007)}]{2007ApJ...656..186B}
{Bresolin}, F. 2007, \apj, 656, 186

\bibitem[{{Brinchmann} {et~al.}(2004){Brinchmann}, {Charlot}, {White},
  {Tremonti}, {Kauffmann}, {Heckman}, \& {Brinkmann}}]{2004MNRAS.351.1151B}
{Brinchmann}, J., {Charlot}, S., {White}, S.~D.~M., {et~al.} 2004, \mnras, 351,
  1151

\bibitem[{{Brinchmann} {et~al.}(2008){Brinchmann}, {Pettini}, \&
  {Charlot}}]{2008MNRAS.385..769B}
{Brinchmann}, J., {Pettini}, M., \& {Charlot}, S. 2008, \mnras, 385, 769

\bibitem[{{Bruzual} \& {Charlot}(2003)}]{2003MNRAS.344.1000B}
{Bruzual}, G. \& {Charlot}, S. 2003, \mnras, 344, 1000

\bibitem[{{Calzetti}(1997)}]{1997AJ....113..162C}
{Calzetti}, D. 1997, \aj, 113, 162

\bibitem[{{Charlot} \& {Fall}(2000)}]{2000ApJ...539..718C}
{Charlot}, S. \& {Fall}, S.~M. 2000, \apj, 539, 718

\bibitem[{{Charlot} \& {Longhetti}(2001)}]{2001MNRAS.323..887C}
{Charlot}, S. \& {Longhetti}, M. 2001, \mnras, 323, 887

\bibitem[{{Chu} {et~al.}(1999){Chu}, {Weis}, \&
  {Garnett}}]{1999AJ....117.1433C}
{Chu}, Y.-H., {Weis}, K., \& {Garnett}, D.~R. 1999, \aj, 117, 1433

\bibitem[{{Conti}(1991)}]{1991ApJ...377..115C}
{Conti}, P.~S. 1991, \apj, 377, 115

\bibitem[{{Conti} {et~al.}(1989){Conti}, {Garmany}, \&
  {Massey}}]{1989ApJ...341..113C}
{Conti}, P.~S., {Garmany}, C.~D., \& {Massey}, P. 1989, \apj, 341, 113

\bibitem[{{Contini} {et~al.}(2001){Contini}, {Kunth}, {Mas-Hesse}, \&
  {Arribas}}]{Contini+01}
{Contini}, T., {Kunth}, D., {Mas-Hesse}, J., \& {Arribas}, A. 2001, EAS Pub.
  Series, EDP Sciences, Les Ulis, France, 1, 163

\bibitem[{{Copetti} {et~al.}(1986){Copetti}, {Pastoriza}, \&
  {Dottori}}]{1986A&A...156..111C}
{Copetti}, M.~V.~F., {Pastoriza}, M.~G., \& {Dottori}, H.~A. 1986, \aap, 156,
  111

\bibitem[{{Crowther}(2000)}]{2000A&A...356..191C}
{Crowther}, P.~A. 2000, \aap, 356, 191

\bibitem[{{Crowther}(2007)}]{2007ARA&A..45..177C}
{Crowther}, P.~A. 2007, \araa, 45, 177

\bibitem[{{Crowther} \& {Hadfield}(2006)}]{2006A&A...449..711C}
{Crowther}, P.~A. \& {Hadfield}, L.~J. 2006, \aap, 449, 711

\bibitem[{{Crowther} {et~al.}(2004){Crowther}, {Hadfield}, {Schild}, \&
  {Schmutz}}]{2004A&A...419L..17C}
{Crowther}, P.~A., {Hadfield}, L.~J., {Schild}, H., \& {Schmutz}, W. 2004,
  \aap, 419, L17

\bibitem[{{de Mello} {et~al.}(1998){de Mello}, {Schaerer}, {Heldmann}, \&
  {Leitherer}}]{1998ApJ...507..199D}
{de Mello}, D.~F., {Schaerer}, D., {Heldmann}, J., \& {Leitherer}, C. 1998,
  \apj, 507, 199

\bibitem[{{Dopita} {et~al.}(2006){Dopita}, {Fischera}, {Sutherland}, {Kewley},
  {Leitherer}, {Tuffs}, {Popescu}, {van Breugel}, \&
  {Groves}}]{2006ApJS..167..177D}
{Dopita}, M.~A., {Fischera}, J., {Sutherland}, R.~S., {et~al.} 2006, \apjs,
  167, 177

\bibitem[{{Dopita} \& {Sutherland}(1995)}]{1995ApJ...455..468D}
{Dopita}, M.~A. \& {Sutherland}, R.~S. 1995, \apj, 455, 468

\bibitem[{{Dopita} \& {Sutherland}(2003)}]{2003adu..book.....D}
{Dopita}, M.~A. \& {Sutherland}, R.~S. 2003, {Astrophysics of the diffuse
  universe} (Astrophysics of the diffuse universe, Berlin, New York: Springer,
  2003.~Astronomy and astrophysics library, ISBN 3540433627)

\bibitem[{{Eldridge} {et~al.}(2008){Eldridge}, {Izzard}, \&
  {Tout}}]{2008MNRAS.tmp..111E}
{Eldridge}, J.~J., {Izzard}, R.~G., \& {Tout}, C.~A. 2008, \mnras, 111

\bibitem[{{Erb} {et~al.}(2006){Erb}, {Shapley}, {Pettini}, {Steidel}, {Reddy},
  \& {Adelberger}}]{2006ApJ...644..813E}
{Erb}, D.~K., {Shapley}, A.~E., {Pettini}, M., {et~al.} 2006, \apj, 644, 813

\bibitem[{{Esteban} {et~al.}(1992){Esteban}, {Vilchez}, {Smith}, \&
  {Clegg}}]{1992A&A...259..629E}
{Esteban}, C., {Vilchez}, J.~M., {Smith}, L.~J., \& {Clegg}, R.~E.~S. 1992,
  \aap, 259, 629

\bibitem[{{Fernandes} {et~al.}(2004){Fernandes}, {de Carvalho}, {Contini}, \&
  {Gal}}]{2004MNRAS.355..728F}
{Fernandes}, I.~F., {de Carvalho}, R., {Contini}, T., \& {Gal}, R.~R. 2004,
  \mnras, 355, 728

\bibitem[{{Foellmi} {et~al.}(2003{\natexlab{a}}){Foellmi}, {Moffat}, \&
  {Guerrero}}]{2003MNRAS.338..360F}
{Foellmi}, C., {Moffat}, A.~F.~J., \& {Guerrero}, M.~A. 2003{\natexlab{a}},
  \mnras, 338, 360

\bibitem[{{Foellmi} {et~al.}(2003{\natexlab{b}}){Foellmi}, {Moffat}, \&
  {Guerrero}}]{2003MNRAS.338.1025F}
{Foellmi}, C., {Moffat}, A.~F.~J., \& {Guerrero}, M.~A. 2003{\natexlab{b}},
  \mnras, 338, 1025

\bibitem[{{Fukugita} {et~al.}(1996){Fukugita}, {Ichikawa}, {Gunn}, {Doi},
  {Shimasaku}, \& {Schneider}}]{1996AJ....111.1748F}
{Fukugita}, M., {Ichikawa}, T., {Gunn}, J.~E., {et~al.} 1996, \aj, 111

\bibitem[{{G{\'o}mez} {et~al.}(2003){G{\'o}mez}, {Nichol}, {Miller}, {Balogh},
  {Goto}, {Zabludoff}, {Romer}, {Bernardi}, {Sheth}, {Hopkins}, {Castander},
  {Connolly}, {Schneider}, {Brinkmann}, {Lamb}, {SubbaRao}, \&
  {York}}]{2003ApJ...584..210G}
{G{\'o}mez}, P.~L., {Nichol}, R.~C., {Miller}, C.~J., {et~al.} 2003, \apj, 584,
  210

\bibitem[{{Gunn} {et~al.}(1998){Gunn}, {Carr}, {Rockosi}, {Sekiguchi}, {Berry},
  {Elms}, {de Haas}, {Ivezi{\'c}}, {Knapp}, {Lupton}, {Pauls}, {Simcoe},
  {Hirsch}, {Sanford}, {Wang}, {York}, {et~al.}}]{1998AJ....116.3040G}
{Gunn}, J.~E., {Carr}, M., {Rockosi}, C., {et~al.} 1998, \aj, 116, 3040

\bibitem[{{Gunn} {et~al.}(2006){Gunn}, {Siegmund}, {Mannery}, {Owen}, {Hull},
  {Leger}, {Carey}, {Knapp}, {York}, {Boroski}, {Kent}, {Lupton}, {Rockosi},
  {Evans}, {Waddell}, {Anderson}, {Annis}, {Barentine}, {Bartoszek}, {Bastian},
  {Bracker}, {Brewington}, {Briegel}, {Brinkmann}, {Brown}, {Carr},
  {Czarapata}, {Drennan}, {Dombeck}, {Federwitz}, {Gillespie}, {Gonzales},
  {Hansen}, {Harvanek}, {Hayes}, {Jordan}, {Kinney}, {Klaene}, {Kleinman},
  {Kron}, {Kresinski}, {Lee}, {Limmongkol}, {Lindenmeyer}, {Long}, {Loomis},
  {McGehee}, {Mantsch}, {Neilsen}, {Neswold}, {Newman}, {Nitta}, {Peoples},
  {Pier}, {Prieto}, {Prosapio}, {Rivetta}, {Schneider}, {Snedden}, \&
  {Wang}}]{2006AJ....131.2332G}
{Gunn}, J.~E., {Siegmund}, W.~A., {Mannery}, E.~J., {et~al.} 2006, \aj, 131,
  2332

\bibitem[{{Guseva} {et~al.}(2000){Guseva}, {Izotov}, \&
  {Thuan}}]{2000ApJ...531..776G}
{Guseva}, N.~G., {Izotov}, Y.~I., \& {Thuan}, T.~X. 2000, \apj, 531, 776

\bibitem[{{Hadfield} \& {Crowther}(2006)}]{2006MNRAS.368.1822H}
{Hadfield}, L.~J. \& {Crowther}, P.~A. 2006, \mnras, 368, 1822

\bibitem[{{Hadfield} \& {Crowther}(2007)}]{2007MNRAS.381..418H}
{Hadfield}, L.~J. \& {Crowther}, P.~A. 2007, \mnras, 381, 418

\bibitem[{{Hamann} {et~al.}(2006){Hamann}, {Gr{\"a}fener}, \&
  {Liermann}}]{2006A&A...457.1015H}
{Hamann}, W.-R., {Gr{\"a}fener}, G., \& {Liermann}, A. 2006, \aap, 457, 1015

\bibitem[{{Hammer} {et~al.}(2006){Hammer}, {Flores}, {Schaerer},
  {Dessauges-Zavadsky}, {Le Floc'h}, \& {Puech}}]{2006A&A...454..103H}
{Hammer}, F., {Flores}, H., {Schaerer}, D., {et~al.} 2006, \aap, 454, 103

\bibitem[{{Han} {et~al.}(2007){Han}, {Podsiadlowski}, \&
  {Lynas-Gray}}]{2007MNRAS.380.1098H}
{Han}, Z., {Podsiadlowski}, P., \& {Lynas-Gray}, A.~E. 2007, \mnras, 380, 1098

\bibitem[{{Heckman} \& {Leitherer}(1997)}]{1997AJ....114...69H}
{Heckman}, T.~M. \& {Leitherer}, C. 1997, \aj, 114, 69

\bibitem[{{Ho} {et~al.}(1995){Ho}, {Filippenko}, \&
  {Sargent}}]{1995ApJS...98..477H}
{Ho}, L.~C., {Filippenko}, A.~V., \& {Sargent}, W.~L. 1995, \apjs, 98, 477

\bibitem[{{Hogg} {et~al.}(2001){Hogg}, {Finkbeiner}, {Schlegel}, \&
  {Gunn}}]{2001AJ....122.2129H}
{Hogg}, D.~W., {Finkbeiner}, D.~P., {Schlegel}, D.~J., \& {Gunn}, J.~E. 2001,
  \aj, 122, 2129

\bibitem[{{Hunter} {et~al.}(2007){Hunter}, {Dufton}, {Smartt}, {Ryans},
  {Evans}, {Lennon}, {Trundle}, {Hubeny}, \& {Lanz}}]{2007A&A...466..277H}
{Hunter}, I., {Dufton}, P.~L., {Smartt}, S.~J., {et~al.} 2007, \aap, 466, 277

\bibitem[{{Hunter} {et~al.}(2008){Hunter}, {Lennon}, {Dufton}, {Trundle},
  {Sim{\'o}n-D{\'{\i}}az}, {Smartt}, {Ryans}, \& {Evans}}]{2008A&A...479..541H}
{Hunter}, I., {Lennon}, D.~J., {Dufton}, P.~L., {et~al.} 2008, \aap, 479, 541

\bibitem[{{Izotov} {et~al.}(2006){Izotov}, {Stasi{\'n}ska}, {Meynet}, {Guseva},
  \& {Thuan}}]{2006A&A...448..955I}
{Izotov}, Y.~I., {Stasi{\'n}ska}, G., {Meynet}, G., {Guseva}, N.~G., \&
  {Thuan}, T.~X. 2006, \aap, 448, 955

\bibitem[{{Karachentsev} {et~al.}(2003){Karachentsev}, {Sharina}, {Dolphin},
  {Grebel}, {Geisler}, {Guhathakurta}, {Hodge}, {Karachentseva}, {Sarajedini},
  \& {Seitzer}}]{2003A&A...398..467K}
{Karachentsev}, I.~D., {Sharina}, M.~E., {Dolphin}, A.~E., {et~al.} 2003, \aap,
  398, 467

\bibitem[{{Kauffmann} {et~al.}(2006){Kauffmann}, {Heckman}, {De Lucia},
  {Brinchmann}, {Charlot}, {Tremonti}, {White}, \&
  {Brinkmann}}]{2006MNRAS.367.1394K}
{Kauffmann}, G., {Heckman}, T.~M., {De Lucia}, G., {et~al.} 2006, \mnras, 367,
  1394

\bibitem[{{Kauffmann} {et~al.}(2003{\natexlab{a}}){Kauffmann}, {Heckman},
  {Tremonti}, {Brinchmann}, {Charlot}, {White}, {Ridgway}, {Brinkmann},
  {Fukugita}, {Hall}, {Ivezi{\'c}}, {Richards}, \&
  {Schneider}}]{2003MNRAS.346.1055K}
{Kauffmann}, G., {Heckman}, T.~M., {Tremonti}, C., {et~al.} 2003{\natexlab{a}},
  \mnras, 346, 1055

\bibitem[{{Kauffmann} {et~al.}(2003{\natexlab{b}}){Kauffmann}, {Heckman},
  {White}, {Charlot}, {Tremonti}, {Brinchmann}, {Bruzual}, {Peng}, {Seibert},
  {Bernardi}, {Blanton}, {Brinkmann}, {Castander}, {Cs{\'a}bai}, {Fukugita},
  {Ivezic}, {Munn}, {Nichol}, {Padmanabhan}, {Thakar}, {Weinberg}, \&
  {York}}]{2003MNRAS.341...33K}
{Kauffmann}, G., {Heckman}, T.~M., {White}, S.~D.~M., {et~al.}
  2003{\natexlab{b}}, \mnras, 341, 33

\bibitem[{{Kennicutt} {et~al.}(2000){Kennicutt}, {Bresolin}, {French}, \&
  {Martin}}]{2000ApJ...537..589K}
{Kennicutt}, R.~C., {Bresolin}, F., {French}, H., \& {Martin}, P. 2000, \apj,
  537, 589

\bibitem[{{Kennicutt} {et~al.}(2003){Kennicutt}, {Bresolin}, \&
  {Garnett}}]{2003ApJ...591..801K}
{Kennicutt}, R.~C., {Bresolin}, F., \& {Garnett}, D.~R. 2003, \apj, 591, 801

\bibitem[{{Kewley} \& {Dopita}(2002)}]{2002ApJS..142...35K}
{Kewley}, L.~J. \& {Dopita}, M.~A. 2002, \apjs, 142, 35

\bibitem[{{Kewley} {et~al.}(2001){Kewley}, {Dopita}, {Sutherland}, {Heisler},
  \& {Trevena}}]{2001ApJ...556..121K}
{Kewley}, L.~J., {Dopita}, M.~A., {Sutherland}, R.~S., {Heisler}, C.~A., \&
  {Trevena}, J. 2001, \apj, 556, 121

\bibitem[{{Kewley} \& {Ellison}(2008)}]{2008arXiv0801.1849K}
{Kewley}, L.~J. \& {Ellison}, S.~L. 2008, accepted for ApJ, astro-ph/0801.1849

\bibitem[{{Kewley} {et~al.}(2006){Kewley}, {Groves}, {Kauffmann}, \&
  {Heckman}}]{2006MNRAS.372..961K}
{Kewley}, L.~J., {Groves}, B., {Kauffmann}, G., \& {Heckman}, T. 2006, \mnras,
  372, 961

\bibitem[{{Kniazev} {et~al.}(2004){Kniazev}, {Pustilnik}, {Grebel}, {Lee}, \&
  {Pramskij}}]{2004ApJS..153..429K}
{Kniazev}, A.~Y., {Pustilnik}, S.~A., {Grebel}, E.~K., {Lee}, H., \&
  {Pramskij}, A.~G. 2004, \apjs, 153, 429

\bibitem[{{Kobulnicky} \& {Koo}(2000)}]{2000ApJ...545..712K}
{Kobulnicky}, H.~A. \& {Koo}, D.~C. 2000, \apj, 545, 712

\bibitem[{{Kobulnicky} \& {Skillman}(1996)}]{1996ApJ...471..211K}
{Kobulnicky}, H.~A. \& {Skillman}, E.~D. 1996, \apj, 471, 211

\bibitem[{{Kobulnicky} {et~al.}(1997){Kobulnicky}, {Skillman}, {Roy}, {Walsh},
  \& {Rosa}}]{1997ApJ...477..679K}
{Kobulnicky}, H.~A., {Skillman}, E.~D., {Roy}, J.-R., {Walsh}, J.~R., \&
  {Rosa}, M.~R. 1997, \apj, 477, 679

\bibitem[{{Kunth} \& {Joubert}(1985)}]{1985A&A...142..411K}
{Kunth}, D. \& {Joubert}, M. 1985, \aap, 142, 411

\bibitem[{{Kunth} \& {Sargent}(1981)}]{1981A&A...101L...5K}
{Kunth}, D. \& {Sargent}, W.~L.~W. 1981, \aap, 101, L5

\bibitem[{{Kunth} \& {Sargent}(1983)}]{1983ApJ...273...81K}
{Kunth}, D. \& {Sargent}, W.~L.~W. 1983, \apj, 273, 81

\bibitem[{{Kunth} {et~al.}(1981){Kunth}, {Sargent}, \&
  {Kowal}}]{1981A&AS...44..229K}
{Kunth}, D., {Sargent}, W.~L.~W., \& {Kowal}, C. 1981, \aaps, 44, 229

\bibitem[{{Le Borgne} {et~al.}(2003){Le Borgne}, {Bruzual}, {Pell{\'o}},
  {Lan{\c c}on}, {Rocca-Volmerange}, {Sanahuja}, {Schaerer}, {Soubiran}, \&
  {V{\'{\i}}lchez-G{\'o}mez}}]{2003A&A...402..433L}
{Le Borgne}, J.-F., {Bruzual}, G., {Pell{\'o}}, R., {et~al.} 2003, \aap, 402,
  433

\bibitem[{{Legrand} {et~al.}(1997){Legrand}, {Kunth}, {Roy}, {Mas-Hesse}, \&
  {Walsh}}]{1997A&A...326L..17L}
{Legrand}, F., {Kunth}, D., {Roy}, J.-R., {Mas-Hesse}, J.~M., \& {Walsh}, J.~R.
  1997, \aap, 326, L17

\bibitem[{{Leitherer} {et~al.}(1999){Leitherer}, {Schaerer}, {Goldader},
  {Delgado}, {Robert}, {Kune}, {de Mello}, {Devost}, \&
  {Heckman}}]{1999ApJS..123....3L}
{Leitherer}, C., {Schaerer}, D., {Goldader}, J.~D., {et~al.} 1999, \apjs, 123,
  3

\bibitem[{{Liu} {et~al.}(2008){Liu}, {Shapley}, {Coil}, {Brinchmann}, \&
  {Ma}}]{2008arXiv0801.1670L}
{Liu}, X., {Shapley}, A.~E., {Coil}, A.~L., {Brinchmann}, J., \& {Ma}, C.-P.
  2008, accepted for ApJ, astro-ph/0801.1670L

\bibitem[{{L{\'o}pez-S{\'a}nchez} {et~al.}(2007){L{\'o}pez-S{\'a}nchez},
  {Esteban}, {Garc{\'{\i}}a-Rojas}, {Peimbert}, \&
  {Rodr{\'{\i}}guez}}]{2007ApJ...656..168L}
{L{\'o}pez-S{\'a}nchez}, {\'A}.~R., {Esteban}, C., {Garc{\'{\i}}a-Rojas}, J.,
  {Peimbert}, M., \& {Rodr{\'{\i}}guez}, M. 2007, \apj, 656, 168

\bibitem[{{Lowenthal} {et~al.}(1991){Lowenthal}, {Hogan}, {Green}, {Caulet},
  {Woodgate}, {Brown}, \& {Foltz}}]{1991ApJ...377L..73L}
{Lowenthal}, J.~D., {Hogan}, C.~J., {Green}, R.~F., {et~al.} 1991, \apjl, 377,
  L73

\bibitem[{{Lowenthal} {et~al.}(1997){Lowenthal}, {Koo}, {Guzman}, {Gallego},
  {Phillips}, {Faber}, {Vogt}, {Illingworth}, \&
  {Gronwall}}]{1997ApJ...481..673L}
{Lowenthal}, J.~D., {Koo}, D.~C., {Guzman}, R., {et~al.} 1997, \apj, 481, 673

\bibitem[{Lupton {et~al.}(2008)}]{Lupton-Submitted}
Lupton, R. {et~al.} 2008, Submitted to AJ

\bibitem[{{Maeder} \& {Meynet}(1994)}]{1994A&A...287..803M}
{Maeder}, A. \& {Meynet}, G. 1994, \aap, 287, 803

\bibitem[{{Mas-Hesse} \& {Kunth}(1991)}]{1991A&AS...88..399M}
{Mas-Hesse}, J.~M. \& {Kunth}, D. 1991, \aaps, 88, 399

\bibitem[{{Mas-Hesse} \& {Kunth}(1999)}]{1999A&A...349..765M}
{Mas-Hesse}, J.~M. \& {Kunth}, D. 1999, \aap, 349, 765

\bibitem[{{Meynet} {et~al.}(2008){Meynet}, {Ekstrom}, {Maeder}, {Hirschi},
  {Georgy}, \& {Beffa}}]{2008arXiv0802.2805M}
{Meynet}, G., {Ekstrom}, S., {Maeder}, A., {et~al.} 2008, astro-ph/0802.2805

\bibitem[{{Meynet} \& {Maeder}(2003)}]{2003A&A...404..975M}
{Meynet}, G. \& {Maeder}, A. 2003, \aap, 404, 975

\bibitem[{{Meynet} \& {Maeder}(2005)}]{2005A&A...429..581M}
{Meynet}, G. \& {Maeder}, A. 2005, \aap, 429, 581

\bibitem[{{Meynet} {et~al.}(1994){Meynet}, {Maeder}, {Schaller}, {Schaerer}, \&
  {Charbonnel}}]{1994A&AS..103...97M}
{Meynet}, G., {Maeder}, A., {Schaller}, G., {Schaerer}, D., \& {Charbonnel}, C.
  1994, \aaps, 103, 97

\bibitem[{{Nugis} \& {Lamers}(2000)}]{2000A&A...360..227N}
{Nugis}, T. \& {Lamers}, H.~J.~G.~L.~M. 2000, \aap, 360, 227

\bibitem[{{Osterbrock}(1989)}]{1989agna.book.....O}
{Osterbrock}, D.~E. 1989, {Astrophysics of gaseous nebulae and active galactic
  nuclei} (Research supported by the University of California, John Simon
  Guggenheim Memorial Foundation, University of Minnesota, et al.~Mill Valley,
  CA, University Science Books, 1989, 422 p.)

\bibitem[{{Osterbrock} \& {Cohen}(1982)}]{1982ApJ...261...64O}
{Osterbrock}, D.~E. \& {Cohen}, R.~D. 1982, \apj, 261, 64

\bibitem[{{Pagel} {et~al.}(1992){Pagel}, {Simonson}, {Terlevich}, \&
  {Edmunds}}]{1992MNRAS.255..325P}
{Pagel}, B.~E.~J., {Simonson}, E.~A., {Terlevich}, R.~J., \& {Edmunds}, M.~G.
  1992, \mnras, 255, 325

\bibitem[{{Pagel} {et~al.}(1986){Pagel}, {Terlevich}, \&
  {Melnick}}]{1986PASP...98.1005P}
{Pagel}, B.~E.~J., {Terlevich}, R.~J., \& {Melnick}, J. 1986, \pasp, 98, 1005

\bibitem[{{Pettini} \& {Pagel}(2004)}]{2004MNRAS.348L..59P}
{Pettini}, M. \& {Pagel}, B.~E.~J. 2004, \mnras, 348, L59

\bibitem[{{Pier} {et~al.}(2003){Pier}, {Munn}, {Hindsley}, {Hennessy}, {Kent},
  {Lupton}, \& {Ivezi{\'c}}}]{2003AJ....125.1559P}
{Pier}, J.~R., {Munn}, J.~A., {Hindsley}, R.~B., {et~al.} 2003, \aj, 125, 1559

\bibitem[{{Pilyugin}(2000)}]{2000A&A...362..325P}
{Pilyugin}, L.~S. 2000, \aap, 362, 325

\bibitem[{{Pindao} {et~al.}(2002){Pindao}, {Schaerer}, {Gonz{\'a}lez Delgado},
  \& {Stasi{\'n}ska}}]{2002A&A...394..443P}
{Pindao}, M., {Schaerer}, D., {Gonz{\'a}lez Delgado}, R.~M., \&
  {Stasi{\'n}ska}, G. 2002, \aap, 394, 443

\bibitem[{{Prochaska} {et~al.}(2007){Prochaska}, {Chen}, {Dessauges-Zavadsky},
  \& {Bloom}}]{2007ApJ...666..267P}
{Prochaska}, J.~X., {Chen}, H.-W., {Dessauges-Zavadsky}, M., \& {Bloom}, J.~S.
  2007, \apj, 666, 267

\bibitem[{{S{\'a}nchez-Bl{\'a}zquez} {et~al.}(2006){S{\'a}nchez-Bl{\'a}zquez},
  {Peletier}, {Jim{\'e}nez-Vicente}, {Cardiel}, {Cenarro},
  {Falc{\'o}n-Barroso}, {Gorgas}, {Selam}, \& {Vazdekis}}]{2006MNRAS.371..703S}
{S{\'a}nchez-Bl{\'a}zquez}, P., {Peletier}, R.~F., {Jim{\'e}nez-Vicente}, J.,
  {et~al.} 2006, \mnras, 371, 703

\bibitem[{{Schaerer}(1996)}]{1996ApJ...467L..17S}
{Schaerer}, D. 1996, \apjl, 467, L17

\bibitem[{{Schaerer} {et~al.}(1999{\natexlab{a}}){Schaerer}, {Contini}, \&
  {Kunth}}]{1999A&A...341..399S}
{Schaerer}, D., {Contini}, T., \& {Kunth}, D. 1999{\natexlab{a}}, \aap, 341,
  399

\bibitem[{{Schaerer} {et~al.}(1999{\natexlab{b}}){Schaerer}, {Contini}, \&
  {Pindao}}]{1999A&AS..136...35S}
{Schaerer}, D., {Contini}, T., \& {Pindao}, M. 1999{\natexlab{b}}, \aaps, 136,
  35

\bibitem[{{Schaerer} \& {Vacca}(1998)}]{1998ApJ...497..618S}
{Schaerer}, D. \& {Vacca}, W.~D. 1998, \apj, 497, 618

\bibitem[{{Schlegel} {et~al.}(1998){Schlegel}, {Finkbeiner}, \&
  {Davis}}]{1998ApJ...500..525S}
{Schlegel}, D.~J., {Finkbeiner}, D.~P., \& {Davis}, M. 1998, \apj, 500, 525

\bibitem[{{Shapley} {et~al.}(2003){Shapley}, {Steidel}, {Pettini}, \&
  {Adelberger}}]{2003ApJ...588...65S}
{Shapley}, A.~E., {Steidel}, C.~C., {Pettini}, M., \& {Adelberger}, K.~L. 2003,
  \apj, 588, 65

\bibitem[{{Shimasaku} {et~al.}(2001){Shimasaku}, {Fukugita}, {Doi}, {Hamabe},
  {Ichikawa}, {Okamura}, {Sekiguchi}, {Yasuda}, {Brinkmann}, {Csabai},
  {Ichikawa}, {Ivezi{\'c}}, {Kunszt}, {Schneider}, {Szokoly}, {Watanabe}, \&
  {York}}]{2001AJ....122.1238S}
{Shimasaku}, K., {Fukugita}, M., {Doi}, M., {et~al.} 2001, \aj, 122, 1238

\bibitem[{{Smith} {et~al.}(2002{\natexlab{a}}){Smith}, {Tucker}, {Kent},
  {Richmond}, {Fukugita}, {Ichikawa}, {Ichikawa}, {Jorgensen}, {Uomoto},
  {Gunn}, {Hamabe}, {Watanabe}, {Tolea}, {Henden},
  {et~al.}}]{2002AJ....123.2121S}
{Smith}, J.~A., {Tucker}, D.~L., {Kent}, S., {et~al.} 2002{\natexlab{a}}, \aj,
  123, 2121

\bibitem[{{Smith} {et~al.}(1996){Smith}, {Shara}, \&
  {Moffat}}]{1996MNRAS.281..163S}
{Smith}, L.~F., {Shara}, M.~M., \& {Moffat}, A.~F.~J. 1996, \mnras, 281, 163

\bibitem[{{Smith} {et~al.}(2002{\natexlab{b}}){Smith}, {Norris}, \&
  {Crowther}}]{2002MNRAS.337.1309S}
{Smith}, L.~J., {Norris}, R.~P.~F., \& {Crowther}, P.~A. 2002{\natexlab{b}},
  \mnras, 337, 1309

\bibitem[{{Steidel} {et~al.}(2003){Steidel}, {Adelberger}, {Shapley},
  {Pettini}, {Dickinson}, \& {Giavalisco}}]{2003ApJ...592..728S}
{Steidel}, C.~C., {Adelberger}, K.~L., {Shapley}, A.~E., {et~al.} 2003, \apj,
  592, 728

\bibitem[{{Steidel} {et~al.}(2001){Steidel}, {Pettini}, \&
  {Adelberger}}]{2001ApJ...546..665S}
{Steidel}, C.~C., {Pettini}, M., \& {Adelberger}, K.~L. 2001, \apj, 546, 665

\bibitem[{{Tremonti} {et~al.}(2004){Tremonti}, {Heckman}, {Kauffmann},
  {Brinchmann}, {Charlot}, {White}, {Seibert}, {Peng}, {Schlegel}, {Uomoto},
  {Fukugita}, \& {Brinkmann}}]{2004ApJ...613..898T}
{Tremonti}, C.~A., {Heckman}, T.~M., {Kauffmann}, G., {et~al.} 2004, \apj, 613,
  898

\bibitem[{{Tucker} {et~al.}(2006){Tucker}, {Kent}, {Richmond}, {Annis},
  {Smith}, {Allam}, {Rodgers}, {Stute}, {Adelman-McCarthy}, {Brinkmann}, {Doi},
  {Finkbeiner}, {Fukugita}, {Goldston}, {Greenway}, {Gunn}, {Hendry}, {Hogg},
  {Ichikawa}, {Ivezi{\'c}}, {Knapp}, {Lampeitl}, {Lee}, {Lin}, {McKay},
  {Merrelli}, {Munn}, {Neilsen}, {Newberg}, {Richards}, {Schlegel},
  {Stoughton}, {Uomoto}, \& {Yanny}}]{2006AN....327..821T}
{Tucker}, D.~L., {Kent}, S., {Richmond}, M.~W., {et~al.} 2006, Astronomische
  Nachrichten, 327, 821

\bibitem[{{Van Bever} \& {Vanbeveren}(2003)}]{2003A&A...400...63V}
{Van Bever}, J. \& {Vanbeveren}, D. 2003, \aap, 400, 63

\bibitem[{{van der Hucht}(2001)}]{2001NewAR..45..135V}
{van der Hucht}, K.~A. 2001, New Astronomy Review, 45, 135

\bibitem[{{Vanbeveren} {et~al.}(2007){Vanbeveren}, {Van Bever}, \&
  {Belkus}}]{2007ApJ...662L.107V}
{Vanbeveren}, D., {Van Bever}, J., \& {Belkus}, H. 2007, \apjl, 662, L107

\bibitem[{{V{\'a}zquez} {et~al.}(2007){V{\'a}zquez}, {Leitherer}, {Schaerer},
  {Meynet}, \& {Maeder}}]{2007ApJ...663..995V}
{V{\'a}zquez}, G.~A., {Leitherer}, C., {Schaerer}, D., {Meynet}, G., \&
  {Maeder}, A. 2007, \apj, 663, 995

\bibitem[{{Vink} \& {de Koter}(2005)}]{2005A&A...442..587V}
{Vink}, J.~S. \& {de Koter}, A. 2005, \aap, 442, 587

\bibitem[{{Woosley} \& {Bloom}(2006)}]{2006ARA&A..44..507W}
{Woosley}, S.~E. \& {Bloom}, J.~S. 2006, \araa, 44, 507

\bibitem[{{Yin} {et~al.}(2007){Yin}, {Liang}, {Hammer}, {Brinchmann}, {Zhang},
  {Deng}, \& {Flores}}]{2007A&A...462..535Y}
{Yin}, S.~Y., {Liang}, Y.~C., {Hammer}, F., {et~al.} 2007, \aap, 462, 535

\bibitem[{{York} {et~al.}(2000)}]{2000AJ....120.1579Y}
{York}, D.~G. {et~al.} 2000, \aj, 120, 1579

\bibitem[{{Zhang} {et~al.}(2007){Zhang}, {Kong}, {Li}, {Zhou}, \&
  {Cheng}}]{2007ApJ...655..851Z}
{Zhang}, W., {Kong}, X., {Li}, C., {Zhou}, H.-Y., \& {Cheng}, F.-Z. 2007, \apj,
  655, 851

\end{thebibliography}
\bibliographystyle{aa}

%
%

\begin{appendix}

\section{Pipeline processing of the spectra}
\label{sec:pipelining}

The spectra of the DR6 have been re-analysed in a two-step process. 
First, all the spectra were re-analysed using the
pipeline outlined by \citet[][hereafter T04]{2004ApJ...613..898T}
and also discussed in \citet[][hereafter B04]{2004MNRAS.351.1151B}
and summarised below.  This provides good first estimates of continuum
features and emission lines in the SDSS spectra.

However, because spectra showing Wolf-Rayet features often originate
in galaxies with particularly elevated star formation activity, the
spectra of interest to us are more complex than the vast majority of
the SDSS spectra and we have therefore reprocessed the spectra using a
new pipeline to more accurately estimate emission line fluxes for very
strong lines and also to measure many more lines than is
normally done, something that would be very inefficient using the
standard pipeline. Furthermore the normal pipeline reduction applies a
smooth continuum correction to adjust for errors in the
spectrophotometric calibration of spectra, but this smooth correction
might indeed obliterate any Wolf-Rayet feature so we have modified
this procedure somewhat as discussed below.


The pipeline, \texttt{platefit}, developed by T04 proceeds by first
calculating a fit to the absorption line spectrum of a
galaxy. Subsequently a set of strong emission lines are fit to the
absorption line subtracted spectrum. For this latter process an
additional smooth component is calculated to adjust for any
non-perfect spectrophotometric calibration of the spectrum.

The key feature of the emission line fit is that the forbidden and
Balmer lines are grouped in two groups and for each of these groups
the line-width and velocity offset of the lines are assumed to be the
same for all lines. This enables us to measure weak lines that
otherwise would not be accurately constrained.

The disadvantage of the technique is that the multi-parameter fit
might occasionally reach a local minimum rather than the global best
fit solution. This problem worsens as more lines are included in the
fit because the search space for the minimisation problem becomes
higher dimensional.

Thus we have developed a second pipeline which takes the width and
velocity offset of the forbidden and Balmer lines from the
T04 pipelines and carries out fits of a larger set of emission
lines. We adopt the Balmer line widths for the Hydrogen and Helium
recombination lines and the forbidden line widths for all the
forbidden lines. The lines superposed on the broad Wolf-Rayet features
are treated separately as discussed further below.

All the lines considered are then fit jointly where line widths
are set to the previously determined values from the T04 fit --- this
allows us to simplify the fitting routine and get good convergence for
the joint fits. Occasionally the joint fit might not give a perfect
fit for very strong lines, so for each line that has S/N$>10$ we redo
the fit, this time fitting the line (or where appropriate, a blend)
with the line position freely determined as well as the continuum
level, we also do the fit allowing the line-width to vary freely and
choose the statistically best fit given the number of free parameters.
This multi-step process gives excellent fits to all the emission lines
considered here and can be applied automatically to all DR6 spectra.

\section{Fitting models to colours and EW(H$\beta$)}
\label{sec:fitting}

As mentioned in the text, the spectra in our sample do not represent
single burst systems. This means that we need to consider more complex
star formation histories but because the Wolf-Rayet phases are very
short-lasting we also need high time resolution. 

To achieve this we calculated combinations of smoothly varying star
formation histories with superposed bursts using the single stellar
population models discussed in the text. We constructed a grid of
models covering the observed space in $u-g$, $g-r$, $r-i$ and
EW(H$\beta$). This grid was parametrised by the underlying star
formation history, the time at which a burst starts, $t_\mathrm{B}$,
the duration of the burst $\Delta t$ and the fraction of final mass
formed in the burst, $m_\mathrm{R}$.

We use three underlying star formation histories (SFH), a top-hat burst of
duration 1\,Gyr, an exponentially declining SFH with time-constant
1\,Gyr and an exponentially declining SFH with time-constant
15\,Gyr. The latter is near-identical to a constant SFH. 

We calculate models for burst durations between 1\,Myr and 0.5\,Gyr
increasing the duration by a factor of 5 between each step, and
introduce bursts on top of the underlying SFHs at 1, 2, 4 and 6 Gyr
after commencement of star formation. The bursts contribute from 5\%
to 90\% of the final mass in steps of 2.5\%. All of these models are
calculated for the four metallicities available, $Z=0.004, 0.008,
0.02, 0.05$ and we use a time-step of 0.5\,Myr at times later than
10\,Myr after the commencement of the burst, and follow the evolution
at this resolution until 3\,Gyr after the burst.

We then compare the observed $u$, $g$, $r$, $i$ fluxes and
EW(H$\beta$) to the model predictions and assign each model a
likelihood given as $P=\exp(-\chi^2/2)$, similarly to the approach
taken by \citet{2003MNRAS.341...33K}.

In the text we only require the conversion factor from
$L_{\mathrm{Blue bump}}/L_{\mathrm{H}\beta}$ to
$N_{\mathrm{WR}}/N_{\mathrm{O}}$, and this quantity \emph{is} well
constrained by the model fits. However it should be kept in mind that
this quantity might be subject to significant unknown systematic
uncertainties since we rely on the $L_{\mathrm{Blue bump}}$
distribution observed in the local Universe which does not fully cover
the parameters of our sample galaxies.

On the other hand, to get reasonable constraints on the age of the
burst it is furthermore required to apply a prior, requiring that the
model has to contain enough Wolf-Rayet stars to have EW(Blue
bump)$>1$\,\AA. This is sensitive to model assumptions, and age
estimates are poorly constrained at EW(H$\beta$)$<100$\,\AA\ even with
this prior. We have therefore opted not to use these age estimates in
the text.

\end{appendix}

\end{document}